\documentclass[journal]{IEEEtran}

\usepackage{cite}

\ifCLASSINFOpdf
\else
   \usepackage[dvipdfmx]{graphicx}
\fi
\usepackage[cmex10]{amsmath}
\usepackage{amssymb}
\usepackage{balance}
\usepackage{enumitem}

\newtheorem{definition}{Definition}
\newtheorem{assumption}{Assumption}
\newtheorem{lemma}{Lemma}
\newtheorem{proposition}{Proposition}
\newtheorem{remark}{Remark}
\newtheorem{theorem}{Theorem}
\newtheorem{corollary}{Corollary}
\newcommand{\dto}{\overset{\mathrm{d}}{\to}}
\newcommand{\aeq}{\overset{\mathrm{a.s.}}{=}}
\newcommand{\al}{\overset{\mathrm{a.s.}}{<}}
\newcommand{\aleq}{\overset{\mathrm{a.s.}}{\leq}}
\newcommand{\ato}{\overset{\mathrm{a.s.}}{\to}}
\newcommand{\ageq}{\overset{\mathrm{a.s.}}{\geq}}
\newcommand{\ag}{\overset{\mathrm{a.s.}}{>}}

\begin{document}
%
\title{Rigorous Dynamics of Expectation-Propagation-Based Signal Recovery from Unitarily Invariant Measurements}
%
%
%

\author{Keigo~Takeuchi,~\IEEEmembership{Member,~IEEE}
\thanks{
Manuscript received April *, 2017. 
The author was in part supported by the Grant-in-Aid for Exploratory Research 
(JSPS KAKENHI Grant Number 15K13987), Japan and by the Grant-in-Aid 
for Scientific Research~(B) (JSPS KAKENHI Grant Number 18H01441), Japan. 
The material in this paper was presented in part at 2017 IEEE 
International Symposium on Information Theory, Aachen, Germany, Jun.\ 
2017~\cite{Takeuchi17}. 
}
\thanks{K.~Takeuchi is with the Department of Electrical and Electronic Information Engineering, Toyohashi University of Technology, Aichi 441-8580, Japan (e-mail: takeuchi@ee.tut.ac.jp).}
}

\markboth{IEEE transactions on information theory,~Vol.~, No.~,}%
{Takeuchi: Rigorous Dynamics of Expectation-Propagation-Based Signal Recovery from Unitarily Invariant Measurements}
%

\IEEEpubid{0000--0000/00\$00.00~\copyright~2017 IEEE}


\maketitle

\begin{abstract}
Signal recovery from unitarily invariant measurements is investigated 
in this paper. A message-passing algorithm is formulated on the basis of 
expectation propagation (EP). A rigorous analysis is presented for the 
dynamics of the algorithm in the large system limit, where both input and 
output dimensions tend to infinity while the compression rate is kept constant. 
The main result is the justification of state evolution (SE) equations 
conjectured by Ma and Ping. This result implies that the EP-based 
algorithm achieves the Bayes-optimal performance that was originally derived 
via a non-rigorous tool in statistical physics and proved partially in a 
recent paper, when the compression rate is larger than a threshold. 
The proof is based on an extension of a conventional conditioning technique 
for the standard Gaussian matrix to the case of the Haar matrix. 
\end{abstract}

\begin{IEEEkeywords}
Compressed sensing, expectation propagation, unitarily invariant measurements, 
state evolution, Haar matrices. 
\end{IEEEkeywords}

%
\IEEEpeerreviewmaketitle

\section{Introduction}
\subsection{Motivation}
\IEEEPARstart{C}{onsider} the recovery problem of an $N$-dimensional signal 
vector $\boldsymbol{x}$ from a compressed noisy measurement vector 
$\boldsymbol{y}\in\mathbb{C}^{M}$ ($M\leq N$)~\cite{Donoho06,Candes06},  
\begin{equation} \label{system}
\boldsymbol{y} 
= \boldsymbol{A}\boldsymbol{x} + \boldsymbol{w}. 
\end{equation} 
In (\ref{system}), $\boldsymbol{A}\in\mathbb{C}^{M\times N}$ denotes 
a known measurement matrix. The signal vector $\boldsymbol{x}$ is an unknown  
sparse\footnote{
In this paper, a signal $x\in\mathbb{R}$ is called sparse if 
the R\'enyi information dimension~\cite{Wu10} of $x$ is smaller than $1$. 
If $x$ is zero with probability $1-p$, the information dimension is 
at most $p$. If $x$ is discrete, it is zero. 
} vector that is composed of independent and identically distributed (i.i.d.) 
elements. The noise vector $\boldsymbol{w}\in\mathbb{C}^{M}$ is independent of 
the other random variables. 
The goal of compressed sensing is to recovery the sparse vector 
$\boldsymbol{x}$ from the knowledge about $\boldsymbol{y}$ and 
$\boldsymbol{A}$, as well as the statistics of all random variables. 

A breakthrough for signal recovery is to construct message-passing (MP) that is 
Bayes-optimal in the large system limit, where the input and output 
dimensions $N$ and $M$ tend to infinity while the compression rate 
$\delta=M/N$ is kept constant. 
The origin of this approach dates back to the Thouless-Anderson-Palmer (TAP) 
equation~\cite{Thouless77} in statistical physics. Motivated by the TAP 
approach, Kabashima~\cite{Kabashima03} proposed an MP algorithm 
based on approximate belief propagation (BP) in the context of code-division 
multiple-access (CDMA) systems with i.i.d.\ zero-mean measurement matrices. 
When the compression rate is larger than the so-called BP 
threshold~\cite{Takeuchi15}, 
the BP-based algorithm was numerically shown to achieve the Bayes-optimal 
performance in the large system limit, which was originally conjectured by 
Tanaka~\cite{Tanaka02} via the replica method---a non-rigorous tool in 
statistical physics, and proved in \cite{Reeves16,Barbier16} for i.i.d.\ 
zero-mean Gaussian measurements. However, Kabashima~\cite{Kabashima03} 
presented no rigorous analysis on the convergence property of the BP-based 
algorithm. 

In order to resolve lack of a rigorous proof, approximate message-passing 
(AMP) was proposed in \cite{Donoho09} and proved in \cite{Bayati11} to 
achieve the optimal performance for i.i.d.\ zero-mean Gaussian measurements, 
when the compression rate is larger than the BP threshold. Spatially coupled 
measurement matrices are required for achieving the optimal performance 
in the whole regime~\cite{Kudekar11,Krzakala12,Donoho13,Takeuchi15}. 
However, it is recognized that AMP fails to converge when the i.i.d.\ 
zero-mean assumption of measurement matrices is broken~\cite{Caltagirone14}, 
unless damping~\cite{Rangan14} is employed. 

As solutions to this convergence issue, since Opper and Winther's 
pioneering work~\cite[Appendix~D]{Opper05}, as well as \cite{Opper01}, 
several algorithms have been proposed on the basis of expectation 
propagation (EP)~\cite{Cespedes14}, expectation consistent (EC) 
approximations~\cite{Opper05,Kabashima14,Fletcher16}, 
S-transform~\cite{Cakmak14}, vector AMP~\cite{Rangan17}, or 
turbo principle~\cite{Yuan13,Ma15,Ma17,Liu16}. The EP-based 
algorithm~\cite{Cespedes14} is systematically derived from Minka's EP 
framework~\cite{Minka01} by approximating the posterior distribution of 
$\boldsymbol{x}$ with factorized Gaussian distributions. 
The EC-based algorithms~\cite{Opper05,Kabashima14,Fletcher16} are iterative 
algorithms for solving a fixed point (FP) of the EC free energy. 
An algorithm in \cite{Cakmak14} is derived via the S-transform of 
$\boldsymbol{A}^{\mathrm{H}}\boldsymbol{A}$. 
Rangan {\it et al.}~\cite{Rangan17} considered an 
EP-like approximation of the BP algorithm on a factor graph with vector-valued 
nodes. The algorithms~\cite{Yuan13,Ma15,Ma17,Liu16} 
based on turbo principle are derived from a few heuristic assumptions. 
Interestingly, the algorithms in 
\cite{Opper05,Cespedes14,Fletcher16,Rangan17,Ma17} are 
essentially equivalent, with the exception of \cite{Kabashima14,Cakmak14}. 
In this paper, these algorithms for signal recovery 
are simply referred to as EP-based algorithms, 
since we follow the EP-based derivation in \cite{Cespedes14}.   

\IEEEpubidadjcol

Ma {\it et al}.~\cite{Ma15,Ma17} derived state evolution (SE) equations 
of an EP-based algorithm under two heuristic assumptions. By investigating the 
properties of the SE equations, they conjectured that, for unitarily invariant 
measurement matrices, the FPs of the SE equations are the same as the extrema  
of an asymptotic energy function that describes the Bayes-optimal performance 
in the large system limit. The energy function was originally derived in 
\cite{Takeda06,Tulino13} via the replica method, and proved for bounded 
signals in \cite{Barbier18}. In other words, the EP-based 
algorithm was conjectured to achieve the optimal performance in the large 
system limit, when the compression rate is larger than the BP threshold. 
Since the algorithm attempts to solve the minimum of the EC free 
energy~\cite{Opper05}, it is conjectured that the extrema of the EC free energy 
correspond to those of the Bayes-optimal one for unitarily invariant 
measurement matrices. 
The purpose of this paper is to present a rigorous proof for the conjecture.

\subsection{Proof Strategy} \label{sec1b}
The proof strategy is based on a conditioning technique used in 
\cite{Bayati11}, originally proposed by Bolthausen~\cite{Bolthausen14}. 
A challenging part in the proof is to evaluate the 
distribution of an estimation error in each iteration conditioned on 
estimation errors in all preceding iterations. Bayati and 
Montanari~\cite{Bayati11} evaluated the conditional distribution via 
the distribution of the measurement matrix $\boldsymbol{A}$ conditioned on 
the estimation errors in all preceding iterations. When linear detection is 
employed as part of MP, the conditional distribution of 
$\boldsymbol{A}$ can be regarded as the posterior distribution of 
$\boldsymbol{A}$ given linear, noiseless, and compressed observations 
of $\boldsymbol{A}$, determined by the estimation errors in all preceding 
iterations. For i.i.d.\ Gaussian measurement matrices, it is well known that 
the posterior distribution is also Gaussian. The proof in \cite{Bayati11} 
heavily relies on this well-known fact. 

In order to present our proof strategy, assume $M=N$, and that 
$\boldsymbol{A}$ is a Haar matrix~\cite{Hiai00,Tulino04}, which is uniformly 
distributed on the space of all possible $N\times N$ unitary matrices. 
Under appropriate coordinate rotations in the column spaces of 
$\boldsymbol{A}$, it is possible to show that the linear, 
noiseless, and compressed observation of $\boldsymbol{A}$ is equivalent to 
observing {\em part} of the columns in $\boldsymbol{A}$. Since any Haar 
matrix is bi-unitarily invariant~\cite{Hiai00}, the distribution of 
$\boldsymbol{A}$ after the coordinate rotations is the same as the original 
one. Thus, evaluating the conditional distribution of $\boldsymbol{A}$ 
reduces to analyzing the conditional distribution of a Haar matrix given part 
of its columns. This argument was implicitly used in \cite{Bayati11}.   

Evaluation of this conditional distribution is an important part  
in this paper, while this part is not required for i.i.d.\ Gaussian 
measurements. For simplicity, let $N=3$ and fix the first column of 
a Haar matrix $\boldsymbol{A}=(\boldsymbol{a}_{1}, \boldsymbol{a}_{2}, 
\boldsymbol{a}_{3})$. Evaluation of the conditional distribution is equivalent 
to characterizing $\boldsymbol{a}_{2}$ and $\boldsymbol{a}_{3}$ for 
given $\boldsymbol{a}_{1}$. The two vectors must be on a plane perpendicular 
to $\boldsymbol{a}_{1}$. From the orthonormality between 
$\boldsymbol{a}_{2}$ and $\boldsymbol{a}_{3}$, the two vectors are on a unit 
circle that has the center at the intersection of the plane and 
$c\boldsymbol{a}_{1}$ for $c\in\mathbb{C}$.  
Intuitively, $\boldsymbol{a}_{2}$ and $\boldsymbol{a}_{3}$ should be 
Haar-distributed on this unit circle. Generalizing this 
intuition, we find that $\boldsymbol{A}$ given its first $t$ columns should 
have degrees of freedom that are equal to those of an $(N-t)\times(N-t)$ Haar 
matrix. On the basis of this intuition, we evaluate the conditional 
distribution of $\boldsymbol{A}$.

\subsection{Related Work}
A similar paper~\cite{Rangan16} was posted on the arXiv a few months before 
posting the first version~\cite{Takeuchi16} of this paper. Short versions of 
the two papers were published in \cite{Rangan17, Takeuchi17}. 
The posted paper~\cite{Rangan16} addressed real-valued systems, while  
we consider complex-valued systems. Interestingly, the two papers share 
the common proof strategy based on \cite{Bayati11}. However, there is a 
mathematically critical difference between them. 

The main difference is in mathematical treatments on almost sure 
convergence. An empirical convergence based on pseudo-Lipschitz 
functions was considered in \cite{Rangan16}. The approach allows us to 
analyze general Lipschitz-continuous decision functions and general 
pseudo-Lipschitz performance measures, as considered in \cite{Bayati11}. 
However, Rangan {\em et al}.~\cite{Rangan16} omitted the proof of an important 
part on almost sure convergence---required in establishing the empirical 
convergence based on pseudo-Lipschitz functions---as pointed out in 
Appendix~\ref{appen_discussion}.   

In this paper, we present a rigorous proof of the part on almost sure 
convergence. Our approach relies on advanced results in probability theory, 
such as the strong law of large numbers for dependent random 
variables~\cite{Lyons88} and statistical properties of a Haar matrix. 
While this paper considers a Bayes-optimal decision function and the 
mean-square error (MSE), the part on almost sure convergence is proved in a 
general form, as considered in \cite{Rangan16}. Thus, combining 
\cite{Rangan16} and this paper establishes a rigorous proof of SE for 
general decision functions and general performance measures. 

\subsection{Contributions}
The main contribution is the rigorous justification of the 
SE equations for the EP-based algorithm, conjectured in \cite{Ma17}. 
More precisely, we derive SE equations for individual elements of the signal 
vector in the large system limit.  
This implies the achievability of the Bayes-optimal performance proved 
in \cite{Barbier18}, when the compression rate is larger than the BP 
threshold, while the converse theorem is partially open, i.e.\ there are no 
algorithms outperforming the EP-based algorithm in the large system limit 
when unbounded signals are considered. 

The technical novelty is in an extension of the conditioning technique in 
\cite{Bayati11} for i.i.d.\ Gaussian measurement matrices to the case of 
Haar matrices. This paper presents a constructive proof for the conditional 
distribution of a Haar matrix. 
The proposed conditioning technique is applicable to any 
MP algorithm for signal recovery from unitarily invariant 
measurements, such as the AMP, unless the algorithm contains 
nonlinear processing in the measurement vector $\boldsymbol{y}$, e.g.\ 
quantization~\cite{Liu16}. However, whether the obtained SE becomes  
simple depends on the MP algorithm and the statistics of 
$\boldsymbol{A}$~\cite{Takeuchi19}. Thus, it is an important future work 
to design a low-complexity MP algorithm such that simple SE equations are 
obtained for unitarily invariant measurements.   

\subsection{Organization}
The remainder of this paper is organized as follows: After summarizing the 
notation used in this paper, Section~\ref{sec2} presents the definition of 
unitarily invariant matrices and technical results associated with 
Haar matrices. In Section~\ref{sec3}, we introduce assumptions used throughout 
this paper, and then formulate an EP-based algorithm. The main result is 
presented in Section~\ref{sec4}, and proved in Section~\ref{sec5}. 
Several technical results are proved in appendices.  

\subsection{Notation} \label{sec1E}
The notation $\boldsymbol{o}(1)$ denotes a 
vector of which the Euclidean norm converges almost surely toward zero 
in the large system limit. 
For a vector $\boldsymbol{v}\in\mathbb{C}^{N}$, we write the $n$th element 
of $\boldsymbol{v}$ as $v_{n}$. 
For a subset $\mathcal{N}\subset\{1,\ldots,N\}$, the vector 
$\boldsymbol{x}_{\mathcal{N}}$ consists of the elements 
$\{x_{n}: n\in\mathcal{N}\}$, while $\boldsymbol{x}_{\backslash\mathcal{N}}$ 
is obtained by eliminating $\{x_{n}: n\in\mathcal{N}\}$ from 
$\boldsymbol{x}$. For a scalar function 
$f:\mathbb{C}\to\mathbb{C}$, we introduce a convention 
in which $f(\boldsymbol{v})$ denotes the vector obtained by the 
component-wise application of $f$ to $\boldsymbol{v}$, i.e.\ 
$[f(\boldsymbol{v})]_{n}=f(v_{n})$.

For a complex number $z\in\mathbb{C}$ and a matrix 
$\boldsymbol{M}\in\mathbb{C}^{M\times N}$, 
the complex conjugate, transpose, and the conjugate transpose are denoted by 
$z^{*}$, $\boldsymbol{M}^{\mathrm{T}}$, and $\boldsymbol{M}^{\mathrm{H}}$. 
We write the $(m, n)$th element of $\boldsymbol{M}$ as $M_{mn}$. When 
$\boldsymbol{M}$ is Hermitian, $\lambda_{\mathrm{min}}(\boldsymbol{M})$ 
represents the minimum eigenvalue of $\boldsymbol{M}$. For $M\geq N$, 
${\cal U}_{M\times N}$ denotes the space of all possible $M\times N$ matrices 
with orthonormal columns, while ${\cal U}_{M\times N}$ for $M<N$ represents the 
space of all possible $M\times N$ matrices with orthonormal rows.  
When $M=N$ holds, ${\cal U}_{M\times N}$ is written as ${\cal U}_{N}$, which is 
the space of all possible $N\times N$ unitary matrices. 

We write the singular-value decomposition (SVD) of $\boldsymbol{M}$ as 
\begin{equation} 
\boldsymbol{M}
=\boldsymbol{\Phi}_{\boldsymbol{M}}
(\boldsymbol{\Sigma}_{\boldsymbol{M}}, \boldsymbol{O})
\boldsymbol{\Psi}_{\boldsymbol{M}}^{\mathrm{H}}  
\end{equation}  
for $M\leq N$, with $\boldsymbol{\Phi}_{\boldsymbol{M}}\in{\cal U}_{M}$ and 
$\boldsymbol{\Psi}_{\boldsymbol{M}}\in{\cal U}_{N}$. Furthermore,  
$\boldsymbol{\Sigma}_{\boldsymbol{M}}$ is an $M\times M$ positive semi-definite 
diagonal matrix. The unitary matrix $\boldsymbol{\Psi}_{\boldsymbol{M}}$ 
is partitioned as 
$\boldsymbol{\Psi}_{\boldsymbol{M}}
=(\boldsymbol{\Psi}_{\boldsymbol{M}}^{\parallel}, 
\boldsymbol{\Psi}_{\boldsymbol{M}}^{\perp})$, in which 
$\boldsymbol{\Psi}_{\boldsymbol{M}}^{\parallel}\in{\cal U}_{N\times M}$
is composed of the first $M$ columns of $\boldsymbol{\Psi}_{\boldsymbol{M}}$, 
while $\boldsymbol{\Psi}_{\boldsymbol{M}}^{\perp}\in{\cal U}_{N\times (N-M)}$  
consists of the remaining columns. For $M>N$, we write the SVD of 
$\boldsymbol{M}$ as 
\begin{equation} 
\boldsymbol{M}
=\boldsymbol{\Phi}_{\boldsymbol{M}}
\begin{pmatrix}
\boldsymbol{\Sigma}_{\boldsymbol{M}} \\
 \boldsymbol{O}
\end{pmatrix}
\boldsymbol{\Psi}_{\boldsymbol{M}}^{\mathrm{H}},   
\end{equation}  
with $\boldsymbol{\Phi}_{\boldsymbol{M}}\in{\cal U}_{M}$ and 
$\boldsymbol{\Psi}_{\boldsymbol{M}}\in{\cal U}_{N}$. Furthermore,  
$\boldsymbol{\Sigma}_{\boldsymbol{M}}$ is an $N\times N$ positive semi-definite 
diagonal matrix. The unitary matrix $\boldsymbol{\Phi}_{\boldsymbol{M}}
=(\boldsymbol{\Phi}_{\boldsymbol{M}}^{\parallel}, 
\boldsymbol{\Phi}_{\boldsymbol{M}}^{\perp})$ is partitioned in 
the same manner as for $M\leq N$. 

When $\boldsymbol{M}$ is full rank, the pseudo-inverse of $\boldsymbol{M}$ 
is denoted by $\boldsymbol{M}^{\dagger}=(\boldsymbol{M}^{\mathrm{H}}
\boldsymbol{M})^{-1}\boldsymbol{M}^{\mathrm{H}}\in\mathbb{C}^{N\times M}$ 
for $M> N$. Let $\boldsymbol{P}_{\boldsymbol{M}}^{\parallel}$ denote the 
orthogonal projection matrix onto the space spanned by the columns of 
$\boldsymbol{M}$. We have 
$\boldsymbol{P}_{\boldsymbol{M}}^{\parallel} 
= \boldsymbol{\Phi}_{\boldsymbol{M}}^{\parallel}
(\boldsymbol{\Phi}_{\boldsymbol{M}}^{\parallel})^{\mathrm{H}}
=\boldsymbol{M}\boldsymbol{M}^{\dagger}$. The projection matrix 
$\boldsymbol{P}_{\boldsymbol{M}}^{\perp}$ onto the orthogonal complement is 
given by $\boldsymbol{P}_{\boldsymbol{M}}^{\perp}
=\boldsymbol{I}_{M} - \boldsymbol{P}_{\boldsymbol{M}}^{\parallel}$. 
For $M\leq N$, we define $\boldsymbol{M}^{\dagger}=\boldsymbol{M}^{\mathrm{H}}
(\boldsymbol{M}\boldsymbol{M}^{\mathrm{H}})^{-1}$, 
$\boldsymbol{P}_{\boldsymbol{M}}^{\parallel}
=\boldsymbol{\Psi}_{\boldsymbol{M}}^{\parallel}
(\boldsymbol{\Psi}_{\boldsymbol{M}}^{\parallel})^{\mathrm{H}}
=\boldsymbol{M}^{\dagger}\boldsymbol{M}$, and 
$\boldsymbol{P}_{\boldsymbol{M}}^{\perp}=\boldsymbol{I}_{N}
-\boldsymbol{P}_{\boldsymbol{M}}^{\parallel}$. 

The proper complex Gaussian distribution with mean $\boldsymbol{m}$ and 
covariance $\boldsymbol{\Sigma}$ is denoted by 
$\mathcal{CN}(\boldsymbol{m}, \boldsymbol{\Sigma})$. 
The expectation and variance of a random variable $X$ is denoted by 
$\mathbb{E}[X]$ and $\mathbb{V}[X]$, respectively. 
The notation $X\aeq Y$ means that $X$ is 
almost surely equal to $Y$. Similarly, $\ato$, $\ageq$, and 
$\aleq$ indicate that $\to$, $\geq$, and $\leq$ hold almost surely. 
The notation $X\sim Y$ means that $X$ follows the same distribution as $Y$.  
The notation $X|_{Y}$ indicates that we focus on the 
conditional distribution of $X$ given $Y$.

\section{Preliminaries} \label{sec2}
\subsection{Definitions} 
The purpose of this section is to present the strong law of large numbers 
for a Haar matrix. The result corresponds to 
\cite[Lemma 2]{Bayati11} for an i.i.d.\ Gaussian matrix. 
We first present several definitions. 

\begin{definition}
A unitary random matrix $\boldsymbol{U}\in{\cal U}_{n}$ is called a Haar 
matrix if $\boldsymbol{U}$ is uniformly distributed on ${\cal U}_{n}$. 
\end{definition}

An important property of a Haar matrix is bi-unitary 
invariance~\cite{Tulino04}---used throughout this paper. 

\begin{definition}
A random matrix $\boldsymbol{M}$ is said to be bi-unitarily invariant if 
$\boldsymbol{M}\sim\boldsymbol{U}\boldsymbol{M}\boldsymbol{V}$ 
holds for all deterministic unitary matrices $\boldsymbol{U}$ and 
$\boldsymbol{V}$.  
\end{definition}

In this paper, the functions $z^{*}f(x+z)$ and $|x-f(x+z)|^{2}$ of 
$x\in\mathbb{C}$ and $z\in\mathbb{C}$ are considered 
for a Lipschitz-continuous function $f:\mathbb{C}\to\mathbb{C}$. 
To characterize these functions, we follow \cite{Bayati11} to define 
pseudo-Lipschitz functions. 

\begin{definition}
For $k\geq1$, we say that a function $f:\mathbb{C}^{n}\to\mathbb{C}$ is 
pseudo-Lipschitz of order $k$ if there is some Lipschitz constant $L>0$ 
such that 
\begin{equation} \label{pseudo_Lipschitz}
\|f(\boldsymbol{x}) - f(\boldsymbol{y})\| 
\leq L\|\boldsymbol{x}-\boldsymbol{y}|(1+\|\boldsymbol{x}\|^{k-1} 
+ \|\boldsymbol{y}\|^{k-1})
\end{equation}   
holds for all $\boldsymbol{x}, \boldsymbol{y}\in\mathbb{C}^{n}$. 
\end{definition}

Note that any pseudo-Lipschitz function of order~$1$ is Lipschitz-continuous.  
A pseudo-Lipschitz function $f(\boldsymbol{x})$ of order~$k$ is 
${\cal O}(\|\boldsymbol{x}\|^{k})$ as$\|\boldsymbol{x}\|\to\infty$. 

The following proposition is used for further evaluation of the upper 
bound~(\ref{pseudo_Lipschitz}) throughout this paper. 

\begin{proposition} \label{proposition_bound} 
For any $k\geq1$, there is some constant $C>0$ such that 
\begin{equation}
(a+b)^{k}\leq C(a^{k}+b^{k})
\end{equation}
holds for all $a\geq0$ and $b\geq0$. 
\end{proposition}
\begin{IEEEproof}
The inequality follows from a general upper bound $\|\cdot\|_{1}\leq
2^{1-1/k}\|\cdot\|_{k}$ on $\mathbb{C}^{2}$. 
\end{IEEEproof}

\subsection{Results}
We consider an array $\{\boldsymbol{X}_{N}\in\mathbb{C}^{N}\}_{N=1}^{\infty}$ of 
dependent random variables 
$\boldsymbol{X}_{N}=(X_{1,N},\ldots,X_{N,N})^{\mathrm{T}}$. 
An array $\{\boldsymbol{X}_{N}\}$ allows the distribution of each element 
$X_{n,N}$ to change as $N$ grows, while the distribution of each element 
is fixed in a sequence $\boldsymbol{X}\in\mathbb{C}^{N}$. 
We first present the strong law of large numbers for an array 
$\{\boldsymbol{X}_{N}\in\mathbb{C}^{N}\}_{N=1}^{\infty}$ of 
dependent random variables.


\begin{theorem}[Lyons~\cite{Lyons88}]
\label{theorem_SLLN}
Let $\{\boldsymbol{X}_{N}\}$ denote an array of complex 
random variables with finite second moments, and define 
$S_{N}=\sum_{n=1}^{N}X_{n, N}$. If the following assumption holds:     
\begin{equation} \label{SLLN_condition}
\sum_{N=1}^{\infty}\frac{\sqrt{\mathbb{V}[S_{N}]}}{N^{2}}
< \infty, 
\end{equation}
the strong law of large numbers for $T_{N}=(S_{N}-\mathbb{E}[S_{N}])/N$ holds, 
i.e.\  $\lim_{N\to\infty}T_{N}\aeq 0$. 
\end{theorem}
\begin{IEEEproof}
Lyons~\cite[Theorem~6]{Lyons88} proved Theorem~\ref{theorem_SLLN} for a 
sequence of complex random variables, i.e.\ $X_{n, N}=X_{n, N'}$ for all 
$N\neq N'$. However, the proof is applicable to the array case with no changes, 
by defining $X_{n,N}=0$ for $n>N$. Thus, Theorem~\ref{theorem_SLLN} holds.  
\end{IEEEproof}

The condition~(\ref{SLLN_condition}) is satisfied when   
$\mathbb{V}[S_{N}]={\cal O}(N^{\alpha})$ holds for some $\alpha<2$. 
In particular, we have $\alpha=1$ when $\{X_{n, N}\}$ are uncorrelated random 
variables. We next present the strong law of large numbers associated with 
a Haar matrix. 

\begin{lemma} \label{lemma_SLLN} 
For $t'\in\mathbb{N}$, 
suppose that $f_{n}:\mathbb{C}^{t'+1}\to\mathbb{C}$ denote a pseudo-Lipschitz 
function of order $k$ with a Lipschitz constant $L_{n}>0$. 
Let $\boldsymbol{\epsilon}_{N}=(\epsilon_{1,N},\ldots,\epsilon_{N,N})^{\mathrm{T}}
\in\mathbb{C}^{N}$ denote a vector that satisfies  
\begin{equation} \label{assumption_a1}
\lim_{N\to\infty}\frac{1}{N}\sum_{n=1}^{N}L_{n}|\epsilon_{n,N}|^{2}\aeq0, 
\end{equation}
\begin{equation} \label{assumption_a2}
\limsup_{N\to\infty}\frac{1}{N}\sum_{n=1}^{N}L_{n}|\epsilon_{n,N}|^{2k-2}\al \infty. 
\end{equation} 
Suppose that $\boldsymbol{a}_{\tau,N}
=(a_{\tau,1,N},\ldots,a_{\tau,N,N})^{\mathrm{T}}\in\mathbb{C}^{N}$ for 
$\tau=0,\ldots,t'$ satisfies 
\begin{equation} \label{assumption_a3}
\limsup_{N\to\infty}\frac{1}{N}\sum_{n=1}^{N}L_{n}^{i}|a_{\tau,n,N}|^{2k-2}\al \infty
\quad \hbox{for $i=1, 2$.}  
\end{equation} 
For $t>0$, let $\boldsymbol{E}_{N}=(\boldsymbol{e}_{1,N}^{\mathrm{T}},\ldots, 
\boldsymbol{e}_{N,N}^{\mathrm{T}})^{\mathrm{T}}\in\mathbb{C}^{N\times t}$ denote 
a matrix that satisfies   
\begin{equation} \label{assumption_E}
\limsup_{N\to\infty}
\frac{1}{N}\sum_{n=1}^{N}L_{n}\|\boldsymbol{e}_{n,N}\|^{\max\{2, 2k-2\}}\al\infty, 
\end{equation}
\begin{equation} \label{norm_E}
\liminf_{N\to\infty}\lambda_{\mathrm{min}}\left(
 \frac{1}{N}\boldsymbol{E}_{N}^{\mathrm{H}}\boldsymbol{E}_{N}
\right) \ag C 
\end{equation}
for some constant $C>0$. 
Suppose that $\{\boldsymbol{X}_{N}\in\mathbb{C}^{N}\}$ is an array of 
unitarily invariant random variables conditioned on 
$\boldsymbol{\epsilon}_{N}$, $\{\boldsymbol{a}_{\tau,N}\}$, and 
$\boldsymbol{E}_{N}$, 
i.e.\ $\boldsymbol{\Phi}\boldsymbol{X}_{N}\sim\boldsymbol{X}_{N}$ 
conditioned on $\boldsymbol{\epsilon}_{N}$, $\{\boldsymbol{a}_{\tau,N}\}$, and 
$\boldsymbol{E}_{N}$ for any 
deterministic unitary matrix $\boldsymbol{\Phi}\in{\cal U}_{N}$. For 
some $v>0$, postulate the following: 
\begin{equation} \label{convergence_X}
\lim_{N\to\infty}\frac{1}{N}\|\boldsymbol{X}_{N}\|^{2}\aeq v>0. 
\end{equation}
Let $\boldsymbol{z}\sim\mathcal{CN}(\boldsymbol{0},\boldsymbol{I}_{N})$ 
denote a standard complex Gaussian random vector.  
Then, the following two properties hold: 
\begin{enumerate}
\item Postulate the following assumptions:
\begin{itemize}
\item $\boldsymbol{\epsilon}_{N}$ has finite $(2k-2)$th moments and 
vanishing second moments, i.e.\ 
$\mathbb{E}[|\epsilon_{n,N}|^{2}]\to0$ as $N\to\infty$. 
\item $\boldsymbol{a}_{\tau,N}$ has finite $(2k-2)$th moments. 
\item $\boldsymbol{E}_{N}$ has finite $\max\{2, 2k-2\}$th moments. 
\item $\boldsymbol{X}_{N}$ has finite $(\max\{2,2k-2\}+\epsilon)$th moments 
for some $\epsilon>0$.
\end{itemize}
Then, for any $t\geq0$ 
\begin{IEEEeqnarray}{rl} 
\lim_{N\to\infty}\mathbb{E}\Bigl[
f_{n}(a_{n,0,N}, \ldots&, a_{n,t'-1,N},  \nonumber \\
&a_{n,t',N}+\epsilon_{n,N}
+[\boldsymbol{\Phi}_{\boldsymbol{E}_{N}}^{\perp}\boldsymbol{X}_{N-t}]_{n})
\Bigr] \nonumber \\ 
= \mathbb{E}\Bigl[
  f_{n}(a_{n,0,N}, \ldots&, a_{n,t'-1,N}, a_{n,t',N}+\sqrt{v}z_{n})
\Bigr], \label{SLLN_individual}
\end{IEEEeqnarray}  
where the convention $\boldsymbol{\Phi}_{\boldsymbol{E}_{N}}^{\perp}
=\boldsymbol{I}_{N}$ is introduced for $t=0$. 

\item If the sequence of Lipschitz constants satisfies 
\begin{equation} \label{Lipschitz_constant}
\frac{1}{N}\sum_{n=1}^{N}L_{n}^{2}<\infty, 
\end{equation}
then for $t\geq0$ 
\begin{IEEEeqnarray}{rl}
\lim_{N\to\infty}\frac{1}{N}\sum_{n=1}^{N}\Bigl\{
f_{n}(&a_{n,0,N}, \ldots, a_{n,t'-1,N}, \nonumber \\ 
&a_{n,t',N}+\epsilon_{n,N}
+[\boldsymbol{\Phi}_{\boldsymbol{E}_{N}}^{\perp}\boldsymbol{X}_{N-t}]_{n})
 \nonumber \\
- \mathbb{E}_{z_{n}}[f_{n}(a_{n,0,N},\ldots,&a_{n,t'-1,N},a_{n,t',N}+\sqrt{v}z_{n})]
\Bigr\} \aeq 0. \nonumber \\
\label{SLLN} 
\end{IEEEeqnarray}  
\end{enumerate}
\end{lemma}
\begin{IEEEproof}
See Appendix~\ref{proof_lemma_SLLN}. 
\end{IEEEproof}


Lemma~\ref{lemma_SLLN} is used repeatedly to prove the main theorem of this 
paper. 
Finally, we prove the following corollary that is used in the derivation of 
the EP-based algorithm. 

\begin{corollary} \label{corollary_Haar_norm} 
Let $\boldsymbol{a}\in\mathbb{C}^{N}$ denote a vector that satisfies  
$\lim_{N\to\infty}N^{-1}\|\boldsymbol{a}\|^{2}\aeq1$. Suppose that 
$\boldsymbol{D}\in\mathbb{C}^{N\times N}$ is a Hermitian matrix with  
$\lim_{N\to\infty}N^{-1}\mathrm{Tr}(\boldsymbol{D}^{i})\aeq d_{i}$ for $i=1, 2$. 
Let $\boldsymbol{V}\in{\cal U}_{N}$ denote a Haar matrix independent of 
$\boldsymbol{a}$ and $\boldsymbol{D}$. Then, 
\begin{equation}
\lim_{N\to\infty}\frac{1}{N}\boldsymbol{a}^{\mathrm{H}}\boldsymbol{V}^{\mathrm{H}}
\boldsymbol{D}\boldsymbol{V}\boldsymbol{a}
\aeq d_{1}. 
\end{equation}
\end{corollary}
\begin{IEEEproof}
Without loss of generality, we can assume that $\boldsymbol{D}$ is 
diagonal since $\boldsymbol{V}$ is a Haar matrix. 
For $\boldsymbol{X}_{N}=\boldsymbol{V}\boldsymbol{a}$, we have 
\begin{equation}
\frac{1}{N}\boldsymbol{a}^{\mathrm{H}}\boldsymbol{V}^{\mathrm{H}}
\boldsymbol{D}\boldsymbol{V}\boldsymbol{a}
= \frac{1}{N}\sum_{n=1}^{N}f_{n}(X_{n,N}), 
\end{equation}
with $f_{n}(z)=D_{n}|z|^{2}$, in which $D_{n}$ denotes the $n$th diagonal 
element of $\boldsymbol{D}$. 
Since $f_{n}$ is a pseudo-Lipschitz function of order~$2$ with the Lipschitz 
constant $|D_{n}|$, the assumptions on $\boldsymbol{a}$ and 
$\boldsymbol{D}$ imply that all assumptions in Lemma~\ref{lemma_SLLN} 
are satisfied with $v=1$. Thus, we use Lemma~\ref{lemma_SLLN} 
to arrive at 
\begin{equation}
\frac{1}{N}\boldsymbol{a}^{\mathrm{H}}\boldsymbol{V}^{\mathrm{H}}
\boldsymbol{D}\boldsymbol{V}\boldsymbol{a}
\aeq \frac{1}{N}\sum_{n=1}^{N}D_{n}\mathbb{E}[|z_{n}|^{2}] + o(1)
\ato d_{1}
\end{equation}
as $N\to\infty$, which implies Corollary~\ref{corollary_Haar_norm}. 
\end{IEEEproof}

\section{System Model} \label{sec3}
\subsection{Assumptions} 
Assumptions on the measurement model~(\ref{system}) are presented. 

\begin{assumption} \label{assumption_x}
The signal vector $\boldsymbol{x}$ is composed of zero-mean i.i.d.\ 
{\em non}-Gaussian elements with unit variance and finite  
$(2+\epsilon)$th moments for some $\epsilon>0$. 
\end{assumption}

From the strong law of large numbers~\cite{Etemadi81}, 
Assumption~\ref{assumption_x} implies that 
$N^{-1}\|\boldsymbol{x}\|^{2}$ converges almost surely to $1$ as $N\to\infty$. 
The i.i.d.\ assumption for $\boldsymbol{x}$ is implicitly used in the 
derivation of an EP-based algorithm. We require no additional assumptions 
for the prior distribution of each element to prove the main theorem, 
whereas it is practically important to postulate some prior distribution 
indicating the sparsity of~$\boldsymbol{x}$.  

\begin{definition}
A Hermitian random matrix $\boldsymbol{M}$ is said to be unitarily invariant if 
$\boldsymbol{M}\sim\boldsymbol{U}\boldsymbol{M}\boldsymbol{U}^{\mathrm{H}}$ 
holds for any deterministic unitary matrix $\boldsymbol{U}$.  
\end{definition}

\begin{assumption} \label{assumption_A}
The measurement matrix $\boldsymbol{A}$ has the following properties: 
\begin{itemize}
\item $\boldsymbol{A}^{\mathrm{H}}\boldsymbol{A}$ is unitarily invariant. 
\item The empirical eigenvalue distribution of 
$\boldsymbol{A}\boldsymbol{A}^{\mathrm{H}}$ converges almost surely to 
a deterministic distribution $\rho(\lambda)$ with a compact support 
in the large system limit.  
\end{itemize}
\end{assumption}

We write the SVD of $\boldsymbol{A}$ as 
\begin{equation} \label{SVD_A}
\boldsymbol{A}=\boldsymbol{U}(\boldsymbol{\Sigma},\boldsymbol{O})
\boldsymbol{V}^{\mathrm{H}}, 
\end{equation} 
with $\boldsymbol{U}\in{\cal U}_{M}$ and $\boldsymbol{V}\in{\cal U}_{N}$. 
Furthermore, $\boldsymbol{\Sigma}$ is an $M\times M$ positive semi-definite 
diagonal matrix. From Assumption~\ref{assumption_A}, 
$\boldsymbol{V}$ is a Haar matrix and independent of 
$\boldsymbol{U}\boldsymbol{\Sigma}$~\cite{Tulino04}.  

\begin{assumption} \label{assumption_w}
The noise vector $\boldsymbol{w}$ has finite $(2+\epsilon)$th 
moments for some $\epsilon>0$. 
Let $\boldsymbol{D}\in\mathbb{C}^{M\times M}$ denote any Hermitian matrix 
such that $\boldsymbol{D}$ is independent of 
$\boldsymbol{U}^{\mathrm{H}}\boldsymbol{w}$, and that 
$M^{-1}\mathrm{Tr}(\boldsymbol{D}^{2})$ converges almost surely as 
$M\to\infty$. Then,  
\begin{equation}
\lim_{M\to\infty}\frac{1}{M}\left\{
 \boldsymbol{w}^{\mathrm{H}}\boldsymbol{U}
 \boldsymbol{D}\boldsymbol{U}^{\mathrm{H}}\boldsymbol{w} 
 - \sigma^{2}\mathrm{Tr}(\boldsymbol{D})
\right\}
\aeq 0. 
\end{equation} 
\end{assumption}

Assumption~\ref{assumption_w} implies that $\sigma^{2}$ corresponds to 
the noise power $\sigma^{2}\aeq\lim_{M\to\infty}M^{-1}\|\boldsymbol{w}\|^{2}$ 
per element, by selecting $\boldsymbol{D}=\boldsymbol{I}_{M}$. 
Assumption~\ref{assumption_w} 
is satisfied if $\boldsymbol{w}$ is unitarily invariant, e.g.\ 
$\boldsymbol{w}\sim\mathcal{CN}(\boldsymbol{0},\sigma^{2}\boldsymbol{I}_{M})$, 
or if $\boldsymbol{U}$ is a Haar matrix.

\subsection{Expectation Propagation}
We start with an MP algorithm proposed in \cite{Ma17}. 
Let the detector postulate that the noise vector $\boldsymbol{w}$ 
in (\ref{system}) is a circularly symmetric complex Gaussian (CSCG) random 
vector with covariance $\sigma^{2}\boldsymbol{I}_{M}$. This postulation 
needs not be consistent with the true distribution of $\boldsymbol{w}$. 

As derived in Appendix~\ref{sec_deriv_EP}, the MP algorithm for this case 
is based on EP and composed of two modules. 
In iteration~$t$, a first module---called module~A---calculates the 
{\em extrinsic} mean $\boldsymbol{x}_{\mathrm{A}\to \mathrm{B}}^{t}$ and variance 
$v_{\mathrm{A}\to \mathrm{B}}^{t}$ of the signal vector $\boldsymbol{x}$ from 
$\boldsymbol{x}_{\mathrm{B}\to \mathrm{A}}^{t}$ and $v_{\mathrm{B}\to \mathrm{A}}^{t}$ 
provided by the other module---called module~B.    
\begin{equation} \label{module_A_mean}
\boldsymbol{x}_{\mathrm{A}\to \mathrm{B}}^{t} 
= \boldsymbol{x}_{\mathrm{B}\to \mathrm{A}}^{t} 
+ \gamma_{t}\boldsymbol{W}^{t} 
(\boldsymbol{y} - \boldsymbol{A}\boldsymbol{x}_{\mathrm{B}\to \mathrm{A}}^{t}), 
\end{equation}
\begin{equation} \label{module_A_var}
v_{\mathrm{A}\to \mathrm{B}}^{t} = \gamma_{t} - v_{\mathrm{B}\to \mathrm{A}}^{t}. 
\end{equation}
In the initial iteration $t=0$, the prior mean 
$\boldsymbol{x}_{\mathrm{B}\to \mathrm{A}}^{0}=\boldsymbol{0}$ and variance 
$v_{\mathrm{B}\to \mathrm{A}}^{0}=N^{-1}\mathbb{E}[\|\boldsymbol{x}\|^{2}]=1$ 
are used.  

In (\ref{module_A_mean}), the linear minimum mean-square error (LMMSE) filter 
$\boldsymbol{W}^{t}\in\mathbb{C}^{N\times M}$ is given by 
\begin{equation}
\boldsymbol{W}^{t} 
= \boldsymbol{A}^{\mathrm{H}}
\left(
 \sigma^{2}\boldsymbol{I}_{M} + v_{\mathrm{B}\to \mathrm{A}}^{t}\boldsymbol{A}
 \boldsymbol{A}^{\mathrm{H}}
\right)^{-1}. 
\end{equation}
The normalization coefficient\footnote{
$\gamma_{t}^{-1}=N^{-1}\mathrm{Tr}(\boldsymbol{W}^{t}\boldsymbol{A})$ may be 
used in practical situations. 
} $\gamma_{t}$ in (\ref{module_A_mean}) is defined as 
\begin{equation} \label{gamma1}
\frac{1}{\gamma_{t}} 
= \lim_{M=\delta N\to\infty}\frac{1}{N}\mathrm{Tr}(\boldsymbol{W}^{t}\boldsymbol{A})
\aeq \frac{1}{\gamma(v_{\mathrm{B}\to\mathrm{A}}^{t})}   
\end{equation}
due to Assumption~\ref{assumption_A}, with 
\begin{equation} \label{gamma2}
\frac{1}{\gamma(v)}
=\int\frac{\delta\lambda}{\sigma^{2}+v\lambda}d\rho(\lambda), 
\end{equation}
where $\rho(\lambda)$ denotes the asymptotic eigenvalue distribution of 
$\boldsymbol{A}\boldsymbol{A}^{\mathrm{H}}$ in the large system limit. 
The coefficient $\gamma_{t}$ keeps the orthogonality between 
estimation errors in the two modules. 

On the other hand, module~B computes the minimum mean-square error (MMSE) 
estimator and the posterior variance of $\boldsymbol{x}$ 
\begin{equation} \label{estimator}
\tilde{\eta}_{t}(\boldsymbol{x}_{\mathrm{A}\to \mathrm{B}}^{t})
=\mathbb{E}[\boldsymbol{x}|\boldsymbol{x}_{\mathrm{A}\to \mathrm{B}}^{t}],
\end{equation}
\begin{equation} \label{vB}
v_{\mathrm{B}}^{t+1} 
= \frac{1}{N}\left\{
 \mathbb{E}[\|\boldsymbol{x}\|^{2}|\boldsymbol{x}_{\mathrm{A}\to \mathrm{B}}^{t}]
 - \|\tilde{\eta}_{t}(\boldsymbol{x}_{\mathrm{A}\to \mathrm{B}}^{t})\|^{2}
\right\}, 
\end{equation}
given the virtual additive white Gaussian noise (AWGN) observation,  
\begin{equation} \label{virtual_AWGN} 
\boldsymbol{x}_{\mathrm{A}\to \mathrm{B}}^{t}
=\boldsymbol{x} + \boldsymbol{z}^{t}, 
\quad \boldsymbol{z}^{t}\sim\mathcal{CN}(\boldsymbol{0}, 
v_{\mathrm{A}\to \mathrm{B}}^{t}\boldsymbol{I}_{N}).
\end{equation}   
If a termination condition is satisfied, 
module~B outputs $\tilde{\eta}_{t}(\boldsymbol{x}_{\mathrm{A}\to \mathrm{B}}^{t})$ 
as an estimate of $\boldsymbol{x}$. Otherwise, module~B feeds 
the extrinsic mean $\boldsymbol{x}_{\mathrm{B}\to \mathrm{A}}^{t+1}$ and variance 
$v_{\mathrm{B}\to \mathrm{A}}^{t+1}$ of $\boldsymbol{x}$ back to module~A, given by   
\begin{equation} \label{module_B_mean}
\boldsymbol{x}_{\mathrm{B}\to \mathrm{A}}^{t+1} 
= \eta_{t}(\boldsymbol{x}_{\mathrm{A}\to \mathrm{B}}^{t}),  
\end{equation}
\begin{equation} \label{module_B_var}
\frac{1}{v_{\mathrm{B}\to \mathrm{A}}^{t+1}} 
= \frac{1}{v_{\mathrm{B}}^{t+1}} 
- \frac{1}{v_{\mathrm{A}\to \mathrm{B}}^{t}}, 
\end{equation}
where the extrinsic decision function $\eta_{t}:\mathbb{C}\to
\mathbb{C}$ is defined as 
\begin{equation} \label{eta}
\eta_{t}(z) 
= v_{\mathrm{B}\to \mathrm{A}}^{t+1}\left(
 \frac{\tilde{\eta}_{t}(z)}{v_{\mathrm{B}}^{t+1}}
 - \frac{z}{v_{\mathrm{A}\to \mathrm{B}}^{t}}
\right). 
\end{equation}

\begin{remark}
The extrinsic decision function~(\ref{eta}) is zero 
if $\boldsymbol{x}\sim\mathcal{CN}(\boldsymbol{0}, \boldsymbol{I}_{N})$ holds. 
We have postulated 
Assumption~\ref{assumption_x} to avoid a constant decision function. 
\end{remark}

It is not trivial whether the posterior variance~(\ref{vB}) is bounded. 
Therefore, we postulate the following assumption: 
\begin{assumption} \label{assumption_posterior}
Each posterior variance 
$\mathbb{E}[|x_{n}|^{2}|x_{n,\mathrm{A}\to\mathrm{B}}^{t}] 
- |\tilde{\eta}_{t}(x_{n,\mathrm{A}\to\mathrm{B}}^{t})|^{2}$ is almost surely bounded.  
\end{assumption}

Assumption~\ref{assumption_posterior} is a necessary condition for utilizing 
the EP-based algorithm in practical situations. The author believes that 
Assumption~\ref{assumption_posterior} can be proved without additional 
conditions.

We present important properties of the Bayes-optimal decision 
function $\tilde{\eta}_{t}$ in module~B. We start with the definition of 
the Wirtinger derivative of a complex function.  

\begin{definition}[Wirtinger derivative]
For a complex number $z=x+\mathrm{i}y$, the Wirtinger derivative is  
defined as
\begin{equation} \label{Wirtinger_derivative} 
\frac{\partial}{\partial z} 
= \frac{1}{2}\left(
 \frac{\partial}{\partial x}
 - \mathrm{i}\frac{\partial}{\partial y}
\right). 
\end{equation} 
For a complex function $f:\mathbb{C}\to\mathbb{C}$, we write 
$(\partial/\partial z)(\Re[f]+\mathrm{i}\Im[f])$ as 
$\partial f/\partial z$.
\end{definition}

\begin{lemma}[Ma and Ping~\cite{Ma17}] \label{lemma_eta}
Suppose that $z\sim\mathcal{CN}(0, v_{\mathrm{A}\to \mathrm{B}}^{t})$ 
is a CSCG random variable with variance 
$v_{\mathrm{A}\to \mathrm{B}}^{t}$ and independent of $x_{n}$. Then, 
the decision function $\tilde{\eta}_{t}$ is Lipschitz-continuous and satisfies 
\begin{equation} \label{Stein}
\mathbb{E}_{z}\left[
 z^{*}\tilde{\eta}_{t}(x_{n} + z)
\right] 
= v_{\mathrm{A}\to\mathrm{B}}^{t}\mathbb{E}_{z}\left[
 \frac{\partial \tilde{\eta}_{t}}{\partial z}(x_{n} + z) 
\right],  
\end{equation}
\begin{equation}
\mathbb{E}\left[
 z^{*}\tilde{\eta}_{t}(x_{n} + z)
\right] 
= \mathrm{MMSE}(v_{\mathrm{A}\to \mathrm{B}}^{t}) 
\label{eta_tilde}
\end{equation}
for any $n$, where $\mathrm{MMSE}(v_{\mathrm{A}\to \mathrm{B}}^{t})$ denotes 
the MMSE based on an AWGN observation, given by 
\begin{equation} \label{MMSE} 
\mathrm{MMSE}(v_{\mathrm{A}\to \mathrm{B}}^{t})
=\mathbb{E}\left[
 |x_{n}-\tilde{\eta}_{t}(x_{n}+z)|^{2}
\right]. 
\end{equation}
\end{lemma}
\begin{IEEEproof}
See Appendix~\ref{proof_lemma_eta} for the proof based on \cite{Ma17}. 
\end{IEEEproof}

Lemma~\ref{lemma_eta} is used to prove the orthogonality between 
estimation errors in the two modules. The identity~(\ref{Stein}) is a 
generalization of Stein's lemma~\cite{Stein72} to the complex-valued case.

\begin{remark}
As considered in \cite{Rangan17,Ma17}, we can replace the decision 
function $\tilde{\eta}_{t}$ with another suboptimal function. 
Such a replacement may be important 
when the true prior distribution of the signal elements is unknown. 
Nonetheless, for simplicity, we only consider the optimal decision function 
$\tilde{\eta}_{t}$. See \cite{Rangan17} for a generalization of the 
decision function. 
\end{remark}

\subsection{Error Recursion} 
An error recursion for the EP-based algorithm is formulated to analyze 
the convergence property. Let $\boldsymbol{h}_{t}=
\boldsymbol{x}_{\mathrm{A}\to \mathrm{B}}^{t}-\boldsymbol{x}$ and $\boldsymbol{q}_{t}
=\boldsymbol{x}_{\mathrm{B}\to \mathrm{A}}^{t}-\boldsymbol{x}$ 
denote the estimation errors for the extrinsic estimates in modules A and B, 
respectively. 
Substituting the system model~(\ref{system}) into the update 
rule~(\ref{module_A_mean}) of $\boldsymbol{x}_{\mathrm{A}\to\mathrm{B}}^{t}$, and 
using the SVD~(\ref{SVD_A}) and the update rule~(\ref{module_B_mean}) 
of $\boldsymbol{x}_{\mathrm{B}\to\mathrm{A}}^{t}$, 
we obtain the error recursion 
\begin{equation} \label{b}
\boldsymbol{b}_{t} 
= \boldsymbol{V}^{\mathrm{H}}\boldsymbol{q}_{t},   
\end{equation}
\begin{equation} \label{m}
\boldsymbol{m}_{t}
= \boldsymbol{b}_{t} 
- \gamma_{t}\tilde{\boldsymbol{W}}_{t}\{
(\boldsymbol{\Sigma},\boldsymbol{O})\boldsymbol{b}_{t} 
- \tilde{\boldsymbol{w}}\},  
\end{equation}
\begin{equation} \label{h}
\boldsymbol{h}_{t} 
= \boldsymbol{V}\boldsymbol{m}_{t}, 
\end{equation}
\begin{equation} \label{q} 
\boldsymbol{q}_{t+1} 
= \eta_{t}(\boldsymbol{x}+\boldsymbol{h}_{t}) - \boldsymbol{x}, 
\end{equation}
with $\tilde{\boldsymbol{w}}=\boldsymbol{U}^{\mathrm{H}}\boldsymbol{w}$. 
In (\ref{m}), the linear filter $\tilde{\boldsymbol{W}}_{t}$ is given by 
\begin{equation} \label{Wt}
\tilde{\boldsymbol{W}}_{t}
= (\boldsymbol{\Sigma}, \boldsymbol{O})^{\mathrm{H}}\left(
 \sigma^{2}\boldsymbol{I}_{M} + v_{\mathrm{B}\to \mathrm{A}}^{t}\boldsymbol{\Sigma}^{2}
\right)^{-1}. 
\end{equation}
Furthermore, we define $\eta_{-1}(\cdot)=0$ to 
obtain $\boldsymbol{q}_{0}=-\boldsymbol{x}$. 

In analyzing the convergence property, we focus on the distribution of 
the estimation error $\boldsymbol{h}_{t}$ conditioned on the preceding 
iteration history. Thus, it is useful to represent the error recursion 
in the matrix form. Define 
\begin{IEEEeqnarray}{rl}
\boldsymbol{Q}_{t}=&(\boldsymbol{q}_{0},\ldots,\boldsymbol{q}_{t-1})
\in\mathbb{C}^{N\times t}, \nonumber \\
\boldsymbol{B}_{t}=&(\boldsymbol{b}_{0},\ldots,\boldsymbol{b}_{t-1})
\in\mathbb{C}^{N\times t}, \nonumber \\
\boldsymbol{M}_{t}=&(\boldsymbol{m}_{0},\ldots,\boldsymbol{m}_{t-1})
\in\mathbb{C}^{N\times t}, \nonumber \\ 
\boldsymbol{H}_{t}=&(\boldsymbol{h}_{0},\ldots,\boldsymbol{h}_{t-1})
\in\mathbb{C}^{N\times t}. 
\end{IEEEeqnarray} 
The error recursion is represented as 
\begin{equation} \label{recursion1}
\boldsymbol{V}^{\mathrm{H}}\boldsymbol{Q}_{t} 
= \boldsymbol{B}_{t}, 
\end{equation}
\begin{equation} \label{recursion2}
\boldsymbol{M}_{t} 
= \boldsymbol{G}_{t}(\boldsymbol{B}_{t}, \tilde{\boldsymbol{w}}), 
\end{equation}
\begin{equation} \label{recursion3}
\boldsymbol{V}\boldsymbol{M}_{t} = \boldsymbol{H}_{t},
\end{equation} 
\begin{equation} \label{recursion4}
\boldsymbol{Q}_{t+1} 
= \boldsymbol{F}_{t}(\boldsymbol{H}_{t}, \boldsymbol{x}), 
\end{equation}
where the $\tau$th columns of $\boldsymbol{G}_{t}
(\boldsymbol{B}_{t}, \tilde{\boldsymbol{w}})$ and 
$\boldsymbol{F}(\boldsymbol{H}_{t}, \boldsymbol{x})$ are equal to 
the right-hand sides (RHSs) of (\ref{m}) and (\ref{q}) for $t=\tau$, 
respectively.  

The random vectors defined in Section~\ref{sec3} may have elements of which 
the distributions change as $N$ grows. Thus, the subscript $N$ should have 
been added in terms of the mathematical notation. Nonetheless, we have omitted 
the subscript $N$ for notational simplicity. 

\section{Main Result} \label{sec4}
Ma and Ping~\cite{Ma17} conjectured that the following SE equations 
describe the dynamics of the EP-based algorithm in the large system limit:
\begin{equation} \label{module_A_SE}
\bar{v}_{\mathrm{A}\to\mathrm{B}}^{t} 
= \gamma\left(
 \bar{v}_{\mathrm{B}\to\mathrm{A}}^{t}
\right) 
- \bar{v}_{\mathrm{B}\to\mathrm{A}}^{t}, 
\end{equation}
\begin{equation} \label{module_B_SE}
\frac{1}{\bar{v}_{\mathrm{B}\to\mathrm{A}}^{t+1}} 
= \frac{1}{\mathrm{MMSE}(\bar{v}_{\mathrm{A}\to\mathrm{B}}^{t})} 
- \frac{1}{\bar{v}_{\mathrm{A}\to\mathrm{B}}^{t}},
\end{equation}
with $\bar{v}_{\mathrm{B}\to\mathrm{A}}^{0}=1$, in which 
$\gamma(\cdot)$ and $\mathrm{MMSE}(\cdot)$ are given in (\ref{gamma2}) and 
(\ref{MMSE}), respectively. The following theorem justifies their 
conjecture. 

\begin{theorem} \label{theorem_SE}
Define $\bar{v}_{\mathrm{A}\to\mathrm{B}}^{t}$ and $\bar{v}_{\mathrm{B}\to\mathrm{A}}^{t}$ 
via the SE equations~(\ref{module_A_SE}) and (\ref{module_B_SE}). Then, 
the following results hold in the large system limit:  
\begin{equation} \label{MSE_A_to_B_end}
\lim_{M=\delta N\to\infty}\frac{1}{N}\|
\boldsymbol{x}_{\mathrm{A}\to \mathrm{B}}^{t}-\boldsymbol{x}\|^{2}
\aeq \bar{v}_{\mathrm{A}\to\mathrm{B}}^{t}, 
\end{equation}
\begin{equation} \label{MMSE_SE_end}
\lim_{M=\delta N\to\infty}\frac{1}{N}\| 
\tilde{\eta}_{t}(\boldsymbol{x}_{\mathrm{A}\to \mathrm{B}}^{t})-\boldsymbol{x}\|^{2}
\aeq \mathrm{MMSE}(\bar{v}_{\mathrm{A}\to\mathrm{B}}^{t}), 
\end{equation}
\begin{equation} \label{MSE_B_to_A_end}
\lim_{M=\delta N\to\infty}\frac{1}{N}\|
\eta_{t}(\boldsymbol{x}_{\mathrm{A}\to \mathrm{B}}^{t})-\boldsymbol{x}\|^{2}
\aeq \bar{v}_{\mathrm{B}\to\mathrm{A}}^{t+1}. 
\end{equation}
\end{theorem}

The update rules~(\ref{module_A_var}) and (\ref{module_B_var}) in the 
EP-based algorithm have the same representation as that in the SE 
equations~(\ref{module_A_SE}) and (\ref{module_B_SE}). This
implies that the EP-based algorithm predicts the exact dynamics of the 
extrinsic variances in the large system limit. 
The FPs of the SE equations were proved in \cite{Ma17} to 
correspond to those of an asymptotic energy function that describes the 
Bayes-optimal performance~\cite{Takeda06,Tulino13,Barbier18}. 
Thus, the Bayes-optimal performance is achievable 
when the SE equations have a unique FP, or equivalently when the compression 
rate $\delta$ is larger than the BP threshold.    

The following theorem justifies the SE equations~(\ref{module_A_SE}) and 
(\ref{module_B_SE}) in terms of individual MSEs. 

\begin{theorem} \label{theorem_individual_SE}
Define $\bar{v}_{\mathrm{A}\to\mathrm{B}}^{t}$ and $\bar{v}_{\mathrm{B}\to\mathrm{A}}^{t}$ 
via the SE equations~(\ref{module_A_SE}) and (\ref{module_B_SE}). Then, 
for any $n$ 
\begin{equation} \label{MMSE_individual_SE_end}
\lim_{M=\delta N\to\infty}\mathbb{E}[
|\tilde{\eta}_{t}(x_{n,\mathrm{A}\to\mathrm{B}}^{t}) - x_{n}|^{2}]
= \mathrm{MMSE}(\bar{v}_{\mathrm{A}\to\mathrm{B}}^{t}),  
\end{equation} 
\begin{equation} \label{MSE_B_to_A_individual_end}
\lim_{M=\delta N\to\infty}\mathbb{E}[
|\eta_{t}(x_{n,\mathrm{A}\to\mathrm{B}}^{t}) - x_{n}|^{2}]
= \bar{v}_{\mathrm{B}\to\mathrm{A}}^{t+1}. 
\end{equation} 
\end{theorem}
\begin{remark}
For simplicity, the individual MSE for the extrinsic estimate in module~A is 
not analyzed in this paper. Furthermore, we have assumed the i.i.d.\ property 
of the elements of the signal vector $\boldsymbol{x}$. However, 
our proof strategy can be applied to justifying 
that the individual MSE $\mathbb{E}[|x_{n,\mathrm{A}\to\mathrm{B}}^{t} - x_{n}|^{2}]$ 
for module~A converges to $\bar{v}_{\mathrm{A}\to\mathrm{B}}^{t}$ in the large 
system limit. Furthermore, the assumption on $\boldsymbol{x}$ can be relaxed 
to the case of independent but non-identically distributed signals. 
\end{remark}

We shall introduce several notations to present a general theorem, of 
which corollaries are Theorems~\ref{theorem_SE} and 
\ref{theorem_individual_SE}. The random variables in the error 
recursions~(\ref{recursion1})--(\ref{recursion4}) are divided 
into three groups: $\boldsymbol{V}$, 
$\Theta=\{\boldsymbol{\Sigma}, \tilde{\boldsymbol{w}}, \boldsymbol{x}\}$, and 
\begin{IEEEeqnarray}{rl}
\mathcal{X}_{t,t'}
=&\left\{
 \boldsymbol{Q}_{t+1},\boldsymbol{B}_{t'}, \boldsymbol{M}_{t'}, 
 \boldsymbol{H}_{t}\left| 
  \boldsymbol{B}_{t'}^{\mathrm{H}}\boldsymbol{M}_{t}
  =\boldsymbol{Q}_{t'}^{\mathrm{H}}\boldsymbol{H}_{t}, 
 \right.
\right. \nonumber \\
&\left.
 \left.
  \boldsymbol{M}_{t'}
  =\boldsymbol{G}_{t'}(\boldsymbol{B}_{t'}, \tilde{\boldsymbol{w}}), 
  \boldsymbol{Q}_{t+1}=\boldsymbol{F}_{t}(\boldsymbol{H}_{t}, \boldsymbol{x})
 \right.
\right\},
\end{IEEEeqnarray} 
for $t'=t$ or $t'=t+1$, while we define 
$\mathcal{X}_{0,0}=\{\boldsymbol{Q}_{1}\}$ 
and $\mathcal{X}_{0,1}=\{\boldsymbol{Q}_{1},\boldsymbol{B}_{1},
\boldsymbol{M}_{1}| \boldsymbol{M}_{1}=\boldsymbol{G}_{1}
(\boldsymbol{B}_{1}, \tilde{\boldsymbol{w}})\}$. 
See Table~\ref{table1} for the notational conventions used in this paper. 

\begin{table}[t]
\caption{
Notational conventions for $t=0$. 
}
\label{table1}
\begin{center}
\begin{tabular}{|l|}
\hline
$\mathcal{X}_{0,0}=\{\boldsymbol{Q}_{1}\}$, 
$\mathcal{X}_{0,1}=\{\boldsymbol{Q}_{1},\boldsymbol{B}_{1},\boldsymbol{M}_{1}
| \boldsymbol{M}_{1}=\boldsymbol{G}_{1}(\boldsymbol{B}_{1},
\tilde{\boldsymbol{w}})\}$, \\
$\boldsymbol{Q}_{0}=\boldsymbol{O}$, 
$\boldsymbol{B}_{0}=\boldsymbol{O}$, $\boldsymbol{M}_{0}=\boldsymbol{O}$, 
$\boldsymbol{H}_{0}=\boldsymbol{O}$, 
$\boldsymbol{M}_{0}^{\dagger}=\boldsymbol{O}$,
$\boldsymbol{Q}_{0}^{\dagger}=\boldsymbol{O}$, \\
$\boldsymbol{\alpha}_{0}=\boldsymbol{0}$, 
$\boldsymbol{\beta}_{0}=\boldsymbol{0}$,  
$\boldsymbol{\Phi}_{\boldsymbol{O}}^{\perp}=\boldsymbol{I}$, 
$\boldsymbol{\Phi}_{(\boldsymbol{M}, \boldsymbol{O})}^{\perp}
=\boldsymbol{\Phi}_{\boldsymbol{M}}^{\perp}$ for any $\boldsymbol{M}$. \\ 
\hline  
\end{tabular}
\end{center}
\end{table}

The set $\Theta$ is fixed throughout this paper. 
Thus, conditioning on $\Theta$ is omitted. 
The set $\mathcal{X}_{t,t}$ describes the history of all preceding 
iterations just before updating (\ref{b}), while $\mathcal{X}_{t,t+1}$ 
represents the history just before updating (\ref{h}).  
Note that the condition $\boldsymbol{B}_{t'}^{\mathrm{H}}\boldsymbol{M}_{t}
=\boldsymbol{Q}_{t'}^{\mathrm{H}}\boldsymbol{H}_{t}$ is a constraint imposing 
$\boldsymbol{V}\in{\cal U}_{N}$, and follows from (\ref{recursion1}) and 
(\ref{recursion3}). In order to investigate the dynamics of the error 
recursions, the distribution of the Haar matrix $\boldsymbol{V}$ conditioned 
on $\mathcal{X}_{t,t'}$ is analyzed. 

Let $\boldsymbol{m}_{t}^{\perp}=\boldsymbol{P}_{\boldsymbol{M}_{t}}^{\perp}
\boldsymbol{m}_{t}$. Since $\boldsymbol{m}_{t}^{\parallel}
=\boldsymbol{m}_{t}-\boldsymbol{m}_{t}^{\perp}$ is in the space spanned 
by the columns of $\boldsymbol{M}_{t}$, we have 
$(\boldsymbol{m}_{t}^{\parallel})^{\mathrm{H}}\boldsymbol{m}_{t}^{\perp}=0$. 
Furthermore, $\boldsymbol{m}_{t}^{\parallel}$ is represented  
as $\boldsymbol{m}_{t}^{\parallel}=\boldsymbol{M}_{t}\boldsymbol{\alpha}_{t}$, 
with $\boldsymbol{\alpha}_{t} = \boldsymbol{M}_{t}^{\dagger}\boldsymbol{m}_{t}
\in\mathbb{C}^{t}$. Similarly, we define $\boldsymbol{q}_{t'}^{\perp}
=\boldsymbol{P}_{\boldsymbol{Q}_{t'}}^{\perp}\boldsymbol{q}_{t'}$ and 
$\boldsymbol{q}_{t'}^{\parallel}=\boldsymbol{q}_{t'}-\boldsymbol{q}_{t'}^{\perp}
=\boldsymbol{Q}_{t'}\boldsymbol{\beta}_{t'}$, with 
$\boldsymbol{\beta}_{t'}=\boldsymbol{Q}_{t'}^{\dagger}\boldsymbol{q}_{t'}
\in\mathbb{C}^{t'}$. 

For notational convenience, we define the conventions listed in 
Table~\ref{table1}, which imply $\boldsymbol{P}_{\boldsymbol{O}}^{\perp}
=\boldsymbol{I}$, $\boldsymbol{m}_{0}^{\perp}=\boldsymbol{m}_{0}$, and 
$\boldsymbol{q}_{0}^{\perp}=\boldsymbol{q}_{0}$. 

\begin{theorem} \label{main_theorem}
Let $\boldsymbol{D}=\mathrm{diag}\{D_{1},\ldots,D_{N}\}$ denote any 
$N\times N$ real diagonal matrix 
that is a function of $\boldsymbol{\Sigma}$. 
For some $\epsilon>0$, suppose that 
$N^{-1}\mathrm{Tr}(\boldsymbol{D}^{k})$ converges almost 
surely for all $k\in[0, 4+\epsilon)$ as $N\to\infty$. Then, 
the following properties for module~A hold for 
each iteration~$\tau=0,1,\ldots$:   
\begin{enumerate}[label=(A\arabic*)]
\item \label{property_A1}
Define 
\begin{equation} \label{b_tilde}
\tilde{\boldsymbol{b}}_{\tau} 
= \boldsymbol{B}_{\tau}\boldsymbol{\beta}_{\tau} 
+ \boldsymbol{M}_{\tau}\boldsymbol{o}(1) 
+ \boldsymbol{B}_{\tau}\boldsymbol{o}(1) 
+ \boldsymbol{\Phi}_{(\boldsymbol{B}_{\tau}, \boldsymbol{M}_{\tau})}^{\perp}
\boldsymbol{z}_{\tau}, 
\end{equation}
with 
\begin{equation} \label{z}
\boldsymbol{z}_{\tau} 
= \tilde{\boldsymbol{V}}^{\mathrm{H}}
(\boldsymbol{\Phi}_{(\boldsymbol{Q}_{\tau}, \boldsymbol{H}_{\tau})}^{\perp})^{\mathrm{H}}
\boldsymbol{q}_{\tau}, 
\end{equation}
where $\tilde{\boldsymbol{V}}\in{\cal U}_{N-2t}$ is a Haar matrix 
and independent of $\Theta$ and $\mathcal{X}_{\tau,\tau}$. Then, we have 
\begin{equation} \label{b_distribution}
\boldsymbol{b}_{\tau}|_{\Theta,\mathcal{X}_{\tau,\tau}}
\sim \tilde{\boldsymbol{b}}_{\tau} 
\end{equation}
conditioned on $\Theta$ and $\mathcal{X}_{\tau,\tau}$ in the large system 
limit, with 
\begin{equation} \label{mu}
\lim_{M=\delta N\to\infty}\frac{1}{N}\left\{
 \|\boldsymbol{z}_{\tau}\|^{2} - \|\boldsymbol{q}_{\tau}^{\perp}\|^{2} 
\right\}
\aeq 0.
\end{equation}

\item For all $\tau'\leq\tau$,  
\begin{equation} \label{bw}
\lim_{M=\delta N\to\infty}\frac{1}{N}\boldsymbol{b}_{\tau'}^{\mathrm{H}}
\boldsymbol{D}\tilde{\boldsymbol{W}}_{\tau}\tilde{\boldsymbol{w}}
\aeq 0, 
\end{equation}
\begin{equation} \label{bbqq} 
\lim_{M=\delta N\to\infty}\frac{1}{N}\left\{
 \boldsymbol{b}_{\tau'}^{\mathrm{H}}\boldsymbol{D}\boldsymbol{b}_{\tau}
 - \frac{\mathrm{Tr}(\boldsymbol{D})}{N}
 \boldsymbol{q}_{\tau'}^{\mathrm{H}}\boldsymbol{q}_{\tau}
\right\}
\aeq 0, 
\end{equation}
\begin{equation} \label{bm}
\lim_{M=\delta N\to\infty}\frac{1}{N}
\boldsymbol{b}_{\tau'}^{\mathrm{H}}\boldsymbol{m}_{\tau}
\aeq 0,
\end{equation}
where $\tilde{\boldsymbol{W}}_{\tau}$ is given by (\ref{Wt}). 

\item Define $\bar{v}_{\mathrm{A}\to\mathrm{B}}^{\tau}$  
in the SE equations~(\ref{module_A_SE}) and (\ref{module_B_SE}), Then, 
\begin{equation} \label{module_A_var_SE}
\lim_{M=\delta N\to\infty}v_{\mathrm{A}\to\mathrm{B}}^{\tau}
\aeq \bar{v}_{\mathrm{A}\to\mathrm{B}}^{\tau}, 
\end{equation}
\begin{equation} \label{MSE_A_to_B}
\lim_{M=\delta N\to\infty}\frac{1}{N}\|\boldsymbol{m}_{\tau}\|^{2}
\aeq \bar{v}_{\mathrm{A}\to\mathrm{B}}^{\tau}.   
\end{equation}

\item \label{property_A3}
For some $\epsilon>0$ and $C>0$, 
\begin{equation} \label{4th_moment_m}
\lim_{M=\delta N\to\infty}\mathbb{E}\left[
 |m_{\tau,n}|^{2+\epsilon}
\right]<\infty, 
\end{equation} 
\begin{equation} \label{mm}
\limsup_{M=\delta N\to\infty}\frac{1}{N}
\boldsymbol{m}_{\tau}^{\mathrm{H}}\boldsymbol{D}\boldsymbol{m}_{\tau}
\al\infty, 
\end{equation} 
\begin{equation} \label{norm_M}
\liminf_{M=\delta N\to\infty}\lambda_{\mathrm{min}}\left(
 \frac{1}{N}\boldsymbol{M}_{\tau+1}^{\mathrm{H}}\boldsymbol{M}_{\tau+1}
\right)
\ag C.  
\end{equation}
\end{enumerate}

The following properties hold for module~B: 
\begin{enumerate}[label=(B\arabic*)]
\item \label{property_B1} 
Define 
\begin{equation} \label{h_tilde}
\tilde{\boldsymbol{h}}_{\tau} 
= \boldsymbol{H}_{\tau}\boldsymbol{\alpha}_{\tau} 
+ \boldsymbol{Q}_{\tau+1}\boldsymbol{o}(1) 
+ \boldsymbol{H}_{\tau}\boldsymbol{o}(1)  
+ \boldsymbol{\Phi}_{(\boldsymbol{Q}_{\tau+1}, \boldsymbol{H}_{\tau})}^{\perp}
\tilde{\boldsymbol{z}}_{\tau},  
\end{equation}
with 
\begin{equation} \label{z_tilde}
\tilde{\boldsymbol{z}}_{\tau} 
= \tilde{\boldsymbol{V}}
(\boldsymbol{\Phi}_{(\boldsymbol{B}_{\tau+1}, \boldsymbol{M}_{\tau})}^{\perp})^{\mathrm{H}}
\boldsymbol{m}_{\tau}, 
\end{equation}
where $\tilde{\boldsymbol{V}}\in{\cal U}_{N-(2t+1)}$ is a Haar matrix 
and independent of $\Theta$ and $\mathcal{X}_{\tau,\tau+1}$. Then, we have 
\begin{equation} \label{h_distribution}
\boldsymbol{h}_{\tau}|_{\Theta, \mathcal{X}_{\tau,\tau+1}}
\sim \tilde{\boldsymbol{h}}_{\tau}
\end{equation}
conditioned on $\Theta$ and $\mathcal{X}_{\tau,\tau+1}$ 
in the large system limit, with 
\begin{equation} \label{nu}
\lim_{M=\delta N\to\infty}\frac{1}{N}\left\{
 \|\tilde{\boldsymbol{z}}_{\tau}\|^{2}  
 - \|\boldsymbol{m}_{\tau}^{\perp}\|^{2}
\right\} \aeq 0.  
\end{equation}

\item For all $\tau'\leq\tau$,  
\begin{equation} \label{hq} 
\lim_{M=\delta N\to\infty}\frac{1}{N}
\boldsymbol{h}_{\tau}^{\mathrm{H}}\boldsymbol{q}_{\tau'+1} 
\aeq 0.
\end{equation}

\item 
Define $\bar{v}_{\mathrm{A}\to\mathrm{B}}^{\tau}$ and 
$\bar{v}_{\mathrm{B}\to\mathrm{A}}^{\tau+1}$   
in the SE equations~(\ref{module_A_SE}) and (\ref{module_B_SE}), Then, 
\begin{equation} \label{vB_SE} 
\lim_{M=\delta N\to\infty}v_{\mathrm{B}}^{\tau+1}
\aeq \mathrm{MMSE}(\bar{v}_{\mathrm{A}\to\mathrm{B}}^{\tau}), 
\end{equation}
\begin{equation} \label{module_B_var_SE} 
\lim_{M=\delta N\to\infty}v_{\mathrm{B}\to\mathrm{A}}^{\tau+1}
\aeq\bar{v}_{\mathrm{B}\to\mathrm{A}}^{\tau+1}, 
\end{equation}
\begin{equation} \label{MMSE_SE} 
\lim_{M=\delta N\to\infty}\frac{1}{N}\|\tilde{\eta}_{t}
(\boldsymbol{x}+\boldsymbol{h}_{t})-\boldsymbol{x}\|^{2}
\aeq  \mathrm{MMSE}(\bar{v}_{\mathrm{A}\to\mathrm{B}}^{\tau}), 
\end{equation}
\begin{equation} \label{MSE_B_to_A}
\lim_{M=\delta N\to\infty}\frac{1}{N}\|\boldsymbol{q}_{\tau+1}\|^{2}
\aeq \bar{v}_{\mathrm{B}\to\mathrm{A}}^{\tau+1}. 
\end{equation}

\item \label{property_B3}
For some $\epsilon>0$ and $C>0$, 
\begin{equation} \label{4th_moment_q}
\lim_{M=\delta N\to\infty}\mathbb{E}\left[
 |q_{\tau+1,n}|^{2+\epsilon}
\right]<\infty,  
\end{equation} 
\begin{equation} \label{norm_Q}
\liminf_{M=\delta N\to\infty}\lambda_{\mathrm{min}}\left(
 \frac{1}{N}\boldsymbol{Q}_{\tau+2}^{\mathrm{H}}\boldsymbol{Q}_{\tau+2}
\right)
\ag C.
\end{equation}

\item 
Define $\bar{v}_{\mathrm{A}\to\mathrm{B}}^{\tau}$ and 
$\bar{v}_{\mathrm{B}\to\mathrm{A}}^{\tau+1}$   
in the SE equations~(\ref{module_A_SE}) and (\ref{module_B_SE}), Then, 
\begin{equation} \label{MMSE_individual_SE}
\lim_{M=\delta N\to\infty}\mathbb{E}[|\tilde{\eta}_{\tau}
(x_{n}+h_{\tau,n})-x_{n}|^{2}]
\aeq  \mathrm{MMSE}(\bar{v}_{\mathrm{A}\to\mathrm{B}}^{\tau}), 
\end{equation}
\begin{equation} \label{MSE_B_to_A_individual}
\lim_{M=\delta N\to\infty}\mathbb{E}[|q_{\tau+1,n}|^{2}]
\aeq \bar{v}_{\mathrm{B}\to\mathrm{A}}^{\tau+1}. 
\end{equation}
\end{enumerate}
\end{theorem}
\begin{IEEEproof}
See Section~\ref{sec5}. 
\end{IEEEproof}

Ma and Ping~\cite[Assumption 1]{Ma17} postulated that 
$\tilde{\boldsymbol{z}}_{\tau}$ in (\ref{h_tilde}) has independent CSCG 
elements. The assumption is too strong to be 
justified. In fact, the references~\cite{Meckes08,Chatterjee08} imply  
that the assumption is not correct, while the assumption holds only for 
finite subsets of the elements of $\tilde{\boldsymbol{z}}_{\tau}$.  
However, the weaker property~\ref{property_B1} is sufficient to prove 
Theorem~\ref{main_theorem}. 

\begin{IEEEproof}[Proof of Theorem~\ref{theorem_SE}]
The property~(\ref{MSE_A_to_B_end}) follows from the definition~(\ref{h}) 
of $\boldsymbol{h}_{t}$ and (\ref{MSE_A_to_B}). 
Furthermore, (\ref{MMSE_SE_end}) and (\ref{MSE_B_to_A_end}) are due to 
(\ref{MMSE_SE}) and (\ref{MSE_B_to_A}), respectively. 
\end{IEEEproof}

\begin{IEEEproof}[Proof of Theorem~\ref{theorem_individual_SE}]
Theorem~\ref{theorem_individual_SE} follows from (\ref{MMSE_individual_SE}) 
and (\ref{MSE_B_to_A_individual}). 
\end{IEEEproof}

\section{Proof of Theorem~\ref{main_theorem}} \label{sec5}
\subsection{Technical Lemma} 
We need to evaluate the two distributions 
$p(\boldsymbol{m}_{t}, \boldsymbol{b}_{t}|\Theta,\mathcal{X}_{t,t})$ and 
$p(\boldsymbol{q}_{t+1}, \boldsymbol{h}_{t}|\Theta,\mathcal{X}_{t,t+1})$. 
The former distribution represents the error recursions (\ref{b}) and 
(\ref{m}) conditioned on the history of all preceding iterations, while 
the latter describes the error recursions (\ref{h}) and (\ref{q}). 
We follow the proof strategy in \cite{Bayati11} to evaluate the two 
distributions via the conditional distribution 
$p(\boldsymbol{V}|\Theta,\mathcal{X}_{t,t'})$ for $t'=t$ or $t'=t+1$.   
See Section~\ref{sec1b} for the main idea in analyzing 
the conditional distributions. 

The following lemma provides a useful representation of 
$\boldsymbol{V}\in{\cal U}_{N}$ conditioned on $\Theta$ and 
$\mathcal{X}_{t,t'}$, and corresponds to \cite[Lemma 10]{Bayati11}. 
See Section~\ref{sec1E} for the notation.  

\begin{lemma} \label{lemma_conditional_distribution}
Suppose that $\boldsymbol{Y}\in\mathbb{C}^{N\times t}$ is full-rank for 
$0<t<N$, and consider the noiseless and compressed observation 
$\boldsymbol{X}\in\mathbb{C}^{N\times t}$ of $\boldsymbol{V}$ given by 
\begin{equation} \label{constraint}
\boldsymbol{X} = \boldsymbol{V}^{\mathrm{H}}\boldsymbol{Y}. 
\end{equation}
Then, the conditional distribution of the Haar matrix $\boldsymbol{V}$ 
given $\boldsymbol{X}$ and $\boldsymbol{Y}$ satisfies 
\begin{equation} \label{V_conditional_distribution} 
\boldsymbol{V}|_{\boldsymbol{X}, \boldsymbol{Y}} 
\sim \boldsymbol{Y}(\boldsymbol{Y}^{\mathrm{T}}\boldsymbol{Y})^{-1}
\boldsymbol{X}^{\mathrm{H}} + \boldsymbol{\Phi}_{\boldsymbol{Y}}^{\perp}
\tilde{\boldsymbol{V}}(\boldsymbol{\Phi}_{\boldsymbol{X}}^{\perp})^{\mathrm{H}}, 
\end{equation}
where $\tilde{\boldsymbol{V}}\in{\cal U}_{N-t}$ is a Haar matrix 
independent of $\boldsymbol{X}$ and $\boldsymbol{Y}$. 
\end{lemma}
\begin{IEEEproof}
See Appendix~\ref{proof_conditional_distribution}. 
\end{IEEEproof}

\begin{proposition} \label{proposition_norm}
Let $\boldsymbol{a}\in\mathbb{C}^{t}$ and 
$\boldsymbol{M}=(\boldsymbol{m}_{0},\ldots,\boldsymbol{m}_{t-1})
\in\mathbb{C}^{N\times t}$. If $N^{-1}\|\boldsymbol{m}_{\tau}\|^{2}$ is bounded  
for any $\tau$ as $N\to\infty$, then 
\begin{equation}
\limsup_{N\to\infty}
\frac{1}{N}\boldsymbol{a}^{\mathrm{H}}\boldsymbol{M}^{\mathrm{H}}\boldsymbol{M}
\boldsymbol{a} 
<\infty. 
\end{equation}
\end{proposition}
\begin{IEEEproof}
Proposition~\ref{proposition_norm} follows from  
$N^{-1}\|\boldsymbol{m}_{\tau}\|^{2}<\infty$ and 
$\|\boldsymbol{M}\boldsymbol{a}\|^{2} 
\leq\|\boldsymbol{M}\|^{2}\|\boldsymbol{a}\|^{2}$. 
\end{IEEEproof}

We are ready to prove Theorem~\ref{main_theorem}. The proof is by induction. 
We first prove the properties of modules~A and B for $\tau=0$. Then, 
the properties are proved for $\tau=t$ under the induction hypotheses for 
all $\tau<t$.  

\subsection{Module~A for $\tau=0$}
\begin{IEEEproof}[Property~\ref{property_A1} for $\tau=0$]
Property~\ref{property_A1} for $\tau=0$ is trivial from the 
definition~(\ref{b}) of $\boldsymbol{b}_{0}$, 
because of the notational convention.  
\end{IEEEproof}

\begin{IEEEproof}[Eq.~(\ref{bw})--(\ref{bm}) for $\tau=0$]
We first prove (\ref{bw}) and (\ref{bbqq}) for $\tau=0$. 
Let $\boldsymbol{X}_{N}=\boldsymbol{b}_{0}$ 
and $f_{n}(z)=z^{*}[\boldsymbol{D}
\tilde{\boldsymbol{W}}_{0}\tilde{\boldsymbol{w}}]_{n}$ 
to have the representation 
\begin{equation}
\frac{1}{N}\boldsymbol{b}_{0}^{\mathrm{H}}\boldsymbol{D}
\tilde{\boldsymbol{W}}_{0}\tilde{\boldsymbol{w}}
= \frac{1}{N}\sum_{n=1}^{N}f_{n}(b_{0,n}). 
\end{equation}
From Property~\ref{property_A1} for $\tau=0$, $\boldsymbol{X}_{N}$ is 
unitarily invariant. The definition~(\ref{b}) of $\boldsymbol{b}_{0}$, 
$\boldsymbol{q}_{0}=-\boldsymbol{x}$, and Assumption~\ref{assumption_x} imply 
the condition~(\ref{convergence_X}) with $v=0$.   
Since $f_{n}$ is Lipschitz-continuous with the Lipschitz constant 
$L_{n}=|[\boldsymbol{D}\tilde{\boldsymbol{W}}_{0}
\tilde{\boldsymbol{w}}]_{n}|$, 
we need to prove the condition~(\ref{Lipschitz_constant}) to use 
Lemma~\ref{lemma_SLLN}. 
Using the definition $\tilde{\boldsymbol{w}}=\boldsymbol{U}^{\mathrm{H}}
\boldsymbol{w}$ and Assumption~\ref{assumption_w} yields  
\begin{IEEEeqnarray}{rl}
&\frac{1}{N}\sum_{n=1}^{N}L_{n}^{2}
= \frac{1}{N}\tilde{\boldsymbol{w}}^{\mathrm{H}}
\tilde{\boldsymbol{W}}_{0}^{\mathrm{H}}\boldsymbol{D}^{2}
\tilde{\boldsymbol{W}}_{0}\tilde{\boldsymbol{w}} \nonumber \\
\aeq& \frac{\sigma^{2}}{N}\mathrm{Tr}\left(
 \boldsymbol{D}^{2}
 \tilde{\boldsymbol{W}}_{0}\tilde{\boldsymbol{W}}_{0}^{\mathrm{H}}
\right) + o(1) \nonumber \\
\leq& \sigma^{2}\left\{
 \frac{\mathrm{Tr}(\boldsymbol{D}^{4})}{N}
\right\}^{1/2} 
\left\{
 \frac{\mathrm{Tr}\{(\tilde{\boldsymbol{W}}_{0}
\tilde{\boldsymbol{W}}_{0}^{\mathrm{H}})^{2}\}}{N}
\right\}^{1/2} + o(1)
\nonumber \\
\al& \infty \label{trace_bound}
\end{IEEEeqnarray}
in the large system limit, 
where the first inequality follows from the Cauchy-Schwarz inequality, 
and where the boundedness is due to the definition of $\boldsymbol{D}$, 
the definition~(\ref{Wt}) of $\tilde{\boldsymbol{W}}_{0}$, and 
Assumption~\ref{assumption_A}. Thus, 
we can use Lemma~\ref{lemma_SLLN} to obtain (\ref{bw}) for $\tau=0$.  
Similarly, we use Lemma~\ref{lemma_SLLN} for $f_{n}(z)=D_{n}|z|^{2}$ 
to have (\ref{bbqq}) for $\tau=0$. 

We next prove (\ref{bm}) for $\tau=0$. Let $k\in[0, 4+\epsilon)$. 
We use H\"older inequality for any $p\in(1, (4+\epsilon)/k)$ to obtain 
\begin{IEEEeqnarray}{rl}
&\frac{1}{N}\mathrm{Tr}\left\{
 \left(
  \boldsymbol{D}\tilde{\boldsymbol{W}}_{0}
  (\boldsymbol{\Sigma}, \boldsymbol{O})
 \right)^{k} 
\right\} \nonumber \\
\leq& \frac{1}{N}\left\{
 \mathrm{Tr}(\boldsymbol{D}^{kp})
\right\}^{1/p}
\left\{
 \mathrm{Tr}\left[
  \left(
   \tilde{\boldsymbol{W}}_{0}
   (\boldsymbol{\Sigma}, \boldsymbol{O})
  \right)^{kq}
 \right]
\right\}^{1/q} \nonumber \\
\al&\infty  \label{trace_bound2} 
\end{IEEEeqnarray}
as $N\to\infty$, with $q^{-1}=1-1/p$, where the boundedness is obtained by 
repeating the proof of the boundedness in (\ref{trace_bound}). Thus, 
we use (\ref{bw}) and (\ref{bbqq}) for $\tau=0$ to have 
\begin{IEEEeqnarray}{rl} 
&\frac{\gamma_{0}}{N}\boldsymbol{b}_{0}^{\mathrm{H}}\boldsymbol{D}
\tilde{\boldsymbol{W}}_{0}
\left\{
 (\boldsymbol{\Sigma}, \boldsymbol{O})\boldsymbol{b}_{0}
 - \tilde{\boldsymbol{w}}
\right\} \nonumber \\
\aeq& \gamma_{0}\frac{\mathrm{Tr}(\boldsymbol{D}
\tilde{\boldsymbol{W}}_{0}(\boldsymbol{\Sigma}, \boldsymbol{O}))}{N}
\frac{1}{N}\boldsymbol{q}_{0}^{\mathrm{H}}\boldsymbol{q}_{0} + o(1).  
\label{bm0}
\end{IEEEeqnarray}

In particular, for $\boldsymbol{D}=\boldsymbol{I}_{N}$ 
we use the definition~(\ref{gamma1}) of $\gamma_{0}$ to obtain 
\begin{equation} \label{trace1}
\frac{\gamma_{0}}{N}
\mathrm{Tr}\left\{
 \tilde{\boldsymbol{W}}_{0}(\boldsymbol{\Sigma}, \boldsymbol{O})
\right\}
\aeq 1 + o(1). 
\end{equation}
Applying (\ref{trace1}) to (\ref{bm0}), we find  
\begin{equation} \label{bm0_tmp}
\frac{\gamma_{0}}{N}\boldsymbol{b}_{0}^{\mathrm{H}}
\tilde{\boldsymbol{W}}_{0}
\left\{
 (\boldsymbol{\Sigma}, \boldsymbol{O})\boldsymbol{b}_{0}
 - \tilde{\boldsymbol{w}}
\right\} 
\aeq \frac{1}{N}\boldsymbol{q}_{0}^{\mathrm{H}}\boldsymbol{q}_{0} + o(1).  
\end{equation}
From the definition~(\ref{m}) of $\boldsymbol{m}_{0}$, (\ref{bbqq}) 
with $\boldsymbol{D}=\boldsymbol{I}_{N}$ for $\tau=0$, and 
(\ref{bm0_tmp}), we arrive at (\ref{bm}) for $\tau=0$. 
\end{IEEEproof}

\begin{IEEEproof}[Eqs.~(\ref{module_A_var_SE})--(\ref{norm_M}) for $\tau=0$]
The almost sure convergence~(\ref{module_A_var_SE}) for $\tau=0$ follows 
from the update rule~(\ref{module_A_var}) of $v_{\mathrm{A}\to\mathrm{B}}^{0}$, 
the definition~(\ref{gamma1}) of $\gamma_{0}$, the SE~(\ref{module_A_SE}) 
for module~A, and 
$v_{\mathrm{B}\to\mathrm{A}}^{0}=\bar{v}_{\mathrm{B}\to\mathrm{A}}^{0}=1$. 

Let us prove (\ref{mm}) for $\tau=0$, before proving (\ref{MSE_A_to_B}). 
Using the definition~(\ref{m}) of $\boldsymbol{m}_{0}$, (\ref{bm0}), 
and Assumption~\ref{assumption_w}, 
as well as (\ref{bbqq}) for $\tau=0$, we have 
\begin{IEEEeqnarray}{rl}
&\frac{\boldsymbol{m}_{0}^{\mathrm{H}}\boldsymbol{D}\boldsymbol{m}_{0}}{N} 
\nonumber \\
\aeq& 
\frac{\mathrm{Tr}(\boldsymbol{D})}{N}
\frac{\boldsymbol{q}_{0}^{\mathrm{H}}\boldsymbol{q}_{0}}{N} 
- 2\gamma_{0}\frac{\mathrm{Tr}(\boldsymbol{D}
\tilde{\boldsymbol{W}}_{0}(\boldsymbol{\Sigma}, \boldsymbol{O}))}{N}
\frac{\boldsymbol{q}_{0}^{\mathrm{H}}\boldsymbol{q}_{0}}{N} 
\nonumber \\
+&\frac{\gamma_{0}^{2}\mathrm{Tr}(\tilde{\boldsymbol{D}})}{N}
\frac{\boldsymbol{q}_{0}^{\mathrm{H}}\boldsymbol{q}_{0}}{N} 
+ \frac{\sigma^{2}\gamma_{0}^{2}}{N}\mathrm{Tr}\left(
 \tilde{\boldsymbol{W}}_{0}^{\mathrm{H}}\boldsymbol{D}\tilde{\boldsymbol{W}}_{0}
\right)
+ o(1) \label{mm0_tmp}
\end{IEEEeqnarray}
in the large system limit, with 
\begin{equation}
\tilde{\boldsymbol{D}} 
= \begin{pmatrix}
\boldsymbol{\Sigma} \\
\boldsymbol{O} 
\end{pmatrix}
\tilde{\boldsymbol{W}}_{0}^{\mathrm{H}}\boldsymbol{D}\tilde{\boldsymbol{W}}_{0}
(\boldsymbol{\Sigma}, \boldsymbol{O}). 
\end{equation}
It is straightforward to confirm the boundedness of (\ref{mm0_tmp}). 
Thus, (\ref{mm}) holds for $\tau=0$. 

In particular, for $\boldsymbol{D}=\boldsymbol{I}_{N}$ we have 
\begin{equation}
\mathrm{Tr}(\tilde{\boldsymbol{D}}) 
= \mathrm{Tr}\left\{
 \left(
  \sigma^{2}\boldsymbol{I}_{M} + v_{\mathrm{B}\to \mathrm{A}}^{0}\boldsymbol{\Sigma}^{2}
 \right)^{-2}\boldsymbol{\Sigma}^{4} 
\right\}, 
\end{equation}
\begin{equation}
\mathrm{Tr}\left(
 \tilde{\boldsymbol{W}}_{0}^{\mathrm{H}}\tilde{\boldsymbol{W}}_{0}
\right)
= \mathrm{Tr}\left\{
 \left(
  \sigma^{2}\boldsymbol{I}_{M} + v_{\mathrm{B}\to \mathrm{A}}^{0}\boldsymbol{\Sigma}^{2}
 \right)^{-2}\boldsymbol{\Sigma}^{2}
\right\}. 
\end{equation}
Applying these results, (\ref{trace1}), and 
$N^{-1}\|\boldsymbol{q}_{0}\|^{2}=N^{-1}\|\boldsymbol{x}\|^{2}\ato 
v_{\mathrm{B}\to\mathrm{A}}^{0}=1$ to (\ref{mm0_tmp}), we obtain 
(\ref{MSE_A_to_B}) for $\tau=0$. 

To prove the moment property~(\ref{4th_moment_m}) for $\tau=0$, we observe  
that $\boldsymbol{b}_{0}$ has finite $(2+\epsilon)$th moments, by using
the definition~(\ref{b}) of $\boldsymbol{b}_{0}$, 
$\boldsymbol{q}_{0}=-\boldsymbol{x}$, and Assumption~\ref{assumption_x}. 
Thus, the moment property~(\ref{4th_moment_m}) 
for $\tau=0$ follows from the definition~(\ref{m}) of $\boldsymbol{m}_{0}$, 
the definition~(\ref{Wt}) of $\tilde{\boldsymbol{W}}_{0}$, 
Assumption~\ref{assumption_A}, and Assumption~\ref{assumption_w}. 

Finally, (\ref{norm_M}) for $\tau=0$ is equivalent to 
$N^{-1}\|\boldsymbol{m}_{0}\|^{2}\aeq\bar{v}_{\mathrm{A}\to\mathrm{B}}^{0}>0$, 
which follows from (\ref{MSE_A_to_B}) for $\tau=0$.  
\end{IEEEproof}

\subsection{Module~B for $\tau=0$}
\begin{IEEEproof}[Property~\ref{property_B1} for $\tau=0$]
We prove (\ref{h_distribution}) for $\tau=0$. 
Applying Lemma~\ref{lemma_conditional_distribution} with 
$\boldsymbol{X}=\boldsymbol{b}_{0}$ and $\boldsymbol{Y}=\boldsymbol{q}_{0}$ 
to the definition~(\ref{h}) of $\boldsymbol{h}_{0}$ yields 
\begin{equation} 
\boldsymbol{h}_{0} 
\sim 
\frac{\boldsymbol{b}_{0}^{\mathrm{H}}\boldsymbol{m}_{0}}
{\|\boldsymbol{q}_{0}\|^{2}}\boldsymbol{q}_{0} 
+ \boldsymbol{\Phi}_{\boldsymbol{q}_{0}}^{\perp}\tilde{\boldsymbol{V}}
(\boldsymbol{\Phi}_{\boldsymbol{b}_{0}}^{\perp})^{\mathrm{H}}\boldsymbol{m}_{0}  
\end{equation}
conditioned on $\Theta$ and $\mathcal{X}_{0,1}$, in which 
$\tilde{\boldsymbol{V}}\in{\cal U}_{N-1}$ is a Haar matrix and independent 
of $\Theta$ and $\mathcal{X}_{0,1}$. 
From $\boldsymbol{q}_{0}=-\boldsymbol{x}$, Assumption~\ref{assumption_x}, 
and (\ref{bm}) for $\tau=0$, we find 
\begin{equation} \label{h_0}
\boldsymbol{h}_{0} 
\sim \boldsymbol{q}_{0}o(1) 
+ \boldsymbol{\Phi}_{\boldsymbol{q}_{0}}^{\perp}\tilde{\boldsymbol{V}}
(\boldsymbol{\Phi}_{\boldsymbol{b}_{0}}^{\perp})^{\mathrm{H}}\boldsymbol{m}_{0} 
\end{equation}
in the large system limit, which implies (\ref{h_distribution}) for $\tau=0$, 
because of the notational convention. 

In order to complete the proof, we shall prove (\ref{nu}) for $\tau=0$.  
Define  
\begin{equation} \label{nu0_tmp}
\nu_{0}
= \frac{1}{N}\boldsymbol{m}_{0}^{\mathrm{H}}
\boldsymbol{P}_{\boldsymbol{b}_{0}}^{\perp}\boldsymbol{m}_{0}.
\end{equation} 
Applying $\boldsymbol{P}_{\boldsymbol{b}_{0}}^{\perp}
=\boldsymbol{I}_{N}-\|\boldsymbol{b}_{0}\|^{-2}\boldsymbol{b}_{0}
\boldsymbol{b}_{0}^{\mathrm{H}}$ to (\ref{nu0_tmp}), and using 
(\ref{bbqq}) and (\ref{bm}) for $\tau=0$, we have 
\begin{equation} \label{nu0}
\nu_{0} 
\aeq \frac{1}{N}
\boldsymbol{m}_{0}^{\mathrm{H}}\boldsymbol{m}_{0} + o(1) 
\aeq \bar{v}_{\mathrm{A}\to\mathrm{B}}^{0} + o(1) 
\end{equation}
in the large system limit, 
where the last equality follows from (\ref{MSE_A_to_B}) for $\tau=0$. 
In particular, we have the convention 
$\boldsymbol{m}_{0}^{\perp}=\boldsymbol{m}_{0}$ to 
find (\ref{nu}) for $\tau=0$. 
Thus, Property~\ref{property_B1} holds for $\tau=0$. 
\end{IEEEproof}

Let $\boldsymbol{X}_{N}=\tilde{\boldsymbol{z}}_{0}$ given in (\ref{z_tilde}), 
$\boldsymbol{a}_{0,N}=\boldsymbol{x}$, 
$\boldsymbol{\epsilon}_{N}=\boldsymbol{q}_{0}o(1)$, and
$\boldsymbol{E}_{N}=\boldsymbol{q}_{0}$. For $k=1$ or $k=2$, 
we prove that all conditions in 
Lemma~\ref{lemma_SLLN} with $v=\bar{v}_{\mathrm{A}\to\mathrm{B}}^{0}$ 
are satisfied for any pseudo-Lipschitz function  
$f_{n}:\mathbb{C}^{2}\to\mathbb{C}$ of order~$k$ with an $n$-independent 
Lipschitz constant $L>0$. Thus, 
\begin{equation} 
\mathbb{E}\left[
 f_{n}(x_{n}, h_{0,n})
\right]
\aeq \mathbb{E}\left[
 f_{n}(x_{n}, \tilde{z}_{0,n})  
\right] + o(1), \label{module_B_identity0_individual}
\end{equation}
\begin{IEEEeqnarray}{rl} 
\frac{1}{N}\sum_{n=1}^{N}f_{n}(x_{n}, h_{0,n})
\aeq& \frac{1}{N}\sum_{n=1}^{N}\mathbb{E}_{\tilde{z}_{0,n}}\left[
 f_{n}(x_{n}, \tilde{z}_{0,n})
\right] + o(1) \nonumber \\
\aeq& \frac{1}{N}\sum_{n=1}^{N}\mathbb{E}\left[
 f_{n}(x_{n}, \tilde{z}_{0,n})  
\right] + o(1)\label{module_B_identity0}
\end{IEEEeqnarray}
in the large system limit, 
with $\tilde{\boldsymbol{z}}_{0}\sim\mathcal{CN}(\boldsymbol{0}, 
\bar{v}_{\mathrm{A}\to\mathrm{B}}^{0}\boldsymbol{I}_{N})$, where the latter 
equality follows from Assumption~\ref{assumption_x}. 

The conditions~(\ref{assumption_a1}), (\ref{assumption_a2}), 
(\ref{assumption_a3}), and (\ref{assumption_E}) follow 
from $\boldsymbol{q}_{0}=-\boldsymbol{x}$, Assumption~\ref{assumption_x}, 
and Theorem~\ref{theorem_SLLN}. 
The condition~(\ref{norm_E}) is due to $N^{-1}\|\boldsymbol{q}_{0}\|^{2}\ato1$. 
The condition~(\ref{convergence_X}) with 
$v=\bar{v}_{\mathrm{A}\to\mathrm{B}}^{0}$ follows from (\ref{MSE_A_to_B}) and 
(\ref{nu}) for $\tau=0$, as well as the convention 
$\boldsymbol{m}_{0}^{\perp}=\boldsymbol{m}_{0}$. 
The moment conditions of $\boldsymbol{a}_{N}$, $\boldsymbol{\epsilon}_{N}$, 
and $\boldsymbol{E}_{N}$ are due to $\boldsymbol{q}_{0}=-\boldsymbol{x}$ 
and Assumption~\ref{assumption_x}. 
The moment condition of $\boldsymbol{X}$ follows from the 
definition~(\ref{z_tilde}) of 
$\tilde{\boldsymbol{z}}_{0}$ and (\ref{4th_moment_m}) for $\tau=0$. 
Thus, all conditions in Lemma~\ref{lemma_SLLN} are satisfied. 

\begin{IEEEproof}[Eqs.~(\ref{vB_SE}) and (\ref{module_B_var_SE}) for $\tau=0$]
We first prove (\ref{vB_SE}) for $\tau=0$. 
From the definition~(\ref{vB}) of the posterior variance $v_{\mathrm{B}}^{1}$, 
we have  
\begin{equation}
v_{\mathrm{B}}^{1}
= \frac{1}{N}\sum_{n=1}^{N}f_{n}(x_{n},h_{0,n}), 
\end{equation}
with $f_{n}(x,z)=\mathbb{V}[x_{n} | x_{n,\mathrm{A}\to\mathrm{B}}^{0}=x+z]$ defined 
via the virtual AWGN observation~(\ref{virtual_AWGN}). 
From Assumption~\ref{assumption_posterior}, 
the posterior variance $\mathbb{V}[x_{n} | x_{n,\mathrm{A}\to\mathrm{B}}^{0}]$ is 
bounded, so that $f_{n}(x,z)$ is a Lipschitz-continuous function with a 
Lipschitz constant $L>0$. We use (\ref{module_B_identity0}) to arrive at 
\begin{equation}
v_{\mathrm{B}}^{1}
\aeq \mathrm{MMSE}(\bar{v}_{\mathrm{A}\to\mathrm{B}}^{0}) + o(1) 
\end{equation}
in the large system limit, where we have used the fact that the expectation 
of the posterior variance is equal to the MMSE~(\ref{MMSE}). 
Thus, (\ref{vB_SE}) holds for $\tau=0$. 

We next prove (\ref{module_B_var_SE}) for $\tau=0$. 
From (\ref{module_A_var_SE}) and (\ref{vB_SE}) for $\tau=0$, 
we observe that $v_{\mathrm{B}\to\mathrm{A}}^{1}$ given in  
(\ref{module_B_var}) converges almost surely to 
$\bar{v}_{\mathrm{B}\to\mathrm{A}}^{1}$ given in (\ref{module_B_SE}) in the large 
system limit. Thus, (\ref{module_B_var_SE}) holds for $\tau=0$. 
\end{IEEEproof}

\begin{IEEEproof}[Eq.~(\ref{hq}) for $\tau=0$]
The Lipschitz-continuity of $\tilde{\eta}_{0}$ proved in   
Lemma~\ref{lemma_eta} implies that 
$f_{n}(x_{n},z)=z^{*}\tilde{\eta}_{0}(x_{n}+z)$ is a pseudo-Lipschitz 
function of order~$2$ with an $n$-independent Lipschitz constant $L>0$. 
From (\ref{module_B_identity0}), we obtain  
\begin{IEEEeqnarray}{rl}
\frac{1}{N}\boldsymbol{h}_{0}^{\mathrm{H}}\tilde{\eta}_{0}(\boldsymbol{x}+
\boldsymbol{h}_{0})
\aeq& \mathbb{E}\left[
  \tilde{z}_{0,n}^{*}\tilde{\eta}_{0}(x_{n}+\tilde{z}_{0,n})
\right] + o(1) 
\end{IEEEeqnarray}
in the large system limit. Using Lemma~\ref{lemma_eta} yields   
\begin{equation} \label{eta_MMSE}
\frac{1}{N}\boldsymbol{h}_{0}^{\mathrm{H}}\tilde{\eta}_{0}(\boldsymbol{x}+
\boldsymbol{h}_{0})
\aeq \mathrm{MMSE}(\bar{v}_{\mathrm{A}\to\mathrm{B}}^{0}) + o(1)
\end{equation}
in the large system limit. Similarly, we obtain 
\begin{equation} \label{hx0}
\frac{1}{N}\boldsymbol{h}_{0}^{\mathrm{H}}\boldsymbol{x}\ato0 
\end{equation}
in the large system limit. 

We use the definition~(\ref{q}) of $\boldsymbol{q}_{1}$, 
the definition~(\ref{eta}) of $\eta_{0}$, (\ref{module_B_var_SE}) for 
$\tau=0$, and (\ref{hx0}) to obtain 
\begin{IEEEeqnarray}{rl}
\frac{1}{N}\boldsymbol{h}_{0}^{\mathrm{H}}\boldsymbol{q}_{1}
\aeq& \bar{v}_{\mathrm{B}\to\mathrm{A}}^{1}\left(
 \frac{\boldsymbol{h}_{0}^{\mathrm{H}}
 \tilde{\eta}_{0}(\boldsymbol{x}+\boldsymbol{h}_{0})}
 {N\mathrm{MMSE}(\bar{v}_{\mathrm{A}\to\mathrm{B}}^{0})}
 - \frac{\|\boldsymbol{h}_{0}\|^{2}}{N\bar{v}_{\mathrm{A}\to\mathrm{B}}^{0}}
\right) + o(1) \nonumber \\
\aeq& o(1) 
\end{IEEEeqnarray}
in the large system limit, where the last equality follows from 
the definition~(\ref{h}) of $\boldsymbol{h}_{0}$, 
(\ref{MSE_A_to_B}) for $\tau=0$, and (\ref{eta_MMSE}). 
Thus, (\ref{hq}) holds for $\tau=0$. 
\end{IEEEproof}

\begin{IEEEproof}[Eqs.~(\ref{MMSE_SE}) and (\ref{MSE_B_to_A}) for $\tau=0$]
We first prove (\ref{MMSE_SE}) for $\tau=0$.  
By repeating the proof of (\ref{hq}) for $\tau=0$, we find 
\begin{equation}
\frac{1}{N}\|\tilde{\eta}_{0}(\boldsymbol{x}+\boldsymbol{h}_{0})
-\boldsymbol{x}\|^{2}
\aeq \mathbb{E}\left[
 |\tilde{\eta}_{0}(x_{n}+\tilde{z}_{0,n}) - x_{n}|^{2}
\right] + o(1) 
\end{equation}
in the large system limit. Since the variance of $\tilde{z}_{0.n}$ is equal 
to $\bar{v}_{\mathrm{A}\to\mathrm{B}}^{0}$, from the definition~(\ref{MMSE}) 
of the MMSE function, we arrive at (\ref{MMSE_SE}) for $\tau=0$. 

We next prove (\ref{MSE_B_to_A}) for $\tau=0$.  
Using the definition~(\ref{q}) of $\boldsymbol{q}_{1}$ and 
the definition~(\ref{eta}) of $\eta_{0}$, and the 
definition~(\ref{module_B_SE}) of $\bar{v}_{\mathrm{B}\to\mathrm{A}}^{1}$, 
as well as (\ref{module_A_var_SE}), (\ref{vB_SE}), and (\ref{module_B_var_SE}) 
for $\tau=0$, we have 
\begin{equation} \label{q1}
\boldsymbol{q}_{1} 
= \frac{\bar{v}_{\mathrm{B}\to \mathrm{A}}^{1}
\{\tilde{\eta}_{0}(\boldsymbol{x}+\boldsymbol{h}_{0}) -\boldsymbol{x}\}}
{\mathrm{MMSE}(\bar{v}_{\mathrm{A}\to\mathrm{B}}^{t})}
- \frac{\bar{v}_{\mathrm{B}\to \mathrm{A}}^{1}}{\bar{v}_{\mathrm{A}\to \mathrm{B}}^{0}}
\boldsymbol{h}_{0}. 
\end{equation}
Applying (\ref{hq}) and (\ref{MMSE_SE}) for $\tau=0$, as well as 
(\ref{eta_MMSE}), (\ref{hx0}), and $N^{-1}\|\boldsymbol{h}_{0}\|^{2}
\ato\bar{v}_{\mathrm{A}\to\mathrm{B}}^{0}$ obtained from the definition~(\ref{h}) 
of $\boldsymbol{h}_{0}$ and (\ref{MSE_A_to_B}) for $\tau=0$, we have 
\begin{equation}
\frac{1}{N}\|\boldsymbol{q}_{1}\|^{2} 
\aeq \frac{(\bar{v}_{\mathrm{B}\to \mathrm{A}}^{1})^{2}}
{\mathrm{MMSE}(\bar{v}_{\mathrm{A}\to \mathrm{B}}^{0})} 
- \frac{(\bar{v}_{\mathrm{B}\to \mathrm{A}}^{1})^{2}} 
{\bar{v}_{\mathrm{A}\to \mathrm{B}}^{0}}
+ o(1)
\ato \bar{v}_{\mathrm{B}\to \mathrm{A}}^{1}
\end{equation}
in the large system limit, where the last equality follows from 
the definition~(\ref{module_B_SE}) of $\bar{v}_{\mathrm{B}\to \mathrm{A}}^{1}$. 
Thus, (\ref{MSE_B_to_A}) holds for $\tau=0$. 
\end{IEEEproof}

\begin{IEEEproof}[Eq.~(\ref{4th_moment_q}) for $\tau=0$]
From the definition~(\ref{q}) of $\boldsymbol{q}_{1}$ 
and Proposition~\ref{proposition_bound}, we have 
\begin{equation}
\mathbb{E}[|q_{1,n}|^{2+\epsilon}]
< C\left(
 \mathbb{E}[|\eta_{0}(x_{n}+h_{0,n})|^{2+\epsilon}] 
 + \mathbb{E}[|x_{n}|^{2+\epsilon}] 
\right)
\end{equation}
for some constant $C>0$. Since Assumption~\ref{assumption_x} implies 
the boundedness of the second term, it is sufficient to prove that 
$\eta_{0}(x_{n}+h_{0,n})$ has a finite $(2+\epsilon)$th moment for some 
$\epsilon>0$.

From Lemma~\ref{lemma_eta} $\tilde{\eta}_{0}$ is Lipschitz-continuous, so 
that $\eta_{0}$ given by (\ref{eta}) is so. Thus, we use 
Proposition~\ref{proposition_bound} to have 
\begin{equation}
\mathbb{E}\left[
 |\eta_{0}(x_{n}+h_{0,n})|^{2+\epsilon}
\right]
\leq L\left(
 1 + \mathbb{E}[|x_{n}|^{2+\epsilon}] + \mathbb{E}[|h_{0,n}|^{2+\epsilon}]
\right),  
\end{equation}
for some $L>0$. The boundedness of $\mathbb{E}[|h_{0,n}|^{2+\epsilon}]$ follows 
from the definition~(\ref{h}) of $\boldsymbol{h}_{0}$ and 
(\ref{4th_moment_m}) for $\tau=0$. 
Thus, (\ref{4th_moment_q}) holds for $\tau=0$. 
\end{IEEEproof}

\begin{IEEEproof}[Eq.~(\ref{norm_Q}) for $\tau=0$]
If $\liminf_{M=\delta N\to\infty}N^{-1}\|\boldsymbol{q}_{1}^{\perp}\|^{2}$ converges 
almost surely to a strictly positive constant, (\ref{norm_Q}) holds 
for $\tau=0$~\cite[Lemmas 8 and 9]{Bayati11}. 
Using (\ref{MSE_B_to_A}) for $\tau=0$, we have 
\begin{equation}
\frac{\|\boldsymbol{q}_{1}^{\perp}\|^{2}}{N}
= \frac{\boldsymbol{q}_{1}^{\mathrm{H}}
P_{\boldsymbol{q}_{0}}^{\perp}\boldsymbol{q}_{1}}{N}  
\aeq \frac{\mathbb{E}[\|\boldsymbol{q}_{1}\|^{2}]}{N} 
- \left|
 \frac{\sqrt{N}(\boldsymbol{\Phi}_{\boldsymbol{q}_{0}}^{\parallel})^{\mathrm{H}}
 \boldsymbol{q}_{1}}{N}
\right|^{2} + o(1), 
\end{equation}
where $\boldsymbol{q}_{1}$ in the first term is given by 
$\boldsymbol{q}_{1}
=\eta_{0}(\boldsymbol{x}+\tilde{\boldsymbol{z}}_{0})-\boldsymbol{x}$. 

Let $f_{n}(x_{n},z)=\sqrt{N}[\boldsymbol{\Phi}_{\boldsymbol{q}_{0}}^{\parallel}]_{n}^{*}
\{\eta(x_{n}+z)-x_{n}\}$. The function $f_{n}$ is a Lipschitz-continuous  
function with the Lipschitz constant 
$L_{n}=L\sqrt{N}|[\boldsymbol{\Phi}_{\boldsymbol{q}_{0}}^{\parallel}]_{n}|$ for some 
$L>0$. The normalization $\|\boldsymbol{\Phi}_{\boldsymbol{q}_{0}}^{\parallel}\|^{2}=1$ 
implies $N^{-1}\sum_{n=1}^{N}L_{n}^{2}=L$, so that we can use 
(\ref{module_B_identity0}) to obtain 
\begin{equation}
\frac{\sqrt{N}(\boldsymbol{\Phi}_{\boldsymbol{q}_{0}}^{\parallel})^{\mathrm{H}}
 \boldsymbol{q}_{1}}{N}
\aeq \frac{\mathbb{E}\{
(\boldsymbol{\Phi}_{\boldsymbol{q}_{0}}^{\parallel})^{\mathrm{H}}
\mathbb{E}_{\tilde{\boldsymbol{z}}_{0}}[\boldsymbol{q}_{1}]\}}{\sqrt{N}} + o(1). 
\end{equation}
Using the Cauchy-Schwarz inequality yields 
\begin{equation}
\left|
 \mathbb{E}\left[
  (\boldsymbol{\Phi}_{\boldsymbol{q}_{0}}^{\parallel})^{\mathrm{H}}
  \mathbb{E}_{\tilde{\boldsymbol{z}}_{0}}[\boldsymbol{q}_{1}]
 \right]
\right|  
\leq \mathbb{E}\left\{
 \|\mathbb{E}_{\tilde{\boldsymbol{z}}_{0}}[\boldsymbol{q}_{1}]\|
\right\}
\leq \mathbb{E}\left[
 \|\boldsymbol{q}_{1}\|
\right],  
\end{equation}
where the latter inequality follows from Jensen's inequality. 
Thus, we obtain 
\begin{equation}
\frac{\|\boldsymbol{q}_{1}^{\perp}\|^{2}}{N}
\ageq\frac{1}{N}\left\{
 \mathbb{E}[\|\boldsymbol{q}_{1}\|^{2}] 
 - \left(
  \mathbb{E}\left[
   \|\boldsymbol{q}_{1}\|
  \right]
 \right)^{2}
\right\} + o(1) 
\end{equation}
which is strictly positive in the large system limit. 
Thus, (\ref{norm_Q}) holds for $\tau=0$. 
\end{IEEEproof}

\begin{IEEEproof}[Eqs.~(\ref{MMSE_individual_SE}) and 
(\ref{MSE_B_to_A_individual}) for $\tau=0$]
From (\ref{MMSE_SE}) and (\ref{MSE_B_to_A}) for $\tau=0$, 
we may conclude (\ref{MMSE_individual_SE}) and (\ref{MSE_B_to_A_individual}) 
for $\tau=0$, since $\boldsymbol{x}$ and $\boldsymbol{h}_{0}$ have 
identically distributed elements in the large system limit. Nonetheless, 
we present a generic proof applicable to the non-identically-distributed case. 

We only prove (\ref{MMSE_individual_SE}) for $\tau=0$, since 
(\ref{MSE_B_to_A_individual}) can be proved in the same manner. 
Lemma~\ref{lemma_eta} implies that $|\tilde{\eta}_{0}(x_{n}+h_{0,n})-x_{n}|^{2}$ 
is a pseudo-Lipschitz function of order $2$. 
We use (\ref{module_B_identity0_individual}) to have  
\begin{equation}
\mathbb{E}[|\tilde{\eta}_{0}
(x_{n}+h_{0,n})-x_{n}|^{2}]
\to \mathrm{MMSE}(\bar{v}_{\mathrm{A}\to\mathrm{B}}^{0})
\end{equation}
in the large system limit.  
\end{IEEEproof}

We have proved that Theorem~\ref{main_theorem} holds for $\tau=0$. 
Next, we assume that Theorem~\ref{main_theorem} is correct for all 
$\tau<t$, and prove that Theorem~\ref{main_theorem} holds for 
$\tau=t$. 

\subsection{Module~A by Induction}
\begin{IEEEproof}[Property~\ref{property_A1} for $\tau=t$]
We prove (\ref{b_distribution}) for $\tau=t$. 
Let $\boldsymbol{Y}=(\boldsymbol{Q}_{t}, \boldsymbol{H}_{t})$ and 
$\boldsymbol{X}=(\boldsymbol{B}_{t}, \boldsymbol{M}_{t})$ in 
Lemma~\ref{lemma_conditional_distribution}. 
The induction hypotheses~(\ref{norm_M}) and (\ref{norm_Q}) $\tau<t$ 
imply that $\boldsymbol{M}_{t}$ and $\boldsymbol{Q}_{t}$ are full rank. 
From the definition~(\ref{recursion3}) of $\boldsymbol{H}_{t}$ and 
the induction hypothesis~(\ref{hq}) for $\tau<t$, we find that 
$\boldsymbol{Y}$ is full rank. Using the definition~(\ref{b}) of 
$\boldsymbol{b}_{t}$ and Lemma~\ref{lemma_conditional_distribution} yields 
\begin{equation}   \label{bt}
\boldsymbol{b}_{t} 
\sim (\boldsymbol{B}_{t}, \boldsymbol{M}_{t})
(\boldsymbol{Q}_{t}, \boldsymbol{H}_{t})^{\dagger}\boldsymbol{q}_{t}
+ \boldsymbol{\Phi}_{(\boldsymbol{B}_{\tau}, \boldsymbol{M}_{\tau})}^{\perp}
\boldsymbol{z}_{t} 
\end{equation}
conditioned on $\Theta$ and $\mathcal{X}_{t,t}$, 
with $\boldsymbol{z}_{t}$ defined in (\ref{z}). 

We evaluate the first term on the RHS of (\ref{bt}). Using 
the induction hypothesis~(\ref{hq}) for $\tau<t$ yields 
\begin{IEEEeqnarray}{rl}
(\boldsymbol{Q}_{t}, \boldsymbol{H}_{t})^{\dagger}
=& \frac{1}{N}\begin{pmatrix}
 N^{-1}\boldsymbol{Q}_{t}^{\mathrm{H}}\boldsymbol{Q}_{t} & 
 N^{-1}\boldsymbol{Q}_{t}^{\mathrm{H}}\boldsymbol{H}_{t} \\ 
 N^{-1}\boldsymbol{H}_{t}^{\mathrm{H}}\boldsymbol{Q}_{t} &
 N^{-1}\boldsymbol{H}_{t}^{\mathrm{H}}\boldsymbol{H}_{t}
\end{pmatrix}^{-1}
\begin{pmatrix}
\boldsymbol{Q}_{t}^{\mathrm{H}} \\
\boldsymbol{H}_{t}^{\mathrm{H}}
\end{pmatrix} \nonumber \\
\aeq& \begin{pmatrix}
\boldsymbol{Q}_{t}^{\dagger} + o(N^{-1})\boldsymbol{H}_{t}^{\mathrm{H}} 
\\
\boldsymbol{H}_{t}^{\dagger} + o(N^{-1})\boldsymbol{Q}_{t}^{\mathrm{H}}
\end{pmatrix}. \label{QH_dagger}
\end{IEEEeqnarray}
Substituting (\ref{QH_dagger}) into the first term, and using the 
same induction hypothesis again, we obtain 
\begin{equation} \label{bt_first_term}
(\boldsymbol{B}_{t}, \boldsymbol{M}_{t})
(\boldsymbol{Q}_{t}, \boldsymbol{H}_{t})^{\dagger}\boldsymbol{q}_{t}
\aeq \boldsymbol{B}_{t}\boldsymbol{\beta}_{t}
+ \boldsymbol{B}_{t}\boldsymbol{o}(1) 
+ \boldsymbol{M}_{t}\boldsymbol{o}(1), 
\end{equation}
which implies (\ref{b_distribution}) for $\tau=t$. 

We next prove (\ref{mu}) for $\tau=t$. 
Repeating the derivation of (\ref{bt_first_term}) with (\ref{QH_dagger}) 
yields 
\begin{IEEEeqnarray}{rl}
&\frac{1}{N}\|\boldsymbol{z}_{t}\|^{2}
=\frac{1}{N}\boldsymbol{q}_{t}^{\mathrm{H}}
\boldsymbol{P}_{(\boldsymbol{Q}_{t}, \boldsymbol{H}_{t})}^{\perp}
\boldsymbol{q}_{t} \nonumber \\
\aeq& \frac{\boldsymbol{q}_{t}^{\mathrm{H}}}{N}
\left\{
 \boldsymbol{P}_{\boldsymbol{Q}_{t}}^{\perp} 
 - \boldsymbol{P}_{\boldsymbol{H}_{t}}^{\parallel} 
 + o(1)\frac{\boldsymbol{Q}_{t}\boldsymbol{H}_{t}^{\mathrm{H}}}{N}
 + o(1)\frac{\boldsymbol{H}_{t}\boldsymbol{Q}_{t}^{\mathrm{H}}}{N}
\right\}
\boldsymbol{q}_{t} \nonumber \\
\aeq& \frac{1}{N}\|\boldsymbol{q}_{t}^{\perp}\|^{2} 
+ o(1), 
\end{IEEEeqnarray}
where the last equality follows from the induction hypothesis~(\ref{hq}) 
for $\tau<t$. Thus, (\ref{mu}) holds for $\tau=t$. 
\end{IEEEproof}

Let $\boldsymbol{X}_{N}=\boldsymbol{z}_{t}$, and $\boldsymbol{\epsilon}_{N}
=\boldsymbol{B}_{t}\boldsymbol{o}(1) + \boldsymbol{M}_{t}\boldsymbol{o}(1)$, 
$\boldsymbol{a}_{1,N}=\boldsymbol{B}_{t}\boldsymbol{\beta}_{t}$, 
and $\boldsymbol{E}_{N}=(\boldsymbol{B}_{t}, \boldsymbol{M}_{t})$ 
in Lemma~\ref{lemma_SLLN}.  
We prove that, for $k=1$ or $k=2$, all conditions in Lemma~\ref{lemma_SLLN} 
with $v=\mu_{t}=\lim_{M=\delta N\to\infty}
N^{-1}\|\boldsymbol{q}_{t}^{\perp}\|^{2}$ are satisfied for 
any pseudo-Lipschitz function $f_{n}$ of order $k$ with a 
Lipschitz-constant $L_{n}$, in which 
$\lim_{N\to\infty}N^{-1}\sum_{n=1}^{N}L_{n}^{2}<\infty$ holds for $k=1$ and 
in which $\lim_{N\to\infty}N^{-1}\sum_{n=1}^{N}L_{n}^{4}<\infty$ holds for $k=2$, 
so that 
\begin{IEEEeqnarray}{rl} 
&\frac{1}{N}\sum_{n=1}^{N}
f_{n}(b_{t,n}) \nonumber \\
\aeq& \frac{1}{N}\sum_{n=1}^{N}\mathbb{E}\left[
 \left.
  f_{n}([\boldsymbol{B}_{t}\boldsymbol{\beta}_{t}]_{n}+z_{t,n})
 \right| \Theta, \mathcal{X}_{t,t} 
\right] + o(1) \label{module_A_identity}
\end{IEEEeqnarray}
conditioned on $\Theta$ and $\mathcal{X}_{t,t}$ in the large system limit, 
with $\boldsymbol{z}_{t}\sim\mathcal{CN}(\boldsymbol{0}, 
\mu_{t}\boldsymbol{I}_{N})$. 

The conditions~(\ref{assumption_a1}) and (\ref{assumption_E}) follow from 
Proposition~\ref{proposition_norm},
the induction hypotheses~(\ref{bbqq}), (\ref{mm}), and 
(\ref{MSE_B_to_A}) for $\tau<t$. 
For $k=1$, the conditions~(\ref{assumption_a2}) and (\ref{assumption_a3}) 
are trivial. For $k=2$, the condition~(\ref{assumption_a2}) 
is due to (\ref{assumption_a1}). 
The condition~(\ref{assumption_a3}) follows from 
Proposition~\ref{proposition_norm}, the induction 
hypotheses~(\ref{bbqq}), (\ref{MSE_B_to_A}) for $\tau<t$, as well as from 
the boundedness of $\|\boldsymbol{\beta}_{t}\|$,  
\begin{IEEEeqnarray}{l}
\|\boldsymbol{\beta}_{t}\|^{2}
=\frac{\boldsymbol{q}_{t}^{\mathrm{H}}\boldsymbol{Q}_{t}}{N}
\left(
 \frac{1}{N}\boldsymbol{Q}_{t}^{\mathrm{H}}\boldsymbol{Q}_{t}
\right)^{-2}
\frac{\boldsymbol{Q}_{t}^{\mathrm{H}}\boldsymbol{q}_{t}}{N} 
\nonumber \\
\al C\left\|
 \frac{1}{N}\boldsymbol{Q}_{t}^{\mathrm{H}}\boldsymbol{q}_{t} 
\right\|^{2}
\leq \frac{C}{N}\sum_{\tau=0}^{t-1}\|\boldsymbol{q}_{\tau}\|^{2} 
\frac{\|\boldsymbol{q}_{t}\|^{2}}{N}
\al\infty \label{beta_bound} 
\end{IEEEeqnarray}  
for some constant $C>0$, where the first two inequalities follow from 
the induction hypothesis~(\ref{norm_Q}) for $\tau=t-2$ and 
from the Cauchy-Schwarz inequality, respectively, and where the boundedness 
is due to the induction hypothesis~(\ref{MSE_B_to_A}) for $\tau<t$. 

The condition~(\ref{norm_E}) follows from the induction hypotheses 
(\ref{bm}), (\ref{norm_M}), and (\ref{norm_Q}) for $\tau<t$, 
as well as the definition (\ref{recursion1}) of $\boldsymbol{B}_{t}$. 
The condition~(\ref{convergence_X}) with 
$v=\mu_{t}$ follows from (\ref{mu}) for $\tau=t$. 
We have proved all assumptions in Lemma~\ref{lemma_SLLN}. 
Thus, (\ref{module_A_identity}) holds. 

\begin{IEEEproof}[Eqs.~(\ref{bw})--(\ref{bm}) for $\tau=t$]
We first prove (\ref{bw}) for $\tau=t$. 
Define the Lipschitz-continuous function $f_{n}(z)=z^{*}
[\boldsymbol{D}\tilde{\boldsymbol{W}}_{t}\tilde{\boldsymbol{w}}]_{n}$. 
We note that $\tilde{\boldsymbol{W}}_{t}$ given in (\ref{Wt}) is independent of 
$v_{\mathrm{B}\to\mathrm{A}}^{t}$ in the large system limit, because of 
the induction hypothesis~(\ref{module_B_var_SE}) for $\tau=t-1$. 
Repeating the proof of (\ref{trace_bound}), we find that 
$N^{-1}\|\boldsymbol{D}\tilde{\boldsymbol{W}}_{t}
\tilde{\boldsymbol{w}}\|^{2}$ is almost surely bounded as $N\to\infty$. 
Thus, we can use (\ref{module_A_identity}) to obtain 
\begin{equation}
\frac{1}{N}\boldsymbol{b}_{\tau}^{\mathrm{H}}
\boldsymbol{D}\tilde{\boldsymbol{W}}_{t}\tilde{\boldsymbol{w}}
\aeq \frac{1}{N}\boldsymbol{\beta}_{\tau}^{\mathrm{H}}\boldsymbol{B}_{\tau}^{\mathrm{H}}
\boldsymbol{D}\tilde{\boldsymbol{W}}_{t}\tilde{\boldsymbol{w}} + o(1)
\ato 0,
\end{equation}
where the last convergence follows from the induction 
hypothesis~(\ref{bw}) for all $\tau<t$. Thus,  
(\ref{bw}) holds for $\tau=t$. 

We next prove (\ref{bbqq}) for $\tau=t$. 
From (\ref{module_A_identity}) for $f_{n}(z)
=D_{n}|z|^{2}$, we have  
\begin{equation}
\frac{1}{N}\boldsymbol{b}_{\tau}^{\mathrm{H}}\boldsymbol{D}
\boldsymbol{b}_{t}
\aeq \frac{1}{N}\mathbb{E}\left[
 \left.
  \boldsymbol{b}_{\tau}^{\mathrm{H}}\boldsymbol{D}
  (\boldsymbol{B}_{t}\boldsymbol{\beta}_{t}
  + \boldsymbol{z}_{t})
 \right| \Theta, \mathcal{X}_{t,t}
\right] + o(1)
\end{equation}
in the large system limit for all $\tau\leq t$, where $\boldsymbol{b}_{\tau}$ 
is replaced by $\boldsymbol{B}_{t}\boldsymbol{\beta}_{t}
+ \boldsymbol{z}_{t}$ for $\tau=t$. For $\tau<t$, 
we have 
\begin{equation}
\frac{1}{N}\boldsymbol{b}_{\tau}^{\mathrm{H}}\boldsymbol{D}
\boldsymbol{b}_{t}
\aeq \frac{1}{N}\boldsymbol{b}_{\tau}^{\mathrm{H}}\boldsymbol{D}
\boldsymbol{B}_{t}\boldsymbol{\beta}_{t} + o(1). 
\end{equation}
Using the induction hypothesis~(\ref{bbqq}) for $\tau<t$,
$\boldsymbol{q}_{t}^{\parallel}=\boldsymbol{Q}_{t}\boldsymbol{\beta}_{t}$, 
and $\boldsymbol{q}_{\tau'}^{\mathrm{H}}\boldsymbol{q}_{t}^{\perp}=0$ yields 
(\ref{bbqq}) for $\tau=t$ and $\tau'<t$. 

For $\tau=t$, we obtain    
\begin{equation}
\frac{1}{N}\boldsymbol{b}_{t}^{\mathrm{H}}\boldsymbol{D}
\boldsymbol{b}_{t}
\aeq \frac{1}{N}\boldsymbol{\beta}_{t}^{\mathrm{H}}\boldsymbol{B}_{t}^{\mathrm{H}}
\boldsymbol{D}\boldsymbol{B}_{t}\boldsymbol{\beta}_{t} 
+ \frac{\mu_{t}}{N}\mathrm{Tr}(\boldsymbol{D}) + o(1)
\end{equation}
in the large system limit. The induction hypothesis~(\ref{bbqq}) for $\tau<t$ 
implies that the fist term converges almost surely to 
$\lim_{M=\delta N\to\infty}N^{-1}
\|\boldsymbol{q}_{t}^{\parallel}\|^{2}N^{-1}\mathrm{Tr}(\boldsymbol{D})$. 
Thus, (\ref{bbqq}) holds for $\tau=\tau'=t$. 

Finally, we prove (\ref{bm}) for $\tau=t$. Repeating the proof of 
(\ref{trace_bound2}) yields the boundedness of $N^{-1}\mathrm{Tr}\{
(\boldsymbol{D}\tilde{\boldsymbol{W}}_{t}(\boldsymbol{\Sigma}, \boldsymbol{O})
)^{k}\}$ for $k\in[0, 4+\epsilon)$. 
Thus, we can use (\ref{bw}) and (\ref{bbqq}) to find 
\begin{IEEEeqnarray}{rl} 
&\frac{\gamma_{t}}{N}
\boldsymbol{b}_{\tau}^{\mathrm{H}}\boldsymbol{D}
\tilde{\boldsymbol{W}}_{t}\left\{
 (\boldsymbol{\Sigma}, \boldsymbol{O})\boldsymbol{b}_{t} 
 - \tilde{\boldsymbol{w}}
\right\} \nonumber \\
\aeq& \gamma_{t}\frac{\mathrm{Tr}\{\boldsymbol{D}
\tilde{\boldsymbol{W}}_{t}(\boldsymbol{\Sigma}, \boldsymbol{O})\}}{N}
\frac{1}{N}\boldsymbol{q}_{\tau}^{\mathrm{H}}\boldsymbol{q}_{t} 
+ o(1). \label{bm_tmp}
\end{IEEEeqnarray}
In particular, for $\boldsymbol{D}=\boldsymbol{I}_{N}$ we find that 
the almost sure convergence~(\ref{bm}) for $\tau=t$ follows from 
the definition~(\ref{gamma1}) of $\gamma_{t}$, 
the definition~(\ref{Wt}) of $\tilde{\boldsymbol{W}}_{t}$, and 
Assumption~\ref{assumption_A}, as well as the boundedness of 
$N^{-1}\boldsymbol{q}_{\tau}^{\mathrm{H}}\boldsymbol{q}_{t}$, obtained from 
the Cauchy-Schwarz inequality and the induction hypothesis~(\ref{MSE_B_to_A}) 
for $\tau<t$. 
\end{IEEEproof}

\begin{IEEEproof}[Eqs.~(\ref{module_A_var_SE})--(\ref{norm_M}) for $\tau=t$]
The almost sure convergence~(\ref{module_A_var_SE}) for $\tau=t$ follows from 
the definition~(\ref{module_A_var}) of $v_{\mathrm{A}\to\mathrm{B}}^{t}$, 
the definition~(\ref{gamma1}) of $\gamma_{t}$, the 
definition~(\ref{module_A_SE}) of $\bar{v}_{\mathrm{A}\to\mathrm{B}}^{t}$, and 
the induction hypothesis~(\ref{module_B_var_SE}) for $\tau=t-1$. 

The properties (\ref{MSE_A_to_B}) and (\ref{mm}) for $\tau=t$ are obtained 
by repeating the proofs of (\ref{MSE_A_to_B}) and (\ref{mm}) for $\tau=0$. 
The moment property~(\ref{4th_moment_m}) for $\tau=t$ follows from the 
definition~(\ref{m}) of $\boldsymbol{m}_{t}$, the definition~(\ref{Wt}) of 
$\tilde{\boldsymbol{W}}_{t}$, Assumption~\ref{assumption_A}, and 
Assumption~\ref{assumption_w}, since we have already proved the 
boundedness of the $(2+\epsilon)$th moments of $\boldsymbol{b}_{t}$. 

Finally, we prove (\ref{norm_M}) for $\tau=t$. 
The induction hypothesis~(\ref{norm_M}) for 
$\tau<t$ implies that (\ref{norm_M}) holds for $\tau=t$ if 
$\liminf_{M=\delta N\to\infty}N^{-1}\|\boldsymbol{m}_{t}^{\perp}\|^{2}$ converges 
almost surely to a strictly positive constant. 
We use the definition~(\ref{m}) of $\boldsymbol{m}_{t}$, (\ref{MSE_A_to_B}), 
and (\ref{module_A_identity}) for 
$f_{n}(z)=[\sqrt{N}(\boldsymbol{\Phi}_{\boldsymbol{M}_{t}}^{\parallel})^{\mathrm{H}}
\{\boldsymbol{I}_{N} - \gamma_{t}\tilde{\boldsymbol{W}}_{t}
(\boldsymbol{\Sigma}, \boldsymbol{O})\} 
]_{\tau,n}z$ for $\tau<t$ to obtain 
\begin{equation}
\frac{\|\boldsymbol{m}_{t}^{\perp}\|^{2}}{N}
\aeq \frac{\mathbb{E}_{\boldsymbol{z}_{t}}[\|\boldsymbol{m}_{t}\|^{2}]}{N} 
- \left\|
 \mathbb{E}_{\boldsymbol{z}_{t}}\left[
  (\boldsymbol{\Phi}_{\boldsymbol{M}_{t}}^{\parallel})^{\mathrm{H}}
  \frac{\boldsymbol{m}_{t}}{\sqrt{N}}
 \right] 
\right\|^{2} + o(1). 
\end{equation}
By repeating the proof of (\ref{norm_Q}) for $\tau=0$, we arrive at 
\begin{equation}
\frac{\|\boldsymbol{m}_{t}^{\perp}\|^{2}}{N}
\ageq \frac{1}{N}\left(
 \mathbb{E}_{\boldsymbol{z}_{t}}[\|\boldsymbol{m}_{t}\|^{2}] 
 - \|\mathbb{E}_{\boldsymbol{z}_{t}}[\boldsymbol{m}_{t}]\|^{2}
\right) + o(1),
\end{equation}
which is strictly positive in the large system limit. 
Thus, (\ref{norm_M}) holds for $\tau=t$. 
\end{IEEEproof}

\subsection{Module~B by Induction}
\begin{IEEEproof}[Property~\ref{property_B1} for $\tau=t$]
Let us prove (\ref{h_distribution}) for $\tau=t$. 
Using (\ref{h}) and 
Lemma~\ref{lemma_conditional_distribution} with 
$\boldsymbol{Y}=(\boldsymbol{Q}_{t+1}, \boldsymbol{H}_{t})$ and 
$\boldsymbol{X}=(\boldsymbol{B}_{t+1}, \boldsymbol{M}_{t})$ yields 
\begin{equation}
\boldsymbol{h}_{t} 
\sim (\boldsymbol{Q}_{t+1}, \boldsymbol{H}_{t})
(\boldsymbol{B}_{t+1}, \boldsymbol{M}_{t})^{\mathrm{\dagger}}\boldsymbol{m}_{t}  
+ \boldsymbol{\Phi}_{(\boldsymbol{Q}_{t+1}, \boldsymbol{H}_{t})}^{\perp}
\tilde{\boldsymbol{z}}_{t} \label{ht}
\end{equation}
conditioned on $\Theta$ and $\mathcal{X}_{t,t+1}$, with 
$\tilde{\boldsymbol{z}}_{t}$ given in (\ref{z_tilde}), where we have used 
the identity $\boldsymbol{X}^{\mathrm{H}}\boldsymbol{X}=\boldsymbol{Y}^{\mathrm{H}}
\boldsymbol{Y}$. Repeating the proof of Property~\ref{property_A1} 
for $\tau=t$, 
we arrive at Property~\ref{property_B1} for $\tau=t$. 
\end{IEEEproof}

Let $\boldsymbol{X}_{N}=\tilde{\boldsymbol{z}}_{t}$ given in (\ref{z_tilde}), 
$\boldsymbol{a}_{0,N}=\boldsymbol{x}$, 
$\boldsymbol{a}_{\tau+1,N}=\boldsymbol{h}_{\tau}$ for $\tau<t$, 
$\boldsymbol{a}_{t+1,N}=\boldsymbol{H}_{t}\boldsymbol{\alpha}_{t}$, 
$\boldsymbol{\epsilon}_{N}=
\boldsymbol{Q}_{t+1}\boldsymbol{o}(1) + \boldsymbol{H}_{t}\boldsymbol{o}(1)$, 
and $\boldsymbol{E}_{N}=(\boldsymbol{Q}_{t+1}, \boldsymbol{H}_{t})$. 
For $k=1$ or $k=2$, 
let $f_{n}:\mathbb{C}^{t+2}\to\mathbb{C}$ denote a pseudo-Lipschitz function of 
order~$k$ with an $n$-independent Lipschitz constant $L>0$. 
We shall prove 
\begin{IEEEeqnarray}{rl}  
&\mathbb{E}[f_{n}(x_{n}, h_{0,n},\ldots, h_{t,n})] \nonumber \\ 
\aeq& \mathbb{E}[g_{n}(x_{n}, h_{0,n},\ldots,h_{t-1,n})] + o(1),  
\label{module_B_identity_tmp_individual}
\end{IEEEeqnarray}
\begin{IEEEeqnarray}{rl}  
&\frac{1}{N}\sum_{n=1}^{N}f_{n}(x_{n}, h_{0,n},\ldots, h_{t,n}) \nonumber \\ 
\aeq& \frac{1}{N}\sum_{n=1}^{N}
g_{n}(x_{n}, h_{0,n},\ldots,h_{t-1,n}) + o(1) \label{module_B_identity_tmp}
\end{IEEEeqnarray}
conditioned on $\Theta$ and $\mathcal{X}_{t,t+1}$ in the large system limit, 
with 
\begin{IEEEeqnarray}{rl}
&g_{n}(x_{n}, h_{0,n},\ldots,h_{t-1,n}) \nonumber \\ 
=& \mathbb{E}_{\tilde{z}_{t,n}}\left[
 f_{n}(x_{n}, h_{0,n},\ldots, h_{t-1,n}, 
 [\boldsymbol{H}_{t}\boldsymbol{\alpha}_{t}]_{n}  + \tilde{z}_{t,n})
\right], \label{g}
\end{IEEEeqnarray}
where $\tilde{\boldsymbol{z}}_{t}\sim\mathcal{CN}(\boldsymbol{0}, 
\nu_{t}\boldsymbol{I}_{N})$ is a CSCG vector with 
$\nu_{t}=\lim_{M=\delta N\to\infty}N^{-1}\|\boldsymbol{m}_{t}^{\perp}\|^{2}$. 

It is sufficient to confirm that all conditions in Lemma~\ref{lemma_SLLN} 
hold. The conditions~(\ref{assumption_a1}) and (\ref{assumption_E}) follow 
from the definition~(\ref{h}) of $\boldsymbol{h}_{t}$, 
the induction hypotheses~(\ref{MSE_A_to_B}), 
(\ref{MSE_B_to_A}) for $\tau<t$, and Proposition~\ref{proposition_norm}. 
The conditions~(\ref{assumption_a2}) and (\ref{assumption_a3}) are 
trivial for $k=1$. For $k=2$, the condition~(\ref{assumption_a2}) is due to 
(\ref{assumption_a1}). The condition~(\ref{assumption_a3})  
follows from Assumption~\ref{assumption_x} when $\tau=0$, 
the definition~(\ref{h}) of $\boldsymbol{h}_{\tau}$ and the induction 
hypothesis~(\ref{MSE_A_to_B}) when $\tau=1,\ldots,t-1$, and from the 
boundedness of $\|\boldsymbol{\alpha}_{t}\|^{2}$ when $\tau=t$, obtained by 
repeating the proof of (\ref{beta_bound}) with the induction 
hypotheses~(\ref{MSE_A_to_B}) and (\ref{norm_M}) for $\tau<t$. 
The condition~(\ref{norm_E}) is due to the induction 
hypotheses~(\ref{norm_M}), (\ref{hq}), and (\ref{norm_Q}) for $\tau<t-1$, 
as well as the definition~(\ref{recursion3}) of $\boldsymbol{H}_{t}$.    
The condition~(\ref{convergence_X}) follows from (\ref{nu}) for $\tau=t$. 

The moment conditions of $\boldsymbol{\epsilon}_{N}$, $\boldsymbol{a}_{\tau,N}$, 
and $\boldsymbol{E}_{N}$ follow from Assumption~\ref{assumption_x}, 
the induction hypothesis~(\ref{4th_moment_q}) for $\tau<t$, and 
the boundedness of the $(2+\epsilon)$th moments of $\boldsymbol{h}_{\tau}$ 
for $\tau<t$, of which the last is due to the definition~(\ref{h}) of 
$\boldsymbol{h}_{\tau}$ and the induction hypothesis~(\ref{4th_moment_m}) 
for $\tau< t$.  The moment condition of $\boldsymbol{X}_{N}$ is 
due to (\ref{4th_moment_m}) for $\tau=t$ and the definition~(\ref{z_tilde}) of 
$\tilde{\boldsymbol{z}}_{t}$. 
Thus, all conditions in Lemma~\ref{lemma_SLLN} hold. 

Define $\boldsymbol{h}_{\tau}^{\mathrm{G}}$ recursively as 
\begin{equation} \label{h_G}
\boldsymbol{h}_{\tau}^{\mathrm{G}} 
= \boldsymbol{H}_{\tau}^{\mathrm{G}}\boldsymbol{\alpha}_{\tau} 
+ \tilde{\boldsymbol{z}}_{\tau},  
\end{equation}
with $\boldsymbol{H}_{\tau}^{\mathrm{G}}=(\boldsymbol{h}_{0}^{\mathrm{G}},\ldots,
\boldsymbol{h}_{\tau-1}^{\mathrm{G}})$, where 
$\{\tilde{\boldsymbol{z}}_{\tau}\sim\mathcal{CN}(\boldsymbol{0}, 
\nu_{t}\boldsymbol{I}_{N})\}$ are independent CSCG vectors.  
By definition, $\boldsymbol{h}_{\tau}^{\mathrm{G}}$ 
conditioned on $\{\boldsymbol{\alpha}_{\tau}\}$ and $\{\nu_{\tau}\}$ is 
a CSCG vector. Comparing the definition~(\ref{h_tilde}) of 
$\tilde{\boldsymbol{h}}_{\tau}$ and the definition~(\ref{h_G}) of 
$\boldsymbol{h}_{\tau}^{\mathrm{G}}$, 
from the definition~(\ref{h}) of $\boldsymbol{h}_{t}$ and 
(\ref{MSE_A_to_B}) for $\tau=t$ we find  
$N^{-1}\mathbb{E}[\|\boldsymbol{h}_{t}^{\mathrm{G}}\|^{2}]\to 
\bar{v}_{\mathrm{A}\to\mathrm{B}}^{t}$ in the large system limit. 

It is straightforward to confirm that the function~(\ref{g}) is 
pseudo-Lipschitz of order~$k$ with an $n$-independent Lipschitz constant. 
Thus, we can repeat the argument in (\ref{module_B_identity_tmp_individual}) 
and (\ref{module_B_identity_tmp}) to arrive at 
\begin{IEEEeqnarray}{rl}  
&\mathbb{E}[f_{n}(x_{n}, h_{0,n},\ldots, h_{t,n})] \nonumber \\ 
\aeq& \mathbb{E}[f_{n}(x_{n}, h_{0,n}^{\mathrm{G}},\ldots,h_{t-1,n}^{\mathrm{G}})] 
+ o(1),  \label{module_B_identity_individual}
\end{IEEEeqnarray}
\begin{IEEEeqnarray}{rl}  
&\frac{1}{N}\sum_{n=1}^{N}f_{n}(x_{n},h_{0,n},\ldots,h_{t,n}) 
\nonumber \\ 
\aeq&\frac{1}{N}\sum_{n=1}^{N}\mathbb{E}\left[
 f_{n}(x_{n,N},h_{0,n}^{\mathrm{G}},\ldots,h_{t,n}^{\mathrm{G}})
\right] + o(1). \label{module_B_identity}
\end{IEEEeqnarray} 

\begin{IEEEproof}[Eqs.~(\ref{vB_SE}) and (\ref{module_B_var_SE}) for $\tau=t$]
Repeating the proofs of (\ref{vB_SE}) and (\ref{module_B_var_SE}) 
for $\tau=0$ with (\ref{module_B_identity}), we arrive at (\ref{vB_SE}) 
and (\ref{module_B_var_SE}) for $\tau=t$.  
\end{IEEEproof}

\begin{IEEEproof}[Eq.~(\ref{hq}) for $\tau=t$]
For $\tau<t$, we use the definition~(\ref{b}) of $\boldsymbol{b}_{t}$ and 
the definition~(\ref{h}) of $\boldsymbol{h}_{t}$ to obtain 
\begin{equation}
\frac{1}{N}\boldsymbol{h}_{t}^{\mathrm{H}}
\boldsymbol{q}_{\tau+1} 
= \frac{1}{N}\boldsymbol{m}_{t}^{\mathrm{H}}\boldsymbol{b}_{\tau+1}
\ato o(1)
\end{equation}
in the large system limit, where the last convergence follows from 
(\ref{bm}) for $\tau=t$ and $\tau'=\tau+1\leq t$. 

For $\tau=t$, we use (\ref{module_B_identity}) for the pseudo-Lipschitz 
function $f_{n}(x_{n},h_{t,n})=h_{t,n}^{*}\{\eta_{t}(x_{n}+h_{t,n})-x_{n}\}$ 
of order~$2$ to have 
\begin{equation} \label{hqt}
\frac{1}{N}\boldsymbol{h}_{t}^{\mathrm{H}}\boldsymbol{q}_{t+1} 
\aeq \frac{1}{N}\mathbb{E}\left[
 (\boldsymbol{h}_{t}^{\mathrm{G}})^{\mathrm{H}}
 \eta_{t}(\boldsymbol{x}+\boldsymbol{h}_{t}^{\mathrm{G}})
\right] + o(1)
\end{equation}
in the large system limit. 
Since $\boldsymbol{h}_{t}^{\mathrm{G}}$ has independent CSCG elements 
with variance $\bar{v}_{\mathrm{A}\to\mathrm{B}}^{t}$, we repeat the proof of 
(\ref{hq}) for $\tau=0$ to obtain (\ref{hq}) for $\tau=\tau'=t$. 
\end{IEEEproof}

\begin{IEEEproof}[Eqs.~(\ref{MMSE_SE})--(\ref{4th_moment_q}) for $\tau=t$]
Repeat the proofs of (\ref{MMSE_SE}), (\ref{MSE_B_to_A}), and 
(\ref{4th_moment_q}) for $\tau=0$ with (\ref{module_B_identity}). 
\end{IEEEproof}

\begin{IEEEproof}[Eq.~(\ref{norm_Q}) for $\tau=t$]
Repeat the proof of (\ref{norm_M}) for $\tau=t$ with 
(\ref{module_B_identity_tmp}).  
\end{IEEEproof}

\begin{IEEEproof}[Eqs.~(\ref{MMSE_individual_SE})--(\ref{MSE_B_to_A_individual}) 
for $\tau=t$]
Repeat the proofs of (\ref{MMSE_individual_SE})--(\ref{MSE_B_to_A_individual}) 
for $\tau=0$ with (\ref{module_B_identity_individual}).  
\end{IEEEproof}

\appendices 

\section{Proof of Lemma~\ref{lemma_SLLN}} 
\label{proof_lemma_SLLN}
\subsection{Technical Results}
Consider $t=0$. Since 
$\boldsymbol{X}_{N}\in\mathbb{C}^{N}$ is unitarily 
invariant, we use the SVD of $\boldsymbol{X}_{N}$ to obtain 
$\boldsymbol{X}_{N}=\boldsymbol{\Phi}_{\boldsymbol{X}_{N}}^{\parallel}
\|\boldsymbol{X}_{N}\|$, 
in which $\boldsymbol{\Phi}_{\boldsymbol{X}_{N}}^{\parallel}
\in{\cal U}_{N\times 1}$ is 
Haar-distributed and independent of the singular value 
$\|\boldsymbol{X}_{N}\|$~\cite{Tulino04}. Furthermore, 
$\boldsymbol{u}\sim\mathcal{CN}(\boldsymbol{0}, 
\boldsymbol{I}_{N})$ is unitarily invariant, so that its SVD is given by 
$\boldsymbol{u}
=\boldsymbol{\Phi}_{\boldsymbol{u}}^{\parallel}\|\boldsymbol{u}\|$, 
in which $\boldsymbol{\Phi}_{\boldsymbol{u}}^{\parallel}
\in{\cal U}_{N\times1}$ is Haar-distributed and independent of 
$\|\boldsymbol{u}\|$. Since 
$\boldsymbol{\Phi}_{\boldsymbol{X}_{N}}^{\parallel}\sim
\boldsymbol{\Phi}_{\boldsymbol{u}}^{\parallel}$ holds, 
we have the following representation: 
\begin{equation} \label{representation0}
\boldsymbol{\epsilon}_{N}+\boldsymbol{X}_{N}
\sim \boldsymbol{\epsilon}_{N}
+\frac{\|\boldsymbol{X}_{N}\|}{\|\boldsymbol{u}\|}\boldsymbol{u}.  
\end{equation}

Let $\mathcal{N}=\{1,\ldots,t\}$ for $t>0$. 
We repeat the same argument to obtain 
\begin{IEEEeqnarray}{rl}  
\boldsymbol{\epsilon}_{N}
+\boldsymbol{\Phi}_{\boldsymbol{E}_{N}}^{\perp}\boldsymbol{X}_{N-t}
\sim& \boldsymbol{\epsilon}_{N}
+\frac{\|\boldsymbol{X}_{N-t}\|}{\|\boldsymbol{u}_{\backslash\mathcal{N}}\|}
\boldsymbol{\Phi}_{\boldsymbol{E}_{N}}^{\perp}\boldsymbol{u}_{\backslash\mathcal{N}} 
\nonumber \\ 
=& \tilde{\boldsymbol{\epsilon}}_{N} 
+ \frac{\|\boldsymbol{X}_{N-t}\|}{\|\boldsymbol{u}_{\backslash\mathcal{N}}\|}
\boldsymbol{z}, 
\label{representation}
\end{IEEEeqnarray} 
with $\boldsymbol{z}=\boldsymbol{\Phi}_{\boldsymbol{E}_{N}}\boldsymbol{u}
\sim\mathcal{CN}(\boldsymbol{0},\boldsymbol{I}_{N})$,  
$\tilde{\boldsymbol{\epsilon}}_{N}=\boldsymbol{\epsilon}_{N}
-\boldsymbol{E}_{N}\boldsymbol{\delta}_{N}$, and 
\begin{equation} \label{epsilon}
\boldsymbol{\delta}_{N}
=\frac{\|\boldsymbol{X}_{N-t}\|}{\|\boldsymbol{u}_{\backslash\mathcal{N}}\|}
(\boldsymbol{E}_{N}^{\mathrm{H}}\boldsymbol{E}_{N})^{-1}
\boldsymbol{E}_{N}^{\mathrm{H}}
\boldsymbol{\Phi}_{\boldsymbol{E}_{N}}^{\parallel}
\boldsymbol{u}_{\mathcal{N}}. 
\end{equation}
Introducing the convention $\boldsymbol{u}_{\backslash\mathcal{N}}=\boldsymbol{u}$, 
$\tilde{\boldsymbol{\epsilon}}_{N}=\boldsymbol{\epsilon}_{N}$, and 
$\boldsymbol{\Phi}_{\boldsymbol{E}_{N}}=\boldsymbol{I}_{N}$ 
for $t=0$, we arrive at the unified representation~(\ref{representation}) 
for $t\geq0$. 

We first prove that $\|\boldsymbol{\delta}_{N}\|^{2}$ given in (\ref{epsilon}) 
converges almost surely to zero as $N\to\infty$. 
From the assumption~(\ref{convergence_X}) and 
$(N-t)^{-1}\|\boldsymbol{u}_{\backslash\mathcal{N}}\|^{2}\ato1$, we have 
\begin{equation}
\|\boldsymbol{\delta}_{N}\|^{2} 
\aeq v\boldsymbol{u}_{\mathcal{N}}^{\mathrm{H}}
(\boldsymbol{\Phi}_{\boldsymbol{E}_{N}}^{\parallel})^{\mathrm{H}}
\boldsymbol{E}_{N}(\boldsymbol{E}_{N}^{\mathrm{H}}\boldsymbol{E}_{N})^{-2}
\boldsymbol{E}_{N}^{\mathrm{H}}
\boldsymbol{\Phi}_{\boldsymbol{E}_{N}}^{\parallel}
\boldsymbol{u}_{\mathcal{N}} + o(1). 
\end{equation}
Using $\boldsymbol{E}_{N}
=\boldsymbol{\Phi}_{\boldsymbol{E}_{N}}^{\parallel}
\boldsymbol{\Sigma}_{\boldsymbol{E}_{N}}
\boldsymbol{\Psi}_{\boldsymbol{E}_{N}}^{\mathrm{H}}$ and  
the assumption~(\ref{norm_E}) yields 
\begin{equation}
\|\boldsymbol{\delta}_{N}\|^{2} 
\aeq \frac{v}{N}\boldsymbol{u}_{\mathcal{N}}^{\mathrm{H}}
\left(
 \frac{1}{N}\boldsymbol{\Sigma}_{\boldsymbol{E}_{N}}^{2}
\right)^{-1}\boldsymbol{u}_{\mathcal{N}} + o(1) 
\al \frac{C}{N}\|\boldsymbol{u}_{\mathcal{N}}\|^{2}   
\end{equation}
for some constant $C>0$. For any $a>0$, we utilize Chebyshev's inequality 
to obtain 
\begin{equation}
\sum_{N=t+1}^{\infty}\mathrm{Pr}\left(
 \frac{\|\boldsymbol{u}_{\mathcal{N}}\|^{2}}{N}>a
\right)
\leq \frac{\mathbb{E}[\|\boldsymbol{u}_{\mathcal{N}}\|^{4}]}{a^{2}}
\sum_{N=t+1}^{\infty}\frac{1}{N^{2}}<\infty. 
\end{equation}
Thus, the Borel-Cantelli lemma implies that 
$\|\boldsymbol{\delta}_{N}\|^{2}$ converges 
almost surely to zero as $N\to\infty$.

Before proving Lemma~\ref{lemma_SLLN}, 
we prove several technical results.  
\begin{proposition} \label{proposition2}
Let $\boldsymbol{z}\sim\mathcal{CN}(\boldsymbol{0},\boldsymbol{I}_{N})$. 
For any $k\geq0$, 
\begin{equation}
\mathbb{E}\left[
 \|\boldsymbol{z}\|^{-k} 
\right] 
\leq \frac{1+o(1)}{N^{k/2}} 
\quad \hbox{as $N\to\infty$.} 
\end{equation}
\end{proposition}
\begin{IEEEproof}
By definition, $2\|\boldsymbol{z}\|^{2}$ follows the chi-square distribution 
with $2N$ degrees of freedom. Let $\Gamma(x)$ denote the gamma function. 
For $N>k/2$, we use the probability density 
function of the chi-square distribution to have 
\begin{IEEEeqnarray}{rl}
&\mathbb{E}\left[
 \frac{1}{\|\boldsymbol{z}\|^{k}}
\right]
= 2^{k/2}\int_{0}^{\infty}\frac{1}{x^{k/2}}\frac{x^{N-1}e^{-x/2}}{2^{N}\Gamma(N)}dx 
\nonumber \\
=& \int_{0}^{\infty}\frac{x^{N-k/2-1}e^{-x}}{\Gamma(N)}dx
= \frac{\Gamma(N-k/2)}{\Gamma(N)}, 
\end{IEEEeqnarray}
where the last equality follows from the definition of the gamma function. 
Using $\Gamma(x+1)=x\Gamma(x)$ and Gautschi's inequality 
$\Gamma(x+s)/\Gamma(x)\leq x^{s}$ for all $x>0$ and $s\in[0, 1]$, we have 
\begin{IEEEeqnarray}{rl}
\mathbb{E}\left[
 \frac{1}{\|\boldsymbol{z}\|^{k}}
\right]
=& \frac{\Gamma(N-k/2)}{(N-1)\cdots(N-\lceil k/2\rceil)
\Gamma(N-\lceil k/2\rceil)}
\nonumber \\ 
\leq& \frac{1}{N^{k/2}}\frac{N^{\lceil k/2\rceil}}{\prod_{i=1}^{\lceil k/2\rceil}(N-i)} 
\end{IEEEeqnarray}
for $N>\lceil k/2\rceil$. 
Since the latter factor tends to $1$ as $N\to\infty$, 
Proposition~\ref{proposition2} holds. 
\end{IEEEproof}

\begin{proposition} \label{proposition3}
Let $\boldsymbol{z}\sim\mathcal{CN}(\boldsymbol{0},\boldsymbol{I}_{N})$. 
For any $k\geq0$, 
\begin{equation}
\mathbb{E}\left[
 \left|
  N - \|\boldsymbol{z}\|^{2}
 \right|^{k}
\right]
={\cal O}(N^{k/2}) 
\quad \hbox{as $N\to\infty$.} 
\end{equation}
\end{proposition}
\begin{IEEEproof}
Let $Z_{N}=N^{-1/2}\sum_{n=1}^{N}(|z_{n}|^{2}-1)$. By definition, we have 
\begin{equation}
\frac{1}{N^{k/2}}\mathbb{E}\left[
 \left|
  N - \|\boldsymbol{z}\|^{2}
 \right|^{k}
\right]
= \mathbb{E}\left[
 |Z_{N}|^{k}
\right]. 
\end{equation}
The central limit theorem implies that $Z_{N}$ converges in distribution to 
a zero-mean Gaussian random variable $Z$ as $N\to\infty$. 
Furthermore, the sequence $\{|Z_{N}|^{k}\}$ is uniformly 
integrable~\cite{Billingsley95} since 
the $(k+1)$th moment of $Z_{N}$ is bounded. Thus, we arrive at 
\begin{equation}
\lim_{N\to\infty}\frac{1}{N^{k/2}}\mathbb{E}\left[
 \left|
  N - \|\boldsymbol{z}\|^{2}
 \right|^{k}
\right]
= \mathbb{E}[|Z|^{k}]<\infty, 
\end{equation}
which implies Proposition~\ref{proposition3}. 
\end{IEEEproof}

\begin{proposition} \label{proposition4}
Let $v_{N}=\|\boldsymbol{X}_{N}\|^{2}/N$, and 
postulate (\ref{convergence_X}) and the moment assumption on 
$\boldsymbol{X}_{N}$ in Lemma~\ref{lemma_SLLN}. For some any $\epsilon>0$,   
\begin{equation}
\lim_{N\to\infty}\mathbb{E}\left[
 \left|
  \sqrt{v}_{N} - \sqrt{v} 
 \right|^{\rho}
\right]
= 0 
\end{equation}
for any $\rho\in[0, \max\{2, 2k-2\}+\epsilon)$. 
\end{proposition}
\begin{IEEEproof}
From (\ref{convergence_X}), $\sqrt{v_{N}}$ converges almost surely to 
$\sqrt{v}$ as $N\to\infty$. Furthermore, $(N^{-1}\|\boldsymbol{X}_{N}\|)^{\rho}$ 
is uniformly integrable for all $\rho\in[0, \max\{2,2k-2\}+\epsilon']$ 
with any $\epsilon'\in(0, \epsilon)$, because of 
the moment assumption on $\boldsymbol{X}_{N}$. 
Thus, Proposition~\ref{proposition4} holds.  
\end{IEEEproof}

Note that Proposition~\ref{proposition4} implies the convergence of the 
$\rho$th moment 
\begin{equation}
\lim_{N\to\infty}\mathbb{E}\left[
 v_{N}^{\rho/2}
\right]
= v^{\rho/2}.
\end{equation}

\subsection{Discussion} \label{appen_discussion}
From the almost sure convergence $\|\boldsymbol{\delta}_{N}\|^{2}\ato0$, 
as well as $\|\boldsymbol{X}_{N}\|^{2}/\|\boldsymbol{u}\|^{2}\ato v$, 
Rangan {\em et al}.~\cite[Proof of Lemma 5]{Rangan16} concluded 
Lemma~\ref{lemma_SLLN}. However, what they have proved should be regarded as 
not the almost sure convergence but as the convergence in probability. 

For simplicity, we assume $t=0$, $f_{n}(z)=z$, and $\boldsymbol{\epsilon}_{N}
=\boldsymbol{0}$. Furthermore, let $S_{N}=N^{-1}\sum_{n=1}^{N}X_{n,N}$ and 
$\tilde{S}_{N}=(\|\boldsymbol{X}_{N}\|/\|\boldsymbol{u}\|)N^{-1}\sum_{n=1}^{N}
u_{n}$. From (\ref{representation0}), for any $\epsilon>0$ and 
$\epsilon'>0$ we have 
\begin{IEEEeqnarray}{rl}
\mathrm{Pr}\left(
 |S_{N}|>\epsilon
\right)
=& \mathrm{Pr}(\mathcal{E}_{N,\epsilon'})\mathrm{Pr}\left(
 \left.
  |\tilde{S}_{N}| >\epsilon
 \right| \mathcal{E}_{N,\epsilon'}
\right) \nonumber \\
&+ \mathrm{Pr}(\mathcal{E}_{N,\epsilon'}^{\complement})\mathrm{Pr}\left(
 \left.
  |\tilde{S}_{N}| >\epsilon
 \right| \mathcal{E}_{N,\epsilon'}^{\complement}
\right),
\end{IEEEeqnarray}
with 
\begin{equation}
\mathcal{E}_{N,\epsilon'}
=\left\{
 \left|
  \frac{\|\boldsymbol{X}_{N}\|^{2}}{\|\boldsymbol{u}\|^{2}} - v
 \right|\leq\epsilon'
\right\}. 
\end{equation}
The almost sure convergence 
$\|\boldsymbol{X}_{N}\|^{2}/\|\boldsymbol{u}\|^{2}\ato v$ implies that 
the second term tends to zero as $N\to\infty$. Using Chebyshev's inequality 
for the first term yields 
\begin{equation}
\mathrm{Pr}(\mathcal{E}_{N,\epsilon'})\mathrm{Pr}\left(
 \left.
  |\tilde{S}_{N}| >\epsilon
 \right| \mathcal{E}_{N,\epsilon'}
\right)
< \frac{\epsilon'+v}{N\epsilon^{2}}\to0. 
\end{equation}
Thus, we arrive at the convergence in probability 
$\mathrm{Pr}(|S_{N}|>\epsilon)\to0$ as $N\to\infty$. 

However, it is not straightforward to prove the almost sure convergence. 
To construct a simple counterexample, suppose that $p_{N,\epsilon}=\mathrm{Pr}
(|\tilde{S}_{N}|>\epsilon)$ is ${\cal O}(N^{-1})$. Then, we find 
\begin{equation}
\sum_{N=1}^{\infty}\mathrm{Pr}\left(
 |S_{N}|>\epsilon
\right)
= \sum_{N=1}^{\infty}p_{N,\epsilon}=\infty. 
\end{equation}
While we do not introduce any statistical properties of 
$\{\boldsymbol{X}_{N}\}$ with respect to $N$, we assume the independence of  
$\{S_{N}\}$ to construct a counterexample. Then, from the second Borel-Cantelli 
lemma we can conclude that $S_{N}$ does not converge almost surely to zero. 
This counterexample implies that we need information about the convergence 
speed of $p_{N,\epsilon}$ to establish the almost sure convergence in 
(\ref{SLLN}). Instead of evaluating the actual convergence speed of 
$p_{N,\epsilon}$, we use Theorem~\ref{theorem_SLLN} to prove the almost sure 
convergence directly. 

\subsection{Proof of (\ref{SLLN_individual})}
Since $\boldsymbol{\epsilon}_{N}$ has vanishing second moments and 
finite $(2k-2)$th moments and 
since $\boldsymbol{E}_{N}$ has finite $\max\{2, 2k-2\}$th moments, 
the almost sure convergence $\|\boldsymbol{\delta}_{N}\|^{2}\ato0$ implies 
that $\tilde{\boldsymbol{\epsilon}}_{N}=\boldsymbol{\epsilon}_{N}
-\boldsymbol{E}_{N}\boldsymbol{o}(1)$ has vanishing second moments and 
finite $(2k-2)$th moments. Furthermore, we only prove the case $t'=1$ with 
$a_{n,1,N}=0$ since an extension of the proof to the general case is 
straightforward. For notational simplicity, we write 
$\boldsymbol{X}_{N-t}$ and $a_{n,0,N}$ as $\boldsymbol{X}$ and $a_{n,N}$. 

Let 
\begin{IEEEeqnarray}{rl} 
Y_{n,N}^{1}
=& f_{n}\left(
 a_{n,N}, \epsilon_{n,N}
 +[\boldsymbol{\Phi}_{\boldsymbol{E}_{N}}^{\perp}\boldsymbol{X}]_{n} 
\right) \nonumber \\
&- f_{n}\left(
 a_{n,N}, \tilde{\epsilon}_{n,N} + v_{N-t}^{1/2}z_{n}
\right), \label{Y}
\end{IEEEeqnarray} 
\begin{equation} \label{Y2}
Y_{n,N}^{2}
= f_{n}\left(
 a_{n,N}, \tilde{\epsilon}_{n,N} + v_{N-t}^{1/2}z_{n}  
\right) 
- f_{n}\left(
 a_{n,N}, v_{N-t}^{1/2}z_{n} 
\right), 
\end{equation}
\begin{equation} \label{Z}
Y_{n,N}^{3}
= f_{n}\left(
 a_{n,N}, v_{N-t}^{1/2}z_{n}  
\right) 
- f_{n}\left(
 a_{n,N}, v^{1/2}z_{n} 
\right), 
\end{equation}
with $v_{N-t}=\|\boldsymbol{X}_{N-t}\|^{2}/N$. It is sufficient to prove 
\begin{equation} \label{expectation_Y}
\mathbb{E}\left[ 
 \left|
  Y_{n,N}^{1}
 \right|
\right]
= {\cal O}\left(
 \frac{L_{n}}{\sqrt{N}} 
\right), 
\end{equation}
\begin{equation} \label{expectation_a}
\mathbb{E}\left[
 \left| 
  Y_{n,N}^{2} 
 \right|
\right] = o(L_{n}),  
\end{equation}
\begin{equation} \label{expectation_Z}
\mathbb{E}\left[
 \left|
  Y_{n,N}^{3}
 \right|
\right]
= o(L_{n}) 
\end{equation}
as $N\to\infty$. 

Let $\mathcal{E}=\{\|\boldsymbol{X}\|, \boldsymbol{a}_{N}, 
\boldsymbol{\epsilon}_{N}, \boldsymbol{E}_{N}\}$. 
We first evaluate the conditional expectation 
$\mathbb{E}[|Y_{n,N}^{1}| | \mathcal{E}]$ 
to prove (\ref{expectation_Y}). Using the 
representation~(\ref{representation}), 
the pseudo-Lipschitz property of $f_{n}$, and 
Proposition~\ref{proposition_bound} yields 
\begin{IEEEeqnarray}{rl}
&\mathbb{E}\left[
 \left|
  |Y_{n,N}^{1}| 
 \right| \mathcal{E} 
\right] \nonumber \\ 
\leq& L_{n}\mathbb{E}\left[
 \left|
  \frac{1}{\|\boldsymbol{u}_{\backslash\mathcal{N}}\|} 
  - \frac{1}{\sqrt{N-t}}
 \right|\|\boldsymbol{X}\||z_{n}|\Biggl\{
 1 + |a_{n,N}|^{k-1}  
\right. \nonumber \\
+&\left.
 \left.
  \left. 
   |\tilde{\epsilon}_{n,N}|^{k-1}
   + \frac{\|\boldsymbol{X}\|^{k-1}|z_{n}|^{k-1}}
   {\|\boldsymbol{u}_{\backslash\mathcal{N}}\|^{k-1}} 
   + v_{N-t}^{\frac{k-1}{2}}|z_{n}|^{k-1}
  \right\}
 \right| \mathcal{E} 
\right] 
\label{expectation_Y_tmp}
\end{IEEEeqnarray}
for some $L_{n}>0$. Using the following upper bound: 
\begin{IEEEeqnarray}{rl}
\left|
 \frac{1}{\|\boldsymbol{u}_{\backslash\mathcal{N}}\|} 
 - \frac{1}{\sqrt{N-t}} 
\right|
=& \frac{|N-t-\|\boldsymbol{u}_{\backslash\mathcal{N}}\|^{2}|}
{\|\boldsymbol{u}_{\backslash\mathcal{N}}\|\sqrt{N-t}
(\sqrt{N-t}+\|\boldsymbol{u}_{\backslash\mathcal{N}}\|)}
\nonumber \\
<& \frac{|N-t-\|\boldsymbol{u}_{\backslash\mathcal{N}}\|^{2}|}{N-t}
\frac{1}{\|\boldsymbol{u}_{\backslash\mathcal{N}}\|}, \label{difference_bound}
\end{IEEEeqnarray}
we have 
\begin{IEEEeqnarray}{rl}
&\mathbb{E}\left[
 \left. 
  \left|
   \frac{1}{\|\boldsymbol{u}_{\backslash\mathcal{N}}\|} 
   - \frac{1}{\sqrt{N-t}}
  \right|\|\boldsymbol{X}\||z_{n}||a_{n,N}|^{k-1}
 \right| \mathcal{E} 
\right] \nonumber \\
<& v_{N-t}^{1/2}|a_{n,N}|^{k-1}\mathbb{E}\left[
 \left. 
  \frac{|N-t-\|\boldsymbol{u}_{\backslash\mathcal{N}}\|^{2}|}{\sqrt{N-t}}
  \frac{|z_{n}|}{\|\boldsymbol{u}_{\backslash\mathcal{N}}\|}
 \right| \mathcal{E} 
\right]. 
\end{IEEEeqnarray}
To evaluate the conditional expectation,  
we use the Cauchy-Schwarz inequality repeatedly to obtain
\begin{IEEEeqnarray}{rl}
&\mathbb{E}\left[
 \left.
  \frac{|N-t-\|\boldsymbol{u}_{\backslash\mathcal{N}}\|^{2}|}{\sqrt{N-t}}
  \frac{|z_{n}|}{\|\boldsymbol{u}_{\backslash\mathcal{N}}\|}
 \right| \mathcal{E} 
\right] \nonumber \\
\leq& \left\{
 \mathbb{E}\left[
   \frac{|N-t-\|\boldsymbol{u}_{\backslash\mathcal{N}}\|^{2}|^{2}}{N-t}
 \right]
\right\}^{1/2}\left\{
 \mathbb{E}\left[
  \left.
   \frac{|z_{n}|^{2}}{\|\boldsymbol{u}_{\backslash\mathcal{N}}\|^{2}}
  \right| \mathcal{E} 
 \right]
\right\}^{1/2} \nonumber \\
\leq& C\left\{
 \mathbb{E}\left[
  \frac{|N-t-\|\boldsymbol{u}_{\backslash\mathcal{N}}\|^{2}|^{2}}{N-t}
 \right]
\right\}^{1/2}
\left\{
 \mathbb{E}\left[
  \frac{1}{\|\boldsymbol{u}_{\backslash\mathcal{N}}\|^{4}}
 \right]
\right\}^{1/4} \nonumber \\ 
=& {\cal O}(N^{-1/2}) \label{second_term}
\end{IEEEeqnarray}
for some $C>0$, where the last follows from 
Propositions~\ref{proposition2} and \ref{proposition3}. 

We repeat the same argument in evaluating the remaining terms 
in (\ref{expectation_Y_tmp}). 
We only present evaluation of the fourth term, 
since the other terms can be bounded in the 
same manner. Applying the upper bound~(\ref{difference_bound}) and 
the Cauchy-Schwarz inequality, we obtain 
\begin{IEEEeqnarray}{rl}
&\mathbb{E}\left[
 \left. 
  \left|
   \frac{1}{\|\boldsymbol{u}_{\backslash\mathcal{N}}\|} 
   - \frac{1}{\sqrt{N-t}}
  \right|\frac{\|\boldsymbol{X}\|^{k}|z_{n}|^{k}}
  {\|\boldsymbol{u}_{\backslash\mathcal{N}}\|^{k-1}}
 \right| \mathcal{E} 
\right] \nonumber \\
<& \frac{v_{N-t}^{k/2}}{\sqrt{N-t}}\mathbb{E}\left[
 \left. 
  \frac{|N-t - \|\boldsymbol{u}_{\backslash\mathcal{N}}\|^{2}|}{\sqrt{N-t}}
  \frac{(N-t)^{k/2}|z_{n}|^{k}}
 {\|\boldsymbol{u}_{\backslash\mathcal{N}}\|^{k}}
 \right| \mathcal{E} 
\right]
\nonumber \\ 
<& \frac{Cv_{N-t}^{k/2}}{\sqrt{N-t}}\left\{ 
 \mathbb{E}\left[
  \frac{|N-t - \|\boldsymbol{u}_{\backslash\mathcal{N}}\|^{2}|^{2}}{N-t}
 \right]
\right\}^{\frac{1}{2}}\left\{
 \mathbb{E}\left[
  \frac{(N-t)^{2k}}{\|\boldsymbol{u}_{\backslash\mathcal{N}}\|^{4k}}
 \right]
\right\}^{\frac{1}{4}}
 \nonumber \\
\aeq& {\cal O}(v_{N}^{k/2}N^{-1/2}), \label{expectation_bound}
\end{IEEEeqnarray}
for some $C>0$, where the last follows from 
Propositions~\ref{proposition2} and \ref{proposition3}.  
Evaluating the remaining terms on the RHS of 
(\ref{expectation_Y_tmp}) in the same manner, we arrive at 
\begin{IEEEeqnarray}{rl}
\mathbb{E}\left[
 \left.
  |Y_{n,N}^{1}|
 \right| \mathcal{E} 
\right] 
\aeq {\cal O}&\left\{
 \frac{L_{n}v_{N}^{1/2}}{\sqrt{N}}(1 + |a_{n,N}|^{k-1} 
\right. \nonumber \\
&\left.
 + |\tilde{\epsilon}_{n,N}|^{k-1} 
 + v_{N}^{(k-1)/2})
\right\}.  
\end{IEEEeqnarray}

Using the Cauchy-Schwarz inequality to evaluate the expectation 
over $\mathcal{E}$, we obtain 
\begin{IEEEeqnarray}{rl}
\mathbb{E}\left[
 |Y_{n,N}^{1}|
\right]
=& {\cal O}\left(
 \frac{L_{n}}{\sqrt{N}}\left\{
  \mathbb{E}[v_{N}^{1/2}] 
  + \left(
   \mathbb{E}[v_{N}]\mathbb{E}[|a_{n,N}|^{2k-2}]
  \right)^{1/2}
 \right.
\right. \nonumber \\
&\left.
 \left.  
  + \left(
   \mathbb{E}[v_{N}]\mathbb{E}[|\tilde{\epsilon}_{n,N}|^{2k-2}]
  \right)^{1/2}
  + \mathbb{E}[v_{N}^{k/2}] 
 \right\}
\right),
\end{IEEEeqnarray}
which reduces to (\ref{expectation_Y}), because of 
Proposition~\ref{proposition4} and the moment properties of 
$a_{n,N}$ and $\tilde{\epsilon}_{n,N}$. 

We next prove (\ref{expectation_a}). Using the definition~(\ref{Y2}) 
of $Y_{n,N}^{2}$, the pseudo-Lipschitz property of $f_{n}$, and 
Proposition~\ref{proposition_bound} yields 
\begin{IEEEeqnarray}{rl}
\frac{\mathbb{E}[ |Y_{n,N}^{2}| | \mathcal{E}]}{L_{n}}  
\leq& |\tilde{\epsilon}_{n,N}|\left(
 1 + |a_{n,N}|^{k-1} + |\tilde{\epsilon}_{n,N}|^{k-1}
\right) \nonumber \\
&+ v_{N-t}^{(k-1)/2}|\tilde{\epsilon}_{n,N}|\mathbb{E}\left[
 |z_{n}|^{k-1}
\right] \label{expectation_Y2} 
\end{IEEEeqnarray}
for some $L_{n}>0$. Using the Cauchy-Schwarz inequality and 
Proposition~\ref{proposition_norm}, we have  
\begin{IEEEeqnarray}{rl}
\frac{\mathbb{E}[ |Y_{n,N}^{2}| ]}{L_{n}}
\aleq& C\left(
 \mathbb{E}\left[
  |\tilde{\epsilon}_{n,N}|^{2} 
 \right]
\right)^{1/2} \nonumber \\
\cdot&\left(
 \mathbb{E}\left[
  1 + |a_{n,N}|^{2k-2} + |\tilde{\epsilon}_{n,N}|^{2k-2} 
 \right]
\right)^{1/2} \nonumber \\ 
&+ C\left(
 \mathbb{E}[v_{N-t}^{k-1}]\mathbb{E}\left[
  |\tilde{\epsilon}_{n,N}|^{2}
 \right]
\right)^{1/2}
\end{IEEEeqnarray}
for some $C>0$. 
Since $|\tilde{\epsilon}_{n,N}|$ has a vanishing second moment and finite 
$(2k-2)$th moment and since $|a_{n,N}|$ has a finite $(2k-2)$th moment, 
we use Proposition~\ref{proposition4} to arrive at (\ref{expectation_a}). 

Finally, we prove (\ref{expectation_Z}). Using the definition~(\ref{Z}) 
of $Y_{n,N}^{3}$, the pseudo-Lipschitz property of $f_{n}$, and 
Proposition~\ref{proposition_bound} yields  
\begin{IEEEeqnarray}{rl}
\left|
 Y_{n,N}^{3}
\right|
\leq& L_{n}\left|
 v_{N-t}^{1/2} - \sqrt{v} 
\right||z_{n}| 
\Bigl\{
 1 + |a_{n,N}|^{k-1}  
\nonumber \\
&\left.
 + v_{N-t}^{(k-1)/2}|z_{n}|^{k-1}
 + v^{(k-1)/2}|z_{n}|^{k-1} 
\right\}. 
\end{IEEEeqnarray}
Evaluating the conditional expectation yields 
\begin{IEEEeqnarray}{rl} 
\mathbb{E}\left[
 \left.
  |Y_{n,N}^{3}|
 \right| \mathcal{E} 
\right] 
\aleq& CL_{n}\left|
 v_{N-t}^{1/2} - \sqrt{v} 
\right| \nonumber \\ 
&\cdot(1 + |a_{n,N}|^{k-1} + v_{N-t}^{(k-1)/2}) 
 \label{expectation_Z_conditional}
\end{IEEEeqnarray}
for some $C>0$. Using the Cauchy-Schwarz inequality to evaluate the 
expectation of the second term, we have   
\begin{IEEEeqnarray}{rl}
&\left(
 \mathbb{E}\left[
  \left|
   v_{N-t}^{1/2} - \sqrt{v} 
  \right||a_{n,N}|^{k-1}
 \right]
\right)^{2}  \nonumber \\ 
\leq& \mathbb{E}\left[
 \left|
  v_{N-t}^{1/2} - \sqrt{v} 
 \right|^{2}
\right]\mathbb{E}\left[
 |a_{n,N}|^{2k-2} 
\right]\to0,
\end{IEEEeqnarray}
where the convergence follows from Proposition~\ref{proposition4} and 
the moment assumption of $a_{n,N}$. For the last term, 
we let $\epsilon'\in(0,\epsilon/k)$ to have
\begin{IEEEeqnarray}{rl}
&\mathbb{E}\left[
 \left\{
  \left|
   v_{N-t}^{1/2} - \sqrt{v} 
  \right|v_{N-t}^{(k-1)/2}
 \right\}^{1+\epsilon'}
\right] \nonumber \\ 
<&\mathbb{E}\left[
 v_{N-t}^{k(1+\epsilon')/2}
\right]
+ v^{\frac{1+\epsilon'}{2}}\mathbb{E}\left[
 v_{N-t}^{(k-1)(1+\epsilon')/2}
\right]<\infty, 
\end{IEEEeqnarray}
where the boundedness follows from Proposition~\ref{proposition4}. 
In other words, the last term on the upper 
bound~(\ref{expectation_Z_conditional}) is uniformly integrable 
over $\|\boldsymbol{X}\|$. Thus, we use the assumption~(\ref{convergence_X}) 
to arrive at (\ref{expectation_Z}).

\subsection{Proof of (\ref{SLLN})}
Since $\boldsymbol{\epsilon}_{N}$ satisfies the assumptions 
(\ref{assumption_a1}) and (\ref{assumption_a2}), and since 
$\boldsymbol{E}_{N}$ satisfies the assumption~(\ref{assumption_E}), 
the almost sure convergence $\|\boldsymbol{\delta}_{N}\|^{2}\ato0$ implies 
that $\tilde{\boldsymbol{\epsilon}}_{N}=\boldsymbol{\epsilon}_{N}
-\boldsymbol{E}_{N}\boldsymbol{o}(1)$ satisfies (\ref{assumption_a1}) and 
(\ref{assumption_a2}) with $\boldsymbol{\epsilon}_{N}$ replaced by 
$\tilde{\boldsymbol{\epsilon}}_{N}$. Furthermore, we only prove 
the case $t'=1$ with $a_{n,1,N}=0$ since an extension of the proof to the 
general case is straightforward. For notational simplicity, we write 
$\boldsymbol{X}_{N-t}$ and $a_{n,0,N}$ as $\boldsymbol{X}$ and $a_{n,N}$, and 
omit the tilde on $\tilde{\epsilon}_{n,N}$. 

From the definitions~(\ref{Y}) and (\ref{Y2}) of $Y_{n,N}^{1}$ and 
$Y_{n,N}^{2}$, we need to prove 
\begin{equation} \label{convergence_Y}
\lim_{N\to\infty}\frac{1}{N}\sum_{n=1}^{N}Y_{n,N}^{1} \aeq 0, 
\end{equation}
\begin{equation} \label{convergence_Y2}
\lim_{N\to\infty}\frac{1}{N}\sum_{n=1}^{N}Y_{n,N}^{2} \aeq 0, 
\end{equation}
\begin{IEEEeqnarray}{rl}
\lim_{N\to\infty}\frac{1}{N}\sum_{n=1}^{N}&\left\{
 f_{n}\left(
  a_{n,N}, v_{N-t}^{1/2}z_{n} 
 \right)
\right. \nonumber \\
&\left. - \mathbb{E}_{z_{n}}\left[
  f_{n}\left(
   a_{n,N}, \sqrt{v}z_{n}  
  \right)
 \right]
\right\}\aeq 0. \label{convergence_Y2_SLLN}
\end{IEEEeqnarray}

Let us prove the first convergence~(\ref{convergence_Y}).  
From the representation~(\ref{representation}) and Theorem~\ref{theorem_SLLN}, 
it is sufficient to prove that $Y_{n,N}^{1}$ given in (\ref{Y}) satisfies 
\begin{equation} \label{Y_covariance}
\lim_{N\to\infty}\frac{1}{N}\sum_{n,n'=1}^{N}
 \mathbb{E}\left[
 \left.
  |Y_{n,N}^{1}Y_{n',N}^{1}| 
 \right| \mathcal{E}
\right]\al\infty, 
\end{equation}
with $\mathcal{E}=\{\|\boldsymbol{X}\|, \boldsymbol{a}_{N}, 
\boldsymbol{\epsilon}_{N}, \boldsymbol{E}_{N}\}$. 

Repeating the derivation of (\ref{expectation_Y_tmp}), we have 
\begin{IEEEeqnarray}{l}
\frac{\mathbb{E}[|Y_{n,N}^{1}||Y_{n',N}^{1}| | \mathcal{E}]}{L_{n}L_{n'}}
\leq \mathbb{E}\left[
 \left|
  \frac{1}{\|\boldsymbol{u}_{\backslash\mathcal{N}}\|} 
  - \frac{1}{\sqrt{N-t}}
 \right|^{2} 
\right. \nonumber \\
\cdot\|\boldsymbol{X}\|^{2}|z_{n}||z_{n'}|\left\{
 1  + |a_{n,N}|^{k-1} + |\epsilon_{n,N}|^{k-1}
 + v_{N-t}^{\frac{k-1}{2}}|z_{n}|^{k-1}
\right. \nonumber \\
+\frac{\|\boldsymbol{X}\|^{k-1}|z_{n}|^{k-1}}
{\|\boldsymbol{u}_{\backslash\mathcal{N}}\|^{k-1}}
\Biggr\}
\Biggl\{
1 + |a_{n',N}|^{k-1} + |\epsilon_{n',N}|^{k-1}
\nonumber \\
\left.
 \left.
  \left.
   +\frac{\|\boldsymbol{X}\|^{k-1}|z_{n'}|^{k-1}}
   {\|\boldsymbol{u}_{\backslash\mathcal{N}}\|^{k-1}} 
   + v_{N-t}^{(k-1)/2}|z_{n'}|^{k-1}
  \right\}
 \right| \mathcal{E} 
\right].
\label{covariance} 
\end{IEEEeqnarray}
Let 
\begin{equation}
A_{n,n'}
=L_{n}L_{n'}\left|
 \frac{1}{\|\boldsymbol{u}_{\backslash\mathcal{N}}\|} 
 - \frac{1}{\sqrt{N-t}}
\right|^{2}\frac{\|\boldsymbol{X}\|^{2k}|z_{n}|^{k}|z_{n'}|^{k}}
 {\|\boldsymbol{u}_{\backslash\mathcal{N}}\|^{2k-2}},
\end{equation}
\begin{IEEEeqnarray}{l}
B_{n,n'} 
= L_{n}L_{n'}\left|
 \frac{1}{\|\boldsymbol{u}_{\backslash\mathcal{N}}\|} 
 - \frac{1}{\sqrt{N-t}}
\right|^{2}\|\boldsymbol{X}\|^{2}|z_{n}||z_{n'}| \nonumber \\
\cdot(|a_{n,N}|^{k-1}+|\epsilon_{n,N}|^{k-1})
(|a_{n',N}|^{k-1}+|\epsilon_{n',N}|^{k-1}).  
\end{IEEEeqnarray}
We only evaluate the conditional expectation of $A_{n,n'}$ and $B_{n,n'}$, since 
the other terms can be evaluated in the same manner. 
Using the upper bound~(\ref{difference_bound}) yields 
\begin{IEEEeqnarray}{rl}
\mathbb{E}[A_{n,n'}| \mathcal{E}]
<& L_{n}L_{n'}v_{N-t}^{k}\mathbb{E}\left[
 \frac{|N-t-\|\boldsymbol{u}_{\backslash\mathcal{N}}\|^{2}|^{2}}{(N-t)^{2}}
\right. \nonumber \\
&\cdot\left.
 \left. 
  (N-t)^{k}\frac{|z_{n}|^{k}|z_{n'}|^{k}}
  {\|\boldsymbol{u}_{\backslash\mathcal{N}}\|^{2k}}
 \right| \mathcal{E} 
\right]. 
\end{IEEEeqnarray}
Repeating the proof of (\ref{second_term}), we find that 
the last factor is ${\cal O}(N^{-1})$. Thus, we obtain 
\begin{equation}
\frac{1}{N}\sum_{n,n'=1}^{N}\mathbb{E}[A_{n,n'}| \mathcal{E}] 
\aeq {\cal O}\left\{
 v_{N}^{k}\left(
  \frac{1}{N}\sum_{n=1}^{N}L_{n}
 \right)^{2}
\right\} 
\aeq {\cal O}(1),  
\end{equation}
because of the assumptions~(\ref{convergence_X}) and 
(\ref{Lipschitz_constant}). 

Similarly, we use the upper bound~(\ref{difference_bound}) to have 
\begin{IEEEeqnarray}{rl}
&\mathbb{E}[B_{n,n'}| \mathcal{E}] 
\nonumber \\
<& L_{n}(|a_{n,N}|^{k-1}+|\epsilon_{n,N}|^{k-1})
L_{n'}(|a_{n',N}|^{k-1}+|\epsilon_{n',N}|^{k-1}) \nonumber \\
&\cdot v_{N-t}\mathbb{E}\left[
 \left. 
  \frac{|N-t-\|\boldsymbol{u}_{\backslash\mathcal{N}}\|^{2}|^{2}}{N-t}
  \frac{|z_{n}||z_{n'}|}{\|\boldsymbol{u}_{\backslash\mathcal{N}}\|^{2}}
 \right| \mathcal{E} 
\right]. 
\end{IEEEeqnarray}
We repeat the proof of (\ref{second_term}) to find that 
the last factor is ${\cal O}(N^{-1})$. Thus, we arrive at 
\begin{IEEEeqnarray}{rl}
&\frac{1}{N}\sum_{n,n'=1}^{N}\mathbb{E}[B_{n,n'}| \mathcal{E}] \nonumber \\
\aeq& {\cal O}\left\{
 v_{N}\left(
  \frac{1}{N}\sum_{n=1}^{N}L_{n}(|a_{n,N}|^{k-1}+|\epsilon_{n,N}|^{k-1})
 \right)^{2}
\right\} \nonumber \\
\aeq& {\cal O}\left\{
 v_{N}\left(
  \frac{1}{N}\sum_{n=1}^{N}L_{n}(|a_{n,N}|^{2k-2}+|\epsilon_{n,N}|^{2k-2})
 \right)^{2}
\right\} \nonumber \\
=& {\cal O}(1), \label{B_bound}
\end{IEEEeqnarray}
where the second equality follows from the Cauchy-Schwarz inequality and 
the assumption~(\ref{Lipschitz_constant}), and where the last is due to 
the assumptions~(\ref{assumption_a2}), 
(\ref{assumption_a3}), and (\ref{convergence_X}). 
Evaluating the conditional expectation of the other terms 
in (\ref{covariance}) in the 
same manner, we arrive at (\ref{Y_covariance}). 
Thus, (\ref{convergence_Y}) holds. 

We next prove the second convergence~(\ref{convergence_Y2}). 
Repeating the proof of (\ref{expectation_Y2}) yields 
\begin{IEEEeqnarray}{rl}
\frac{1}{N}\sum_{n=1}^{N}|Y_{n,N}^{2}|
&\leq \frac{1}{N}\sum_{n=1}^{N}L_{n}|\epsilon_{n,N}|
\Biggl\{1 + |a_{n,N}|^{k-1}
\nonumber \\
+&\left. 
 |\epsilon_{n,N}|^{k-1} + v_{N-t}^{(k-1)/2}|z_{n}|^{k-1} 
\right\}. \label{Y2_bound}
\end{IEEEeqnarray}
Using the Cauchy-Schwarz inequality for the second term yields 
\begin{IEEEeqnarray}{rl}
&\left(
 \frac{1}{N}\sum_{n=1}^{N}L_{n}|\epsilon_{n,N}||a_{n,N}|^{k-1} 
\right)^{2} \nonumber \\
\leq& \frac{1}{N}\sum_{n=1}^{N}L_{n}|\epsilon_{n,N}|^{2} 
\frac{1}{N}\sum_{n=1}^{N}L_{n}|a_{n,N}|^{2k-2}\ato0 
\end{IEEEeqnarray}
as $N\to\infty$, because of the assumptions~(\ref{assumption_a1}) and 
(\ref{assumption_a3}). Similarly, we find that the third term 
converges almost surely to zero as $N\to\infty$. 

We use the Cauchy-Schwarz inequality for the last term on 
the upper bound~(\ref{Y2_bound}) to obtain  
\begin{IEEEeqnarray}{rl}
\frac{v_{N-t}^{(k-1)/2}}{N}\sum_{n=1}^{N}L_{n}|\epsilon_{n,N}|&|z_{n}|^{k-1}
\leq \left(
 \frac{1}{N}\sum_{n=1}^{N}L_{n}|\epsilon_{n,N}|^{2} 
\right)^{1/2} \nonumber \\ 
\cdot v_{N-t}^{\frac{k-1}{2}}
&\left(
 \frac{1}{N}\sum_{n=1}^{N}L_{n}|z_{n}|^{2k-2}
\right)^{\frac{1}{2}}. 
\end{IEEEeqnarray}
The assumptions~(\ref{assumption_a1}) and (\ref{convergence_X}) imply that 
the first and second factors converge almost surely to zero and 
$v^{(k-1)/2}$ as $N\to\infty$, respectively.  Furthermore, from the 
assumption~(\ref{Lipschitz_constant}) we use Theorem~\ref{theorem_SLLN} 
to find 
\begin{equation}
\frac{1}{N}\sum_{n=1}^{N}L_{n}|z_{n}|^{2k-2}
\aeq \frac{\mathbb{E}[|z_{1}|^{2k-2}]}{N}\sum_{n=1}^{N}L_{n} + o(1) 
<\infty. 
\end{equation}
Thus, the last term on the upper bound~(\ref{Y2_bound}) converges almost 
surely to zero as $N\to\infty$. Since the almost sure convergence of the 
remaining terms to zero can be proved in the same manner, we arrive at 
(\ref{convergence_Y2}). 

Finally, we prove the last convergence~(\ref{convergence_Y2_SLLN}). 
We observe that $\{f_{n}(a_{n,N},v_{N-t}^{1/2}z_{n})\}$ are 
conditionally independent given $\mathcal{E}$. Furthermore, 
we use the pseudo-Lipschitz property of $f_{n}$ to obtain  
\begin{IEEEeqnarray}{rl}
&\frac{1}{N}\sum_{n=1}^{N}\mathbb{V}\left[
 \left. 
  f_{n}\left(
   a_{n,N}, v_{N-t}^{1/2}z_{n}
  \right)
 \right| \mathcal{E}
\right] \nonumber \\
\leq& \frac{v_{N-t}}{N}\sum_{n=1}^{N}L_{n}^{2}\mathbb{E}_{z_{n},Z}\left[
 |z_{n}-Z|^{2}\left(
  1 + |a_{n,N}|^{2k-2}
 \right. 
\right. \nonumber \\
&\left.
 \left.
  \left. 
   + v_{N-t}^{k-1}|z_{n}|^{2k-2} + v_{N-t}^{k-1}|Z|^{2k-2} 
  \right)
 \right| \mathcal{E} 
\right]
\al \infty, 
\end{IEEEeqnarray}
where $Z$ is a standard complex Gaussian random variable and independent 
of $z_{n}$, and where the boundedness follows from the 
assumptions~(\ref{assumption_a3}), (\ref{convergence_X}), and 
(\ref{Lipschitz_constant}). 
Thus, we can use Theorem~\ref{theorem_SLLN} to find 
\begin{IEEEeqnarray}{rl}
\lim_{N\to\infty}&\frac{1}{N}\sum_{n=1}^{N}\left\{
 f_{n}\left(
  a_{n,N}, v_{N-t}^{1/2}z_{n}
 \right)
\right. \nonumber \\
&\left.
 - \mathbb{E}\left[
  \left.
   f_{n}\left(
    a_{n,N}, v_{N-t}^{1/2}z_{n}
   \right)
  \right| \mathcal{E} 
 \right]
\right\}\aeq 0. 
\end{IEEEeqnarray}

To obtain (\ref{convergence_Y2_SLLN}), from the definition~(\ref{Z}) of 
$Y_{n,N}^{3}$ we need to prove $N^{-1}\sum_{n=1}^{N}
\mathbb{E}[Y_{n,N}^{3} | \mathcal{E}]\ato0$ as $N\to\infty$. 
Using (\ref{expectation_Z_conditional}) yields 
\begin{IEEEeqnarray}{rl}
\left|
 \frac{1}{N}\sum_{n=1}^{N}
 \mathbb{E}\left[
  \left.
   Y_{n,N}^{3}
  \right| \mathcal{E} 
 \right]
\right|
&\aleq C\left|
 v_{N-t}^{1/2} - \sqrt{v} 
\right| \nonumber \\
\cdot\frac{1}{N}\sum_{n=1}^{N}L_{n}&(1 + |a_{n,N}|^{k-1} + v_{N-t}^{(k-1)/2}), 
\end{IEEEeqnarray}  
which converges almost surely to zero as $N\to\infty$, because of the 
assumptions~(\ref{assumption_a3}), (\ref{convergence_X}), and 
(\ref{Lipschitz_constant}). Thus, (\ref{convergence_Y2_SLLN}) holds. 

\section{Derivation of Message-Passing}
\label{sec_deriv_EP}
EP~\cite{Minka01,Cespedes14} provides a framework for deriving MP algorithms 
that calculate the marginal posterior distribution 
$p(x_{n}|\boldsymbol{y}, \boldsymbol{A})
=\int p(\boldsymbol{x}|\boldsymbol{y}, \boldsymbol{A})
d\boldsymbol{x}_{\backslash n}$, in which $\boldsymbol{x}_{\backslash n}$ is the 
vector obtained by eliminating $x_{n}$ from $\boldsymbol{x}$. 
We consider the large system limit to derive an MP algorithm, which coincides 
with the algorithm derived in a heuristic manner~\cite{Ma17}. 

We approximate the marginal posterior distribution 
$p(x_{n}|\boldsymbol{y}, \boldsymbol{A})$ by a tractable probability density 
function (pdf) $q_{\mathrm{A}}(x_{n})=\int q_{\mathrm{A}}(\boldsymbol{x})
d\boldsymbol{x}_{\backslash n}$, given by  
\begin{equation} \label{q_A}
q_{\mathrm{A}}(\boldsymbol{x})
\propto p(\boldsymbol{y}|\boldsymbol{A},\boldsymbol{x}) 
\prod_{n=1}^{N}q_{\mathrm{B}\to \mathrm{A}}(x_{n}). 
\end{equation}
In (\ref{q_A}), the notation $f(\boldsymbol{x})\propto g(\boldsymbol{x})$ 
means that there is a positive constant $C$ such that 
$f(\boldsymbol{x})=Cg(\boldsymbol{x})$ holds. Furthermore, 
$q_{\mathrm{B}\to \mathrm{A}}(x_{n})$ is a conjugate prior to the 
likelihood $p(\boldsymbol{y}|\boldsymbol{A},\boldsymbol{x})$. 
When the noise vector $\boldsymbol{w}$ in (\ref{system}) is regarded as 
a CSCG random vector with covariance $\sigma^{2}\boldsymbol{I}_{M}$,  
the conjugate prior $q_{\mathrm{B}\to \mathrm{A}}(x_{n})$ is proper complex 
Gaussian, 
\begin{equation} \label{q_B_A}
q_{\mathrm{B}\to \mathrm{A}}(x_{n})
\propto \exp\left(
 -\frac{|x_{n}-x_{n,\mathrm{B}\to \mathrm{A}}|^{2}}{v_{\mathrm{B}\to \mathrm{A}}}
\right), 
\end{equation}
where $x_{n,\mathrm{B}\to \mathrm{A}}$ and $v_{\mathrm{B}\to \mathrm{A}}$ are 
the mean and variance of $q_{\mathrm{B}\to \mathrm{A}}(x_{n})$, respectively. 
In order to derive the MP algorithm proposed in \cite{Ma17}, 
we have selected the identical variance $v_{\mathrm{B}\to \mathrm{A}}$ for all $n$, 
while C\'espedes {\it et al}.~\cite{Cespedes14} selected different values 
for different $n$ to improve the performance for finite-sized systems. 

We first evaluate the marginal pdf $q_{\mathrm{A}}(x_{n})$ in the large system 
limit, defined via (\ref{q_A}).  
Since the conjugate prior~(\ref{q_B_A}) has been selected, 
the joint pdf $q_{\mathrm{A}}(\boldsymbol{x})$ is also Gaussian. 
\begin{equation}
q_{\mathrm{A}}(\boldsymbol{x})
\propto 
\exp\left\{
 -(\boldsymbol{x}-\boldsymbol{x}_{\mathrm{A}})^{\mathrm{H}}
 \boldsymbol{V}_{\mathrm{A}}^{-1}(\boldsymbol{x}-\boldsymbol{x}_{\mathrm{A}})
\right\}, 
\end{equation}
where the mean and covariance are given by 
\begin{equation} \label{x_A}
\boldsymbol{x}_{\mathrm{A}} 
= \boldsymbol{x}_{\mathrm{B}\to \mathrm{A}} 
+ \frac{1}{\sigma^{2}}\boldsymbol{V}_{\mathrm{A}}
\boldsymbol{A}^{\mathrm{H}}
(\boldsymbol{y}-\boldsymbol{A}\boldsymbol{x}_{\mathrm{B}\to \mathrm{A}}), 
\end{equation}
\begin{equation} \label{V_A}
\boldsymbol{V}_{\mathrm{A}} 
= \left(
 \frac{1}{v_{\mathrm{B}\to \mathrm{A}}}\boldsymbol{I}_{N}
 + \frac{1}{\sigma^{2}}\boldsymbol{A}^{\mathrm{H}}\boldsymbol{A} 
\right)^{-1}, 
\end{equation}
respectively. Using the matrix inversion lemma, it is possible to show that 
(\ref{x_A}) and (\ref{V_A}) reduce to 
\begin{equation} \label{x_A2}
\boldsymbol{x}_{\mathrm{A}} 
= \boldsymbol{x}_{\mathrm{B}\to \mathrm{A}} 
+ v_{\mathrm{B}\to \mathrm{A}}\boldsymbol{A}^{\mathrm{H}}\boldsymbol{\Xi}^{-1} 
(\boldsymbol{y}-\boldsymbol{A}\boldsymbol{x}_{\mathrm{B}\to \mathrm{A}}), 
\end{equation}
\begin{equation} \label{V_A_n}
[\boldsymbol{V}_{\mathrm{A}}]_{n,n} 
= v_{\mathrm{B}\to \mathrm{A}} - \boldsymbol{a}_{n}^{\mathrm{H}}
 \boldsymbol{\Xi}^{-1}\boldsymbol{a}_{n}v_{\mathrm{B}\to \mathrm{A}}^{2}, 
\end{equation}
respectively, with 
\begin{equation} \label{Xi}
\boldsymbol{\Xi} 
= \sigma^{2}\boldsymbol{I}_{M} 
+ v_{\mathrm{B}\to \mathrm{A}}\boldsymbol{A}\boldsymbol{A}^{\mathrm{H}}.
\end{equation}

We shall prove that $\boldsymbol{a}_{n}^{\mathrm{H}}\boldsymbol{\Xi}^{-1}
\boldsymbol{a}_{n}$ converges almost surely to 
$\gamma(v_{\mathrm{B}\to \mathrm{A}})^{-1}$ 
in the large system limit for all $n$, in which 
$\gamma(v_{\mathrm{B}\to \mathrm{A}})$ is 
given by (\ref{gamma2}). Applying the SVD~(\ref{SVD_A}) to 
$\boldsymbol{a}_{n}^{\mathrm{H}}\boldsymbol{\Xi}^{-1}\boldsymbol{a}_{n}$, 
defined via (\ref{Xi}), we have  
\begin{equation} \label{aXia}
\boldsymbol{a}_{n}^{\mathrm{H}}
\boldsymbol{\Xi}^{-1}\boldsymbol{a}_{n} 
= \boldsymbol{e}_{n}^{\mathrm{H}}\boldsymbol{V}\boldsymbol{D} 
\boldsymbol{V}^{\mathrm{H}}\boldsymbol{e}_{n}, 
\end{equation}
with 
\begin{equation}
\boldsymbol{D} 
= \begin{pmatrix}
\boldsymbol{\Sigma}\\
\boldsymbol{O}
\end{pmatrix}\left(
 \sigma^{2}\boldsymbol{I}_{M} 
 + v_{\mathrm{B}\to \mathrm{A}}\boldsymbol{\Sigma}^{2}
\right)^{-1}(\boldsymbol{\Sigma}, \boldsymbol{O}). 
\end{equation}
In (\ref{aXia}), $\boldsymbol{e}_{n}$ denotes the $n$th column of 
$\boldsymbol{I}_{N}$.  
Thus, Corollary~\ref{corollary_Haar_norm} and Assumption~\ref{assumption_A} imply that  
$\boldsymbol{a}_{n}^{\mathrm{H}}\boldsymbol{\Xi}^{-1}\boldsymbol{a}_{n}$ converges 
almost surely to $\gamma(v_{\mathrm{B}\to \mathrm{A}})^{-1}$  
in the large system limit. 

This observation indicates that for any $n$ 
the diagonal element~(\ref{V_A_n}) converges almost surely to 
\begin{equation} \label{V_A_n_asym} 
v_{\mathrm{A}} = v_{\mathrm{B}\to \mathrm{A}} 
- \gamma^{-1}(v_{\mathrm{B}\to \mathrm{A}})v_{\mathrm{B}\to \mathrm{A}}^{2}
\end{equation} in the large system limit. Thus,  
the marginal pdf $q_{\mathrm{A}}(x_{n})=\int q_{\mathrm{A}}(\boldsymbol{x})
d\boldsymbol{x}_{\backslash n}$ is the proper complex Gaussian pdf with 
mean $x_{n,\mathrm{A}}=[\boldsymbol{x}_{\mathrm{A}}]_{n}$ and variance 
$v_{\mathrm{A}}$, i.e.\ 
\begin{equation} \label{q_A_n}
q_{\mathrm{A}}(x_{n}) 
\propto \exp\left(
 -\frac{|x_{n}-x_{n,\mathrm{A}}|^{2}}{v_{\mathrm{A}}}
\right). 
\end{equation}

In order to present a crucial step in EP, we define the extrinsic pdf of 
$x_{n}$ as 
\begin{equation} \label{q_A_B}
q_{\mathrm{A}\to \mathrm{B}}(x_{n})
\propto \frac{q_{\mathrm{A}}(x_{n})}{q_{\mathrm{B}\to \mathrm{A}}(x_{n})}. 
\end{equation}
Let $x_{n,\mathrm{B}}$ and $v_{n,\mathrm{B}}$ denote the mean and variance of 
$x_{n}$ with respect to the pdf 
$p_{\mathrm{B}}(x_{n})\propto q_{\mathrm{A}\to \mathrm{B}}(x_{n})p(x_{n})$. 
The crucial step in EP is to update the message 
$q_{\mathrm{B}\to \mathrm{A}}(x_{n})$ so as to satisfy the moment matching 
conditions~\cite{Minka01}, 
\begin{equation} \label{matching_mean}
\mathbb{E}_{q_{\mathrm{B}}}[x_{n}] = x_{n,\mathrm{B}}, 
\end{equation}
\begin{equation} \label{matching_var}
\mathbb{V}_{q_{\mathrm{B}}}[x_{n}] 
= \frac{1}{N}\sum_{n=1}^{N}v_{n,\mathrm{B}} 
\equiv v_{\mathrm{B}}, 
\end{equation}
where the expectations are taken with respect to  
\begin{equation} \label{q_B}
q_{\mathrm{B}}(x_{n})\propto q_{\mathrm{A}\to \mathrm{B}}(x_{n})
q_{\mathrm{B}\to \mathrm{A}}^{\mathrm{new}}(x_{n}). 
\end{equation}
In (\ref{q_B}), the updated pdf $q_{\mathrm{B}\to \mathrm{A}}^{\mathrm{new}}(x_{n})$ 
is given by 
\begin{equation} \label{q_B_A_new}
q_{\mathrm{B}\to \mathrm{A}}^{\mathrm{new}}(x_{n})
\propto \exp\left(
 -\frac{|x_{n}-x_{n,\mathrm{B}\to \mathrm{A}}^{\mathrm{new}}|^{2}}
 {v_{\mathrm{B}\to \mathrm{A}}^{\mathrm{new}}} 
\right). 
\end{equation}

We first derive module~A. Using (\ref{q_B_A}) and (\ref{q_A_n}), we find 
that the extrinsic pdf (\ref{q_A_B}) reduces to  
\begin{equation} \label{q_A_B2}
q_{\mathrm{A}\to \mathrm{B}}(x_{n}) 
\propto \exp\left(
 -\frac{|x_{n}-x_{n,\mathrm{A}\to \mathrm{B}}|^{2}}{v_{\mathrm{A}\to \mathrm{B}}}
\right), 
\end{equation}
with 
\begin{equation} \label{x_A_B}
x_{n,\mathrm{A}\to \mathrm{B}} = v_{\mathrm{A}\to \mathrm{B}}\left(
 \frac{x_{n,\mathrm{A}}}{v_{\mathrm{A}}} - \frac{x_{n,\mathrm{B}\to \mathrm{A}}}
 {v_{\mathrm{B}\to \mathrm{A}}} 
\right), 
\end{equation}
\begin{equation} \label{v_A_B}
\frac{1}{v_{\mathrm{A}\to \mathrm{B}}} = \frac{1}{v_{\mathrm{A}}} 
- \frac{1}{v_{\mathrm{B}\to \mathrm{A}}}.  
\end{equation}
Substituting (\ref{V_A_n_asym}) into (\ref{v_A_B}) yields  
\begin{equation} \label{v_A_B2}
v_{\mathrm{A}\to \mathrm{B}} 
= \gamma(v_{\mathrm{B}\to \mathrm{A}}) - v_{\mathrm{B}\to \mathrm{A}}, 
\end{equation}
which results in the update rule~(\ref{module_A_var}). 
Similarly, Applying (\ref{x_A2}), (\ref{V_A_n_asym}), (\ref{v_A_B}), 
and (\ref{v_A_B2}) to (\ref{x_A_B}), we arrive at 
\begin{equation} \label{x_A_B2}
\boldsymbol{x}_{\mathrm{A}\to \mathrm{B}} 
= \boldsymbol{x}_{\mathrm{B}\to \mathrm{A}} 
+ \gamma(v_{\mathrm{B}\to \mathrm{A}})\boldsymbol{A}^{\mathrm{H}}\boldsymbol{\Xi}^{-1}
(\boldsymbol{y}-\boldsymbol{A}\boldsymbol{x}_{\mathrm{B}\to \mathrm{A}}),  
\end{equation}
which implies the update rule~(\ref{module_A_mean}). 

We next evaluate the moment matching conditions~(\ref{matching_mean}) 
and (\ref{matching_var}) to derive module~B. Substituting 
(\ref{q_B_A_new}) and (\ref{q_A_B2}) into (\ref{q_B}) yields 
\begin{equation}
q_{\mathrm{B}}(x_{n}) 
\propto \exp\left(
 - \frac{|x_{n}-\tilde{x}_{n,\mathrm{B}}|^{2}}{\tilde{v}_{\mathrm{B}}} 
\right), 
\end{equation}
with 
\begin{equation}
\tilde{x}_{n,\mathrm{B}} 
= \tilde{v}_{\mathrm{B}}\left(
 \frac{x_{n,\mathrm{A}\to \mathrm{B}}}{v_{\mathrm{A}\to \mathrm{B}}}
 + \frac{x_{n,\mathrm{B}\to \mathrm{A}}^{\mathrm{new}}}
 {v_{\mathrm{B}\to \mathrm{A}}^{\mathrm{new}}}
\right), 
\end{equation}
\begin{equation}
\frac{1}{\tilde{v}_{\mathrm{B}}} 
= \frac{1}{v_{\mathrm{A}\to \mathrm{B}}} 
+ \frac{1}{v_{\mathrm{B}\to \mathrm{A}}^{\mathrm{new}}}.   
\end{equation}
Using the moment matching conditions~(\ref{matching_mean}) and 
(\ref{matching_var}), we arrive at the update rules (\ref{module_B_mean}) 
and (\ref{module_B_var}) in module~B, 
\begin{equation}
\boldsymbol{x}_{\mathrm{B}\to \mathrm{A}}^{\mathrm{new}}
= v_{\mathrm{B}\to \mathrm{A}}^{\mathrm{new}}\left(
 \frac{\boldsymbol{x}_{\mathrm{B}}}{v_{\mathrm{B}}} 
 - \frac{\boldsymbol{x}_{\mathrm{A}\to \mathrm{B}}}{v_{\mathrm{A}\to \mathrm{B}}}
\right), 
\end{equation}
\begin{equation}
\frac{1}{v_{\mathrm{B}\to \mathrm{A}}^{\mathrm{new}}} 
= \frac{1}{v_{\mathrm{B}}} - \frac{1}{v_{\mathrm{A}\to \mathrm{B}}}.  
\end{equation}

\section{Proof of Lemma~\ref{lemma_eta}}
\label{proof_lemma_eta} 
We utilize the following technical lemma: 
\begin{lemma} \label{lemma_generating_function}
We define the cumulant generating function $\chi_{t}:\mathbb{C}\to\mathbb{R}$ 
of the posterior distribution of $x_{n}$ as 
\begin{equation}
\chi_{t}(z) 
= \frac{v_{\mathrm{A}\to \mathrm{B}}^{t}}{2}\ln\mathbb{E}_{x_{n}}\left[
 \exp\left(
  -\frac{|z-x_{n}|^{2}}{v_{\mathrm{A}\to \mathrm{B}}^{t}}
 \right)
\right] + \frac{|z|^{2}}{2}. 
\end{equation}
Then, $\chi_{t}$ is twice continuously differentiable with respect to 
$\Re[z]$ and $\Im[z]$, and satisfies 
\begin{equation}
\frac{\partial\chi_{t}}{\partial\Re[z]} 
= \Re[\tilde{\eta}_{t}(z)], 
\quad \frac{\partial\chi_{t}}{\partial\Im[z]} 
= \Im[\tilde{\eta}_{t}(z)], 
\end{equation}
\begin{equation}
\frac{v_{\mathrm{A}\to \mathrm{B}}^{t}}{2}
\frac{\partial^{2}\chi_{t}}{\partial\Re[z]^{2}} 
= \mathbb{E}\left[
 \left. 
  (\Re[x_{n}] - \Re[\tilde{\eta}_{t}(z)])^{2}
 \right| 
 z
\right], 
\end{equation}
\begin{equation}
\frac{v_{\mathrm{A}\to \mathrm{B}}^{t}}{2}
\frac{\partial^{2}\chi_{t}}{\partial\Im[z]^{2}} 
= \mathbb{E}\left[
 \left. 
  (\Im[x_{n}] - \Im[\tilde{\eta}_{t}(z)])^{2}
 \right| 
 z
\right],  
\end{equation}
\begin{equation} \label{correlation}
\frac{v_{\mathrm{A}\to \mathrm{B}}^{t}}{2}
\frac{\partial^{2}\chi_{t}}{\partial\Re[z]\partial\Im[z]} 
= \mathbb{E}\left[
 \left. 
  \Re[x_{n}]\Im[x_{n}]
 \right| z
\right]
- \Re[\tilde{\eta}_{t}(z)]\Im[\tilde{\eta}_{t}(z)]. 
\end{equation}
\end{lemma}
\begin{IEEEproof}
The former statement follows from Assumption~\ref{assumption_x} and 
the dominated convergence theorem. The latter statement is obtained 
by calculating the derivatives of $\chi_{t}$ directly. 
\end{IEEEproof}

We first prove the Lipschitz-continuity of $\tilde{\eta}_{t}$. 
We need to prove that all first-order derivatives of $\Re[\tilde{\eta}_{t}]$ 
and $\Im[\tilde{\eta}_{t}]$ are bounded. 
From Assumption~\ref{assumption_posterior} and 
Lemma~\ref{lemma_generating_function}, it is sufficient to confirm that 
(\ref{correlation}) is almost surely bounded. Using the Cauchy-Schwarz 
inequality yields 
\begin{IEEEeqnarray}{rl}
&\left(
  \mathbb{E}\left[
  \left. 
   \Re[x_{n}]\Im[x_{n}]
  \right| z
 \right]
 - \Re[\tilde{\eta}_{t}(z)]\Im[\tilde{\eta}_{t}(z)]
\right)^{2} \nonumber \\
=& \left(
 \mathbb{E}\left[
  \left. 
   (\Re[x_{n}] - \Re[\tilde{\eta}_{t}(z)])
   (\Im[x_{n}] - \Im[\tilde{\eta}_{t}(z)])
  \right| z
 \right]
\right)^{2} \nonumber \\ 
\leq&\mathbb{E}\left[
 \left. 
  (\Re[x_{n}] - \Re[\tilde{\eta}_{t}(z)])^{2}
 \right| 
 z
\right]
\mathbb{E}\left[
 \left. 
  (\Im[x_{n}] - \Im[\tilde{\eta}_{t}(z)])^{2}
 \right| 
 z
\right], \nonumber \\
\end{IEEEeqnarray}
which is almost surely bounded, because of 
Assumption~\ref{assumption_posterior}. 
Thus, $\tilde{\eta}_{t}$ is Lipschitz-continuous. 

We next prove (\ref{Stein}) and (\ref{eta_tilde}). 
For notational convenience, we write $\tilde{\eta}_{t}(x_{n}+z)$ as 
$\tilde{\eta}$. By definition, we have 
\begin{equation}
z^{*}\tilde{\eta}
= \Re[z]\Re[\tilde{\eta}] + \Im[z]\Im[\tilde{\eta}]
+ \mathrm{i}(\Re[z]\Im[\tilde{\eta}] - \Im[z]\Re[\tilde{\eta}]). 
\end{equation}
Since $\Re[z]$ and $\Im[z]$ are independent Gaussian random 
variables with zero-mean and variance $v_{\mathrm{A}\to \mathrm{B}}^{t}/2$, 
using Stein's lemma~\cite{Stein72} yields 
\begin{IEEEeqnarray}{rl}
\mathbb{E}_{z}\left[
 z^{*}\tilde{\eta}
\right]
=& \frac{v_{\mathrm{A}\to \mathrm{B}}^{t}}{2}\mathbb{E}_{z}\left[
 \frac{\partial \Re[\tilde{\eta}]}{\partial\Re[z]}
 + \frac{\partial \Im[\tilde{\eta}]}{\partial\Im[z]}
\right] \nonumber \\ 
&+ \frac{\mathrm{i}v_{\mathrm{A}\to \mathrm{B}}^{t}}{2}\mathbb{E}_{z}\left[
 \frac{\partial \Im[\tilde{\eta}]}{\partial\Re[z]}
 - \frac{\partial \Re[\tilde{\eta}]}{\partial\Im[z]}
\right] \nonumber \\ 
=& v_{\mathrm{A}\to\mathrm{B}}^{t}\mathbb{E}_{z}\left[
 \frac{\partial}{\partial z}\left(
  \Re[\tilde{\eta}] + \mathrm{i}\Im[\tilde{\eta}] 
 \right)
\right], \label{z_eta}
\end{IEEEeqnarray}
where $\partial/\partial z$ denotes the Wirtinger 
derivative~(\ref{Wirtinger_derivative}). This implies that (\ref{Stein}) 
holds. Furthermore, 
applying Lemma~\ref{lemma_generating_function} to the former expression 
in (\ref{z_eta}), we obtain 
\begin{equation}
\mathbb{E}_{z}\left[
 z^{*}\tilde{\eta}
\right]
= \mathbb{E}_{z}\left[
  |x_{n} - \tilde{\eta}_{t}(x_{n}+z)|^{2}
\right]. 
\end{equation}
Taking the expectation of both sides over $x_{n}$, we arrive at 
Lemma~\ref{lemma_eta}.

\section{Proof of Lemma~\ref{lemma_conditional_distribution}}
\label{proof_conditional_distribution}
For $\hat{\boldsymbol{V}}
=\boldsymbol{V}\boldsymbol{\Phi}_{\boldsymbol{X}}\in{\cal U}_{N}$, 
we first prove the identity 
\begin{equation} \label{V_hat_condition}
\hat{\boldsymbol{V}}
= \left(
 \boldsymbol{\Phi}_{\boldsymbol{Y}}^{\parallel}, 
 \boldsymbol{\Phi}_{\boldsymbol{Y}}^{\perp}\tilde{\boldsymbol{V}}
\right), 
\end{equation}
with some unitary matrix $\tilde{\boldsymbol{V}}\in{\cal U}_{N-t}$. 

Since $\boldsymbol{V}$ is unitary, using the constraint~(\ref{constraint}) 
yields $\boldsymbol{X}^{\mathrm{H}}\boldsymbol{X}
=\boldsymbol{Y}^{\mathrm{H}}\boldsymbol{Y}$. 
This implies that 
$\boldsymbol{X}$ and $\boldsymbol{Y}$ have identical singular values and 
right-singular vectors, i.e.\ $\boldsymbol{X}=\boldsymbol{\Phi}_{\boldsymbol{X}}
(\boldsymbol{\Sigma}_{\boldsymbol{X}}, \boldsymbol{O})^{\mathrm{T}}
\boldsymbol{\Psi}_{\boldsymbol{X}}^{\mathrm{H}}$ and 
$\boldsymbol{Y}=\boldsymbol{\Phi}_{\boldsymbol{Y}}
(\boldsymbol{\Sigma}_{\boldsymbol{Y}}, \boldsymbol{O})^{\mathrm{T}}
\boldsymbol{\Psi}_{\boldsymbol{Y}}^{\mathrm{H}}$ with 
$\boldsymbol{\Sigma}_{\boldsymbol{X}}=\boldsymbol{\Sigma}_{\boldsymbol{Y}}$ and 
$\boldsymbol{\Psi}_{\boldsymbol{X}}=\boldsymbol{\Psi}_{\boldsymbol{Y}}$. 
Since $\boldsymbol{\Sigma}_{\boldsymbol{X}}=\boldsymbol{\Sigma}_{\boldsymbol{Y}}$ is 
assumed to be invertible, applying these SVDs to the 
constraint~(\ref{constraint}) yields 
\begin{equation} \label{constraint_tmp}
\boldsymbol{\Phi}_{\boldsymbol{Y}}^{\parallel}
= \boldsymbol{V}\boldsymbol{\Phi}_{\boldsymbol{X}} 
\begin{pmatrix}
\boldsymbol{I}_{t} \\ 
\boldsymbol{O}_{N\times(N-t)}
\end{pmatrix}. 
\end{equation} 

Consider the partition $\hat{\boldsymbol{V}}=(\hat{\boldsymbol{V}}_{0}, 
\hat{\boldsymbol{V}}_{1})$, 
with $\hat{\boldsymbol{V}}_{0}\in\mathbb{C}^{N\times t}$ and 
$\hat{\boldsymbol{V}}_{1}\in\mathbb{C}^{N\times(N-t)}$. 
From (\ref{constraint_tmp}) we have $\hat{\boldsymbol{V}}_{0}
=\boldsymbol{\Phi}_{\boldsymbol{Y}}^{\parallel}$. Thus,  
the orthogonality between the columns of $\hat{\boldsymbol{V}}_{0}$ and 
$\hat{\boldsymbol{V}}_{1}$ implies the structure~(\ref{V_hat_condition}) 
with some matrix $\tilde{\boldsymbol{V}}\in\mathbb{C}^{(N-t)\times(N-t)}$. 
Furthermore, from the orthonormality between the columns of 
$\hat{\boldsymbol{V}}_{1}$ we find that $\tilde{\boldsymbol{V}}$ is 
a unitary matrix. Thus, (\ref{V_hat_condition}) is correct. 

We next prove that (\ref{V_hat_condition}) is equivalent to 
the RHS of (\ref{V_conditional_distribution}). Substituting 
(\ref{V_hat_condition}) into 
$\boldsymbol{V}=\hat{\boldsymbol{V}}
\boldsymbol{\Phi}_{\boldsymbol{X}}^{\mathrm{H}}$ yields 
\begin{equation} \label{V_conditional_distribution_tmp}
\boldsymbol{V}
= \boldsymbol{\Phi}_{\boldsymbol{Y}}^{\parallel}
(\boldsymbol{\Phi}_{\boldsymbol{X}}^{\parallel})^{\mathrm{H}}  
+ \boldsymbol{\Phi}_{\boldsymbol{Y}}^{\perp}
\tilde{\boldsymbol{V}}(\boldsymbol{\Phi}_{\boldsymbol{X}}^{\perp})^{\mathrm{H}}. 
\end{equation}
It is straightforward to confirm that the first term on the RHS of 
(\ref{V_conditional_distribution}) reduces to 
$\boldsymbol{\Phi}_{\boldsymbol{Y}}^{\parallel}
(\boldsymbol{\Phi}_{\boldsymbol{X}}^{\parallel})^{\mathrm{H}}$, by using the 
SVDs of $\boldsymbol{X}$ and $\boldsymbol{Y}$ with 
$\boldsymbol{\Sigma}_{\boldsymbol{X}}=\boldsymbol{\Sigma}_{\boldsymbol{Y}}$ 
and $\boldsymbol{\Psi}_{\boldsymbol{X}}=\boldsymbol{\Psi}_{\boldsymbol{Y}}$.    

To complete the proof of Lemma~\ref{lemma_conditional_distribution}, 
we prove that $\tilde{\boldsymbol{V}}\in{\cal U}_{N-t}$ is a Haar matrix 
independent of $\boldsymbol{X}$ and $\boldsymbol{Y}$. 
Since the Haar matrix $\boldsymbol{V}$ is bi-unitarily invariant, 
we have $\boldsymbol{V}\boldsymbol{\Phi}_{\boldsymbol{X}}\sim \boldsymbol{V}$. 
Thus, without loss of generality, (\ref{constraint_tmp}) allows us to assume 
$\boldsymbol{X}=(\boldsymbol{I}_{t}, \boldsymbol{O})^{\mathrm{T}}$ 
in the constraint~(\ref{constraint}). 
Under this assumption, conditioning on $\boldsymbol{X}$ and $\boldsymbol{Y}$ 
is equivalent to conditioning the first $t$ columns $\boldsymbol{V}_{0}$ of 
$\boldsymbol{V}$. 

Consider the following structure:   
\begin{equation} \label{V_structure}
\boldsymbol{V} 
= \left(
 \boldsymbol{V}_{0}, \boldsymbol{\Phi}_{\boldsymbol{V}_{0}}^{\perp}
 \tilde{\boldsymbol{V}}
\right). 
\end{equation}
We prove that $\boldsymbol{V}$ is Haar-distributed if and only if 
$\tilde{\boldsymbol{V}}$ is a Haar matrix and independent of 
$\boldsymbol{V}_{0}$. Since $\boldsymbol{X}$ and $\boldsymbol{Y}$ depend 
on $\boldsymbol{V}$ only through $\boldsymbol{V}_{0}$, we arrive at 
Lemma~\ref{lemma_conditional_distribution}. 

For any deterministic unitary 
matrix $\boldsymbol{\Phi}\in{\cal U}_{N}$, it is known that the left-invariance 
$\boldsymbol{\Phi}\boldsymbol{V}\sim\boldsymbol{V}$ induces the Haar measure 
on the unitary group of dimension~$N$ satisfying 
$\boldsymbol{V}\sim\boldsymbol{V}^{\mathrm{H}}$, so that 
we have the right-invariance $\boldsymbol{V}\boldsymbol{\Psi}\sim 
\boldsymbol{V}^{\mathrm{H}}\boldsymbol{\Psi}
=(\boldsymbol{\Psi}^{\mathrm{H}}\boldsymbol{V})^{\mathrm{H}}
\sim\boldsymbol{V}^{\mathrm{H}}\sim\boldsymbol{V}$ for any 
deterministic $\boldsymbol{\Psi}\in{\cal U}_{N}$. Thus,  
we only consider the left-invariance 
$\boldsymbol{\Phi}\boldsymbol{V}\sim\boldsymbol{V}$. 

There is some unitary matrix $\boldsymbol{U}_{\boldsymbol{V}_{0}}
\in{\cal U}_{N-t}$ such that 
$\boldsymbol{\Phi}\boldsymbol{\Phi}_{\boldsymbol{V}_{0}}^{\perp}
= \boldsymbol{\Phi}_{\boldsymbol{\Phi}\boldsymbol{V}_{0}}^{\perp}
\boldsymbol{U}_{\boldsymbol{V}_{0}}$ holds, because of 
\begin{equation}
\boldsymbol{\Phi}\boldsymbol{\Phi}_{\boldsymbol{V}_{0}}^{\perp}
(\boldsymbol{\Phi}\boldsymbol{\Phi}_{\boldsymbol{V}_{0}}^{\perp})^{\mathrm{H}}
= \boldsymbol{\Phi}(\boldsymbol{I}_{N} - \boldsymbol{V}_{0}
\boldsymbol{V}_{0}^{\mathrm{H}})\boldsymbol{\Phi}^{\mathrm{H}}
= \boldsymbol{P}_{\boldsymbol{\Phi}\boldsymbol{V}_{0}}^{\perp}. 
\end{equation} 
This implies that (\ref{V_structure}) satisfies 
\begin{equation} 
\boldsymbol{\Phi}\boldsymbol{V}
= \left(
 \boldsymbol{\Phi}\boldsymbol{V}_{0}, 
 \boldsymbol{\Phi}_{\boldsymbol{\Phi}\boldsymbol{V}_{0}}^{\perp}
 \boldsymbol{U}_{\boldsymbol{V}_{0}}\tilde{\boldsymbol{V}}
\right),
\end{equation}
which indicates that $\boldsymbol{\Phi}\boldsymbol{V}\sim\boldsymbol{V}$ 
holds if and only if $(\boldsymbol{\Phi}\boldsymbol{V}_{0}, 
\boldsymbol{U}_{\boldsymbol{V}_{0}}\tilde{\boldsymbol{V}})\sim
(\boldsymbol{V}_{0}, \tilde{\boldsymbol{V}})$ is satisfied. Since 
$\boldsymbol{V}_{0}$ is Haar-distributed, 
$\boldsymbol{\Phi}\boldsymbol{V}\sim\boldsymbol{V}$ holds if and only if   
$\tilde{\boldsymbol{V}}$ is a Haar matrix independent of 
$\boldsymbol{V}_{0}$. Thus, Lemma~\ref{lemma_conditional_distribution} holds.

\balance

\ifCLASSOPTIONcaptionsoff
  \newpage
\fi



\bibliographystyle{IEEEtran}
\bibliography{IEEEabrv,kt-it2017_1}
\end{document}